\documentclass[12pt]{article}
\usepackage{amssymb}
\usepackage{amsmath}
\usepackage[dvips]{graphicx}
\usepackage{epsfig}

\def\ka{g}

\begin{document}

\baselineskip=18pt
\numberwithin{equation}{section}
\allowdisplaybreaks  

\pagestyle{myheadings}

\thispagestyle{empty}

\vspace*{-2cm}
\begin{flushright}
{\tt arXiv:0809.0507v1}\\
ITFA 2008-34\\
\end{flushright}

\vspace*{2.5cm}
\begin{center}
 {\LARGE  The chiral ring of AdS${}_3$/CFT${}_2$ and the attractor mechanism\\}
 \vspace*{1.7cm}
Jan de Boer, Jan Manschot, Kyriakos Papadodimas and Erik Verlinde\\

 \vspace*{1.0cm}
   Institute for Theoretical Physics, University of Amsterdam\\
Valckenierstraat 65, 1018 XE Amsterdam, The Netherlands\\
 \vspace*{0.8cm}
 {\tt J.deBoer, J.Manschot, K.Papadodimas, E.P.Verlinde@ uva.nl}
\end{center}
\vspace*{1.0cm}

\noindent We study the moduli dependence of the chiral ring in ${\cal
  N}=(4,4)$ superconformal field theories, with special emphasis on
those CFT's that are dual to type IIB string theory on
AdS${}_3\times$S${}^3\times$X${}_4$. The chiral primary operators are
sections of vector bundles, whose connection describes the operator
mixing under motion on the moduli space. This connection can be
exactly computed using the constraints from ${\cal N}=(4,4)$
supersymmetry. Its curvature can be determined using the $tt^*$
equations, for which we give a derivation in the physical theory which
does not rely on the topological twisting. We show that for ${\cal
  N}=(4,4)$ theories the chiral ring is covariantly constant over the
moduli space, a fact which can be seen as a non-renormalization
theorem for the three-point functions of chiral primaries in
AdS$_3$/CFT$_2$. From the spacetime point of view our analysis has the
following applications. First, in the case of a D1/D5 black string, we
can see the matching of the attractor flow in supergravity to RG-flow
in the boundary field theory perturbed by irrelevant operators, to
first order away from the fixed point. Second, under spectral flow the
chiral primaries become the Ramond ground states of the CFT. These
ground states represent the microstates of a small black hole in five
dimensions consisting of a D1/D5 bound state. The connection that we
compute can be considered as an example of Berry's phase for the
internal microstates of a supersymmetric black hole.

\newpage
\setcounter{page}{1}

\tableofcontents

\section{Introduction}

The AdS$_3$/CFT$_2$ correspondence \cite{Maldacena:1997re} is one of
the best understood holographic dualities and has been very useful for
the analysis of black holes in string theory. While it has been
studied in great detail by now, most of the computations have been
performed in special weakly-coupled limits. The AdS$_3$/CFT$_2$ is
characterized by a parameter space ${\cal M}$ which corresponds to the
expectation values of the scalar fields in the bulk, or equivalently
to the position on the moduli space of the boundary CFT.  There are
special points on ${\cal M}$ where the boundary CFT is weakly coupled
and others where the holographic dual string theory is in the
perturbative regime. At a generic point on ${\cal M}$, none of the two
descriptions is weakly coupled and it is difficult to make any
explicit computations. Is there anything we can say about the theory
in the interior of its moduli space?

In this paper, whenever we speak of the AdS$_3$/CFT$_2$
correspondence, we will have the duality between type IIB on
AdS${}_3\times$S${}^3\times$X${}_4$ and suitable ${\cal N}=(4,4)$
superconformal field theories in mind. These CFT's are believed to
be related to a sigma model whose target space is a deformation of
the symmetric product $X^N/ S_N $, where $ X=T^4$ or $K3$. This is
a hyperk\"ahler space and such sigma models are indeed compatible
with ${\cal N}=(4,4)$ sypersymmetry.  It is a natural assumption
that at all points on ${\cal M}$ the theory has a boundary
description in terms of an ${\cal N}=(4,4)$ superconformal field
theory, which may be strongly coupled. Such theories have a sector
protected by supersymmetry, the chiral ring\cite{Lerche:1989uy},
which can be studied exactly even away from the weak-coupling
limits. In this paper we analyze the moduli dependence of the
chiral ring of ${\cal
  N}=(4,4)$ superconformal field theories, mainly motivated by its
relevance for the boundary CFT that appears in the AdS$_3$/CFT$_2$
correspondence. Our analysis is exact everywhere on the moduli space,
since we only assume that the ${\cal N}=(4,4)$ superconformal
structure of the theory is preserved and that generically the number
of chiral primaries does not jump as we move on ${\cal M}$.  This
allows us to make some exact statements about the theory in the regime
of strong coupling and for finite $N$.

The chiral ring of a superconformal field theory depends on the
moduli in two ways. First, the chiral primaries mix among
themselves as we change the parameters of the theory. Technically
this means that the chiral primary operators are sections of
vector bundles over the moduli space, which can have nontrivial
curvature. Second, the multiplication between the chiral
primaries, expressed in terms of the structure constants
$C^i_{jk}$, may also be moduli dependent. Supersymmetry imposes
strong constraints on the structure of the chiral ring and the way
it behaves under a change of the coupling constants. The case of
${\cal N}=(2,2)$ superconformal theories has been extensively
studied and the supersymmetry constraints are expressed in terms
of the $tt^*$ equations
\begin{equation}
 R_{i \overline{j}}\equiv [\nabla_i,\nabla_{\overline{j}}]\simeq-
 [C_i,\overline{C_j}]
\label{tteq}
\end{equation}
which give the curvature of the bundles of chiral primaries in
terms of the chiral ring coefficients. These equations were
originally derived by Cecotti and Vafa using a method called
topological anti-topological fusion
\cite{Cecotti:1991me},\cite{Bershadsky:1993cx} which is based on
the topological twisting of the superconformal theory. However as
we show they can also be derived using ordinary conformal
perturbation theory in the untwisted theory\footnote{The original
  derivation is more general since it also works for non-conformal
  ${\cal N}=(2,2)$ theories.}.

The $tt^*$ equations are also relevant for theories with ${\cal
  N}=(4,4)$ supersymmetry, such as the boundary theory in the
  class of
AdS$_3$/CFT$_2$ correspondences we consider here, once we
appropriately project to their ${\cal
  N}=(2,2)$ subalgebras.  In ${\cal N}=(4,4)$ theories a simple
observation leads to the following additional constraint
\begin{equation}
  \nabla C^k_{ij} = 0
\label{crder}
\end{equation}
where $\nabla$ represents the covariant derivative along any
marginal deformation. This is true for the following reason: in an
${\cal
  N}=(2,2)$ theory it is known that the chiral ring coefficients
depend on the moduli holomorphically, so they are independent of
anti-holomorphic deformations
\begin{equation}
  \nabla_{\overline{m}} C^k_{ij}=0
\label{antih}
\end{equation}
while in general $\nabla_m C^k_{ij} \neq 0$. An ${\cal N}=(4,4)$
theory has many inequivalent ${\cal N}=(2,2)$ subalgebras. It can be
shown that in an ${\cal N}=(4,4)$ theory any marginal deformation can
be written as an anti-holomorphic deformation with respect to some
${\cal N}=(2,2)$ subalgebra, and then \eqref{crder} follows from
\eqref{antih}.

The result \eqref{crder} can be interpreted as a non-renormalization
theorem for the 3-point functions of chiral primaries in AdS$_3$/
CFT$_2$, which explains the agreement of computations performed at
different points on the moduli space \cite{Gaberdiel:2007vu},
\cite{Dabholkar:2007ey} and also \cite{Pakman:2007hn},
\cite{Taylor:2007hs}.  This is the analogue of the non-renormalization
theorem \cite{Intriligator:1998ig}, \cite{Intriligator:1999ff} for
3-point functions of chiral primaries in AdS$_5$/CFT$_4$ which
explained the agreement of the weakly and strongly coupled
computations \cite{Lee:1998bxa}. Our arguments do not depend on taking
a large $N$ limit, so the 3-point functions of chiral primaries have
to be (covariantly) constant even at finite $N$. It is easy to show
that more generally {\it extremal} correlators of chiral primaries are
also not renormalized as we change the moduli.

Combining the non-renormalization of the chiral ring coefficients with
the $tt^*$ equations we can derive a stronger statement. By acting
with $\nabla$ on both sides of \eqref{tteq} and using \eqref{crder} we
conclude that the curvature of the bundle of chiral primaries is
covariantly constant
\begin{equation}
  \nabla  R_{i\overline{j}} = 0
\end{equation}
We also know \cite{Seiberg:1988pf},\cite{Cecotti:1990kz} that for
${\cal N}=(4,4)$ theories the moduli space is locally a symmetric
space of the form
\begin{equation}
  {SO(4,n) \over SO(4) \times SO(n)}
\end{equation}
for some $n$.  Bundles with covariantly constant curvature over
symmetric spaces are called {\it homogeneous bundles} and their
geometry is completely determined in terms of some basic
group-theoretic data. In some ${\cal N}=(4,4)$ theories, such
those that arise in the AdS$_3$/CFT$_2$ correspondence, if we know
the number of chiral primaries of a given conformal dimension, it
is rather straightforward to fit them into homogeneous bundles.
Then the exact connection and curvature on these bundles is
determined without any further input from the dynamics of the CFT.
In this sense we can compute the exact mixing of chiral primary
operators as we move on the moduli space, even at strong coupling.

An application of our analysis from the spacetime point of view is
that it realizes a connection between the attractor flow in
supergravity and RG-flow in the boundary field theory, in a
certain toy-model, as we now explain. Extremal black holes in
supergravity exhibit a remarkable phenomenon, called the attractor
mechanism \cite{Ferrara:1995ih}.  The values of many of the scalar
fields near the horizon of the black hole are fixed by its
electric and magnetic charges and completely independent of their
values at spatial infinity.  The same black holes can be described
by appropriate bound states of D-branes. The worldvolume theory of
these branes is an open string theory, which flows to a conformal
field theory at low energies. This raises a natural question,
namely what is the meaning of the attractor flow in the D-brane
picture of the black hole?

As is well known, the AdS throat of the supergravity solution is
holographically dual to the conformal IR fixed point of the effective
field theory describing the excitations on the D-branes that create
the black hole. The AdS/CFT correspondence is derived by taking the
low energy limit which on the supergravity side is equivalent to
keeping only the near horizon AdS geometry. In that region of the
supergravity solution the moduli have already reached their attractor
values. As a result the attractor mechanism is not visible in the
usual AdS/CFT correspondence.

Clearly, to see the attractor flow we have to move outside the AdS
throat towards the asymptotically flat region. This requires an
extension of AdS/CFT beyond the strict $\alpha'\rightarrow 0$
limit, where it turns into a duality between closed string theory
and open string theory. In the open string language the system is
described by a stack of D-branes in flat space, and on the other
hand, in the large $N$ limit, we can consider the closed string
description where we replace the D-branes by a curved closed
string background. On the boundary side going outside the AdS
throat is described by deforming the CFT by irrelevant operators.
From this perspective we expect to see the attractor flow as
RG-flow on the worldvolume theory of the branes towards the IR
fixed point.

It is not easy to make this relation precise, since going outside the
AdS throat means that there is no honest decoupling between open and
closed string modes. In particular, since the open strings living on
the branes are not decoupled from the bulk closed string modes it is
not clear what we mean by the ``boundary theory''.  However as we
approach the IR fixed point, this coupling should become less and less
important. In this sense we expect that at least near the fixed point
it should be possible to describe the theory on the branes in terms of
an effective field theory flowing to a CFT in the IR. In view of these
conceptual difficulties we will only consider the first order
perturbation away from the conformal point towards the UV, which
should correspond to the final stages of the attractor flow.
\begin{figure}
  \epsfig{file=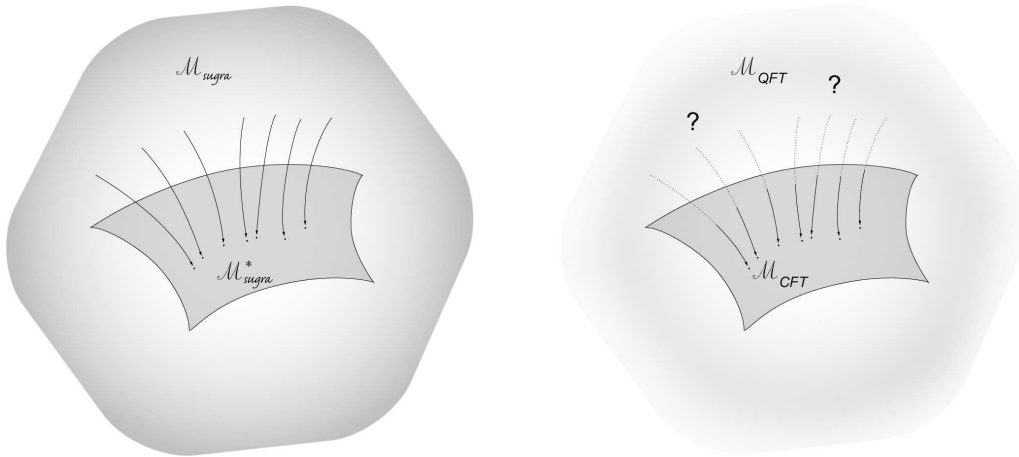,height=6cm,trim= -60 0 0 0}
  \caption{Attractor flow in supergravity (left) and RG-flow on the
    worldvolume of the branes (right).}
  \label{flow}
\end{figure}
More precisely, as shown in figure \ref{flow}, let us call ${\cal
  M}_{sugra}$ the moduli space of supergravity and ${\cal
  M}_{sugra}^*\subset {\cal M}_{sugra}$ the attractor submanifold for
given charges\footnote{We would like to remind that even in Calabi-Yau
  compactifications of type II, while the attractor equations fix the
  vector multiplets to discrete points, the hypermultiplets are
  unfixed, so also in this case there is a continuous family of
  attractor points parametrized by the hypermultiplet moduli space.}.
On the boundary side we have a family of effective quantum field
theories characterized by a moduli space ${\cal M}_{QFT}$ which
flow in the IR to a family of conformal field theories with moduli
space ${\cal
  M}_{CFT}\subset {\cal M}_{QFT}$. According to the AdS/CFT
correspondence the moduli spaces ${\cal
  M}^*_{sugra}$ and ${\cal M}_{CFT}$ should be identical\footnote{This has
  been demonstrated in some examples of AdS$_3$/CFT$_2$
  \cite{Dijkgraaf:1998gf}. We would expect the same for other cases,
  such as the MSW CFT\cite{Maldacena:1997de}. In the case of 4d
  black holes and AdS$_2$/CFT$_1$ the equivalent statement would be
  that the ``moduli space'' of the superconformal quantum mechanics
  must be the same as the hypermultiplet moduli space. It would be
  interesting to give a more precise meaning to this statement.}.
Moreover, matching the final stages of the attractor flow to
RG-flow means that the normal bundle of ${\cal M}_{sugra}^*$
inside ${\cal M} _{sugra}$ should have the same structure as that
of ${\cal M}_{CFT}$ inside ${\cal
  M}_{QFT}$\footnote{\label{corrections}In general the geometry of ${\cal M}_{sugra}^*$
  will receive corrections beyond supergravity, which have to be taken
  into account in order to achieve a precise matching with the CFT
  moduli space. This does
  not happen in the D1/D5 system due to the extended
  supersymmetry.}. In particular, this means that the dimensionality of the bundles
should agree, in other words - we should have the same number of
irrelevant operators as the number of scalar moduli fixed by the
attractor mechanism, and in addition the connection on the two
bundles should be the same.

This picture is easy to check in the simple case of the attractor flow
near an extremal black string in six dimensions. In this case the
boundary CFT is the one appearing in the AdS$_3$/CFT$_2$
correspondence. As we will see the irrelevant operators which preserve
supersymmetry are descendants of certain fields in the chiral
ring. Hence, their number can be counted and moreover the connection
and geometry of their bundle can be exactly computed using our general
analysis. The result that we find on the CFT side agrees with the
predictions from the attractor flow in supergravity.

Finally let us mention another interpretation of the geometry of
the chiral ring that we study in this paper. Spectral flow relates
the chiral primaries of the CFT to Ramond ground states. In the
D1/D5 CFT, the Ramond ground states have the following
interpretation. We consider IIB compactified on $K3 \times S^1$
and a bound state of D1/D5 branes wrapped on the internal
manifold. This looks like a small supersymmetric black hole in
five dimensions. The Ramond ground states of the CFT represent the
internal microstates of the black hole. If we adiabatically change
the moduli of the compactification the microstates will mix among
themselves, as is well known from the non-abelian generalization
of Berry's phase for quantum mechanical systems with degenerate
microstates. The connection for the chiral primaries is related to
the connection of the Ramond ground states over the moduli space,
in other words it yields a geometric phase for the internal
microstates of the black hole.

In the first half of the paper we review background material. In
section \ref{sec:reviewchiral} we review some basic facts about the
chiral ring in superconformal field theories. In section
\ref{sec:cftdeform} we discuss the deformation of conformal field
theories by marginal operators and the associated connections for the
bundle of operators over the moduli space. In section
\ref{sec:review22} we review basic results for the connection of the
bundle of chiral primaries for the case of ${\cal N}=(2,2)$ theories
and show how the $tt^*$ equations follow from conformal perturbation
theory. In section \ref{sec:n4algebra} we introduce the ${\cal
  N}=(4,4)$ algebra and discuss its basic properties.  In section
\ref{sec:mainr} we show that the 3-point functions in ${\cal N}=(4,4)$
theories are covariantly constant and we compute the curvature for the
bundle of chiral primaries. In section \ref{sec:attract} we present
the relevance of our computation for the connection between the
attractor flow and RG-flow.  In section \ref{sec:bhbp} we discuss how
the connection for chiral primaries is related to Berry's phase for
black hole microstates.  In section \ref{sec:conclu} we summarize our
results and discuss some possibilities for future research.

\section{AdS$_3$/CFT$_2$ and its chiral ring}
\label{sec:reviewchiral}
\subsection{Generalities}

We can derive the AdS$_3$/CFT$_2$ correspondence with 16 supercharges
by starting with IIB string theory compactified on $X$, where $X=T^4$
or $K3$ and considering a BPS black string in six dimensions,
consisting of a bound state of D1 strings and D5 branes wrapped on
$X$. By taking a low energy decoupling limit of this system we find
the duality between IIB on
\begin{equation}
  AdS_3 \times S^3 \times X
\end{equation}
and a two dimensional CFT with ${\cal N}=(4,4)$ supersymmetry and
$SU(2)^R_{left}\times SU(2)^R_{right}$ as its current
algebra. Excluding the center of mass degrees of freedom, the level
$k$ and central charge $c$ are given by
\begin{equation}
  k = Q_1Q_5 \qquad c =6Q_1 Q_5.
\label{centralcharge}
\end{equation}
This CFT can be understood as a supersymmetric sigma model whose
target space is a resolution of the symmetric product $X^{N}/S_N$
with $N=Q_1 Q_5$, which is moduli space of instantons of degree
$Q_1$ of a $U(Q_5)$ gauge theory living on $X$.

The AdS$_3$/CFT$_2$ correspondence is characterized by the integer
$c$ and by a set of continuous parameters determined by the
background values of the moduli fields of IIB. In other words, the
correspondence has a moduli space\footnote{Similarly the
AdS$_5$/CFT$_4$ has the
  discrete parameter $N$ and the continuous parameter $\tau = {\theta
    \over 2 \pi} + i {4 \pi \over g_{YM}^2} $.} ${\cal M}$.  This
moduli space is visible on the boundary side as the moduli space
of the conformal field theory ${\cal M}_{CFT}$ and on the bulk
side as the moduli space ${\cal M}_{sugra}^*$ of possible values
of the scalar fields near the horizon of the black string in 6d.
The local structure of the moduli space is exactly computable from
both sides of the duality\cite{Dijkgraaf:1998gf} and it is of the
form
\begin{equation}
  {\cal M}\simeq {SO(4,n) \over SO(4)\times SO(n) }
\end{equation}
where $n=5$ for $X=T^4$ and $n=21$ for $X=K3$.

Notice that this is a local statement. The global structure of ${\cal
  M}$ is more complicated\cite{Dijkgraaf:1998gf},\cite{Seiberg:1999xz}
,\cite{Larsen:1999uk} and there are points where the CFT is
singular. In this paper we will only consider local properties and
ignore all subtleties related to the global structure of the moduli
space and possible monodromies around singularities.

There are points of ${\cal M}$ where the boundary CFT is weakly
coupled. It is believed that there is a point where the CFT can be
described as a symmetric orbifold CFT
\cite{Vafa:1995zh},\cite{Strominger:1996sh}, which is analogous to
the $\lambda\rightarrow 0$ limit in AdS$_5$/CFT$_4$. There are
other points of ${\cal M}$ where the bulk side of the
correspondence is at weak coupling and where it is possible to
perform computations in weakly coupled string theory or
supergravity.

Once we consider the theory away from these special limits, at a
generic point in the interior of the moduli space ${\cal M}$, it
is hard to compute anything exactly since both the bulk and the
boundary sides have coupling constants of order unity. However, as
long as we stay away from singularities, it is reasonable to
assume that at all points of ${\cal M}$ the theory has a boundary
description in terms of a 2-dimensional conformal field theory
with ${\cal N}=(4,4)$ supersymmetry and central charge given by
\eqref{centralcharge}.

In any conformal field theory with extended supersymmetry, and in
particular in the boundary theory of AdS$_3$/CFT$_2$, there is a
protected sector consisting of chiral primary operators. These
operators form a ring under multiplication and their 3-point
functions characterize the structure of the ring
\cite{Lerche:1989uy}.  The chiral primaries in the AdS$_3$/CFT$_2$
correspondence have been identified in the weak-coupling limits of
${\cal M}$. The counting of their degeneracies in the orbifold CFT
limit is in agreement with their counting from supergravity
\cite{Larsen:1998xm,deBoer:1998ip}.  More surprisingly, their
3-point functions, that is the structure of the chiral ring, is
the same at different points of the moduli space
\cite{Gaberdiel:2007vu},
\cite{Dabholkar:2007ey},\cite{Pakman:2007hn},
\cite{Taylor:2007hs}.

Our goal is to compute the moduli dependence of the chiral ring at
a generic point of ${\cal M}$, where no weakly coupled description
of the theory is available. This is possible due to the extended
supersymmetry. As we will see, the chiral ring is covariantly
constant over ${\cal M}$. In particular, we will understand the
non-renormalization theorem for the 3-point functions of chiral
primaries in AdS$_3$/CFT$_2$.  In the rest of this section we will
review some background material.

\subsection{Chiral primaries and the chiral ring}

We start with a quick review of the chiral ring of 2-dimensional
superconformal field theories \cite{Lerche:1989uy}. Ultimately we are
interested in ${\cal N}=(4,4)$ theories, but for simplicity of
notation in this section we will only consider the left-moving part of
an ${\cal N}=(2,2)$ SCFT.

The left-moving currents are the energy momentum tensor $T(z)$, two
supercurrents $G^\pm(z)$ and the $U(1)$ R-current $J(z)$. The
superscript index of the supercurrents denotes their R-charge which is
$\pm 1$. An operator $\phi$ is called superconformal primary if it
satisfies the condition
\begin{equation}
L_n|\phi\rangle = J_n|\phi\rangle =
  G^+_{n-{1\over 2}} |\phi\rangle = G^-_{n-{1\over 2}}
  |\phi\rangle=0,\qquad n> 0.
\end{equation}
If in addition it satisfies
\begin{equation}
  G^+_{-1/2} |\phi\rangle =0
\end{equation}
then it is called {\it chiral primary}. Using the ${\cal N}=2$ algebra
we can show that for such operators we have
\begin{equation}
  (2L_0 - J_0) |\phi\rangle = 0
\label{unibound}
\end{equation}
and therefore the conformal dimension $h$ and the R-charge $q$ are
related as $h = q/2$.  Conversely, in a unitary CFT we can show
that a primary field satisfying \eqref{unibound} will be chiral.
Similarly we define antichiral primary fields $\overline{\phi}$
which satisfy
\begin{equation}
  G^-_{-1/2} |\overline{\phi}\rangle= 0 \qquad \Leftrightarrow \qquad
  (2L_0 + J_0) |\overline{\phi}\rangle =0.
\end{equation}
Their dimension and R-charge are related by $h=-q/2$. Obviously if a
field $\phi$ is chiral, then $\phi^\dagger$ is antichiral.

A remarkable property of chiral primary operators is that they form a
ring.  The OPE of two chiral primaries is nonsingular as can be
demonstrated by $U(1)$ charge conservation and unitarity and has the
form
\begin{equation}
  \phi_i(z)\phi_j(w) = C^k_{ij}\phi_k(w) +...
\end{equation}
where the operator $\phi_k$ is also chiral primary of charge
$q_k=q_i+q_j$.  The constants $C^k_{ij}$ are the structure constants
of the ring.

We define the two point function of chiral primaries on the sphere
which plays the role of Zamolodchikov's metric
\begin{equation}
  \langle \phi_i(0) \overline{\phi_{j}}(\infty)\rangle =
  g_{i\overline{j}},
\end{equation}
and can be nonzero only if the fields have opposite R-charges. We
also have the 3-point functions on the sphere
\begin{equation}
  \langle \phi_i(0)
  \phi_j(1)\overline{\phi_{k}}(\infty)\rangle =
  C_{ij\overline{k}}
\end{equation}
where again from charge conservation it must be of the form
chiral-chiral-antichiral.

Using the OPE of the chiral ring we find the following relation
between the chiral ring coefficients and the 3-point functions
\begin{equation}
  C_{ij\overline{k}} = C_{ij}^lg_{l\overline{k}}.
\end{equation}

Our discussion up to this point has been about the left-moving sector
of an ${\cal N}=2$ theory. When we consider the full ${\cal N}=(2,2)$
theory we can have fields which are chiral on both sides, antichiral
on both, or chiral - antichiral, and we will have four corresponding
rings $(cc),(aa),(ca),(ac)$ which are pairwise complex conjugate.

As we will explain in more detail later, for ${\cal N}=(4,4)$
theories we can use the enhanced R-symmetry $SU(2)^R_{left}\times
SU(2)^R_{right}$ to rotate a chiral field into an antichiral one.
This implies that all four rings are equivalent, so essentially
there is only one ring in an ${\cal N}=(4,4)$ theory.

\subsection{Moduli dependence}

So far we have considered the chiral primaries and their OPEs in a
given SCFT. Usually superconformal field theories come in families,
parametrized by a moduli space ${\cal M}_{CFT}$. Motion along ${\cal
  M}_{CFT}$ is generated by marginal operators. We will consider
perturbations by operators which preserve the ${\cal N}=(2,2)$, or
${\cal N}=(4,4)$, structure and we will stay away from any
singularities on the moduli space, so we will assume that ${\cal
  M}_{CFT}$ is a smooth manifold of fixed dimension, at least locally.

While the dynamics of the CFT depends on the position on ${\cal
  M}_{CFT}$, certain properties of the chiral ring are protected.  For
example the number of chiral primaries of given dimension is generally
constant on ${\cal M}_{CFT}$.  It is possible for chiral primaries to
pair up into long multiplets and leave the BPS spectrum, but this will
happen at special points or submanifolds of the moduli space. We will
restrict our analysis to regions of ${\cal M}_{CFT}$ where this does
not happen. In AdS$_3$/CFT$_2$ this assumption is justified by the
agreement of the counting of chiral primaries in the symmetric
orbifold and the supergravity limits.

In a general ${\cal N}=(2,2)$ SCFT the structure
constants $C^k_{ij}$ and the 2- and 3-point functions
$g_{i\overline{j}}$, $C_{ij\overline{k}}$ are usually nontrivial
functions on ${\cal M}_{CFT}$. The agreement of 3-point functions of
chiral primaries in AdS$_3$/CFT$_2$ at different points of the moduli
space is a strong indication that in this system they are actually
constant on the moduli space. This moduli-independence must be a
consequence of the extended supersymmetry in ${\cal N}=(4,4)$
superconformal field theories, which implies a non-renormalization
theorem for the 3-point functions of chiral primaries in theories of
this type.

\subsection{The bundle of chiral primaries and the chiral ring}

In general comparing the correlation functions of operators at
different points of the moduli space of a theory is not
straightforward due to operator mixing. More precisely, to compare
their correlation functions in a meaningful way, we first have to
verify that the operators under comparison are actually ``the
same'' at the two different points. Since the underlying quantum
field theory is also changing as we vary the moduli, there is no
natural identification of operators at different points of the
moduli space. We could try to label operators by their conformal
dimension and other conserved charges, but in general there is too
large a degeneracy of operators of given charge to uniquely
identify them. Moreover, as we will see later, the correct
identification of operators between different points on the moduli
space is actually path dependent.

Consider the moduli space ${\cal M}_{CFT}$ of a CFT. At each point
$p\in {\cal M}_{CFT}$ we have the vector space $V_q^{(p)}$ of
chiral primary operators of charge $q$. As we argued above we will
assume that the dimension of this space is the same at all points,
however there is no natural identification between the chiral
primaries at different points of ${\cal M}_{CFT}$. This means that
$V_{q}^{(p)}$ is the fiber of a vector bundle
\begin{equation}
  {\cal V}_q
\end{equation}
of chiral primaries of charge $q$ over the moduli space. The chiral
ring coefficients can be thought of as multiplication between bundles
of this form
\begin{equation}
  C^k_{ij}: {\cal V}_{p}\otimes {\cal V}_{q} \rightarrow {\cal V}_{p+q}
\end{equation}
and similarly the three point functions
\begin{equation}
  C_{ij\overline{k}}:
{\cal V}_{p}\otimes {\cal V}_{q} \otimes \overline{{\cal V}_{p+q}}
\rightarrow{\mathbb C}.
\end{equation}
It should be clear that to meaningfully compare the 3-point functions
of chiral primaries at different points, we have to compute the
connection on the bundles ${\cal V}_q$ which will specify how exactly
we can ``parallel transport'' operators from one point to another. The
connection on the bundle of operators over the moduli space is
generally determined by the dynamics of the CFT as we explain in the
next sections. In the special case of chiral primaries in theories
with ${\cal N}=(2,2)$ supersymmetry this computation is simplified and
the connection of the bundles ${\cal V}_q$ can be computed by the
$tt^*$ equations which will be described later.

In this paper we want to compute the geometry of the bundles of chiral
primaries in ${\cal N}=(4,4)$ theories, and in particular for the
theory relevant for AdS$_3$/CFT$_2$.  The first result of our analysis
is to show that the 3-point functions are covariantly constant, that
is they satisfy
\begin{equation}
  \nabla_\mu C_{ij\overline{k}} = 0
\end{equation}
where $\nabla_\mu$ is a covariant derivative\footnote{It should be
  clear that naive expression
\begin{equation}
  \partial_\mu C_{ij\overline{k}} \stackrel{?}{=}  0
\end{equation}
is meaningless since the ordinary, instead of the covariant,
derivative of a geometric object is not an invariant quantity.}
along a tangent direction on ${\cal M}_{CFT}$, associated to the
connection on the bundles ${\cal V}_q$. This is a
non-renormalization theorem for the chiral primary 3-point
functions in AdS$_3$/CFT$_2$ and more generally for any ${\cal
N}=(4,4)$ theory. The second result is the computation of the
connection on the bundles of chiral primaries at a general point
on the moduli space of ${\cal N}=(4,4)$ theories, using the
constraints from supersymmetry which allows us to express them in
terms of the $tt^*$ equations.

\section{Families of conformal field theories and the connection for operators}
\label{sec:cftdeform}

The fact that we have to define a connection on the bundle of
operators over the moduli space of a conformal field theory is
quite general and not specific to theories with supersymmetry. The
most familiar example is the connection for exactly marginal
operators. The marginal operators ${\cal O}_\mu(z,\lambda)$ of a
CFT correspond to tangent vectors on the moduli space at the point
$\lambda\in {\cal
  M}_{CFT}$. Comparing marginal operators at different points of
${\cal M}_{CFT}$ is analogous to comparing tangent vectors at
different points of a manifold, i.e. impossible, unless we first
define a connection which describes their parallel transport. The
moduli space ${\cal M}_{CFT}$ of a conformal field theory has the
structure of a Riemannian manifold.  This structure is defined by the
Zamolodchikov metric $g_{\mu\nu}(\lambda)$ which is given by the
2-point function
\begin{equation}
  \langle {\cal O}_\mu(z,\lambda) {\cal O}_\nu(w,\lambda) \rangle=
          {g_{\mu \nu}(\lambda) \over |z-w|^4}.
\label{2pointz}
\end{equation}
In general the metric $g_{\mu\nu}(\lambda)$ depends on the
position $\lambda \in{\cal M}_{CFT}$ which means that the moduli
space has a non-trivial geometry. We can use the metric to define
a metric-compatible connection for the operators ${\cal
O_\mu}(z,\lambda)$, allowing us to parallel transport and compare
them at different points of ${\cal M}_{CFT}$. So the vector bundle
of marginal operators is isomorphic to the tangent bundle of the
moduli space and the natural connection on it is the Levi-Civita
connection associated to the Zamolodchikov metric. The mixing of
marginal operators under deformations of the theory is expressed
by the equation
\begin{equation}
\delta_\mu {\cal O}_\nu = \Gamma^\kappa_{\nu \mu} {\cal O}_\kappa
\label{mixingmarg}
\end{equation}
where
\begin{equation}
  \Gamma_{\nu\mu}^\kappa = {1\over 2}g^{\kappa \lambda}(\partial_\nu
  g_{\mu \lambda} + \partial_{\mu} g_{\nu \lambda} - \partial_\lambda
  g_{\mu\nu})
\label{christ}
\end{equation}
and $g_{\mu\nu}(\lambda)$ is the Zamolodchikov metric defined in
\eqref{2pointz}.\footnote{In general the marginal operators correspond
  to tangent vectors on the moduli space. The relation between
  operator mixing \eqref{mixingmarg} and the Zamolodchikov metric
  \eqref{2pointz} via \eqref{christ} is true only if we choose a basis
  of marginal operators corresponding to commuting vector fields on
  the moduli space, so that they can be interpreted as derivatives with respect
  to a choice of coordinates. Otherwise they have to be treated in terms of
  a basis of vielbeins and the expression for their mixing has to be written in terms
  of the spin-connection.}

Similar arguments hold for operators of higher conformal
dimension. For simplicity we can assume that at all points of ${\cal
  M}_{CFT}$ we have a set of operators $\{\varphi_I\}$ of conformal
weight $(h,\overline{h})$. If there are no additional conserved
charges distinguishing them, then they will generically mix among
themselves when we move on ${\cal M}_{CFT}$. Under a deformation
generated by a marginal operators ${\cal O}_\mu$ we have the mixing
\begin{equation}
  \delta_\mu \varphi_I = A_{\mu I}^J \,\varphi_J
\label{connf}
\end{equation}
where $A_{\mu I}^J$ plays the role of the connection. Similarly if we
consider an infinitesimal closed loop of deformations spanned by two
marginal operators ${\cal O}_\mu,{\cal O}_\nu$, we have the curvature
\begin{equation}
 (\delta_{\mu}\delta_\nu-\delta_{\nu}\delta_{\mu})\,\varphi_I =
  R_{\mu\nu I}^J \,\varphi_J.
\label{curvdefin}
\end{equation}
So the operators $\{\varphi_I\}$ take values in a vector bundle
over the moduli space, whose connection is $A_{\mu I}^J$ and the
curvature $R_{\mu\nu I}^J$. In what follows we will explain that
there is a natural connection which is completely determined by
the dynamics of the CFT.

\subsection{Deformations of conformal field theories}

Before we proceed, we would like to pause and discuss some
(well-known) subtleties which will clarify the underlying reason for
having a nontrivial connection for the operators in a family of
conformal field theories. Let us start with a given theory
characterized by a set of correlation functions
\begin{equation}
  G_n(x)=\langle \varphi_1(x_1)...\varphi_n(x_n)\rangle
\label{undeformcor}
\end{equation}
which satisfy the axioms of a 2-dimensional CFT.\footnote{Notice that
  up to this point the correlation functions are defined only for
  distinct points, $x_i\neq x_j$.} We consider an operator ${\cal
  O}(z)$ in this theory. From the Lagrangian formulation point of
view, we can deform the theory by adding to the action
\begin{equation}
  S \rightarrow S + {\lambda\over \pi} \int d^2z {\cal O}(z)
\label{deformact}
\end{equation}
where $\lambda$ is a small parameter. The effect of this deformation
is to modify the $n$-point functions
\begin{equation}
  G_n(x) \rightarrow  G_n(x) + \delta G_n(x).
\end{equation}
For deformations of the form \eqref{deformact}, the deformed
$n$-point functions are given, to first order in $\lambda$, in
terms of integrated $(n+1)$-point functions of the original
undeformed theory
\begin{equation}
  \delta G_n(x) \equiv \delta \langle
  \varphi_1(x_1)...\varphi_n(x_n)\rangle \simeq {\lambda \over
    \pi}\int d^2z \langle \varphi_1(x_1)...\varphi_n(x_n){\cal
    O}(z)\rangle
\label{deformcor}
\end{equation}
where the meaning of the symbol $\simeq$ will become clear below.  To
second order in $\lambda$ we have to consider the twice integrated
$(n+2)$-point function of the undeformed theory and so on.

The deformed theory may be a local quantum field theory, but not
necessarily a CFT. By demanding that the deformed correlation
functions satisfy the CFT axioms, we find certain conditions for
the deformation operator ${\cal O}(z)$. To first order in
$\lambda$ the condition is that ${\cal O}(z)$ must be an operator
of dimension $(1,1)$, that is a marginal operator. More
constraints from the requirement of conformal invariance appear at
higher orders in $\lambda$, and if all these are satisfied ${\cal
O}(z)$ is called an exactly marginal operator.

Going back to \eqref{deformcor} we see that in order to compute the
deformed correlators we have to integrate the insertion of ${\cal
  O}(z)$ over $z$, but when $z\rightarrow x_i$ the operator ${\cal
  O}(z)$ will hit the other insertions. This introduces two
subtleties. First, in the original theory the correlators
\eqref{undeformcor} were defined for distinct points, and formally
we may have contact terms when the insertions coincide
\cite{Seiberg:1988pf},\cite{Kutasov:1988xb}. Second, the integral
over $z$ in \eqref{deformcor} will generally diverge because of
short distance singularities between the operator ${\cal O}(z)$
and the other insertions $\varphi_i(x_i)$. So the right hand side
of equation \eqref{deformcor} is not well defined at this stage.
Notice that for large $z$ the correlator decays at least as
$|z|^{-4}$, so there are no IR divergences to worry about.

Actually, the two aforementioned subtleties are related in the
sense that we define the contact terms to precisely cancel the
infinities arising from the integration over $z$ around the
punctures. While the infinities are cancelled in this way, there
may be finite remaining contributions from this subtraction
prescription which are responsible for the nontrivial connection
for the operators of the CFT.

Equivalently we can forget about contact terms, but instead define a
renormalization prescription for the integrated $(n+1)$-point function
which is consistent with locality. Considering \eqref{deformcor}
again, we see that the more precise statement should be
\begin{equation}
  \delta G_n(x) ={\lambda\over \pi}\left[\int d^2z \langle
    \varphi_1(x_1)...\varphi_n(x_n){\cal O}(z)\rangle \right]_{ren}
\label{renormcorf}
\end{equation}
where the subscript $ren$ stands for {\it renormalized}, and its exact
meaning will be explained in the next subsection.

Now if we consider two deformations, one by the operator ${\cal
  O}_\mu$ and one by ${\cal O}_\nu$ then the naive answer
\eqref{deformcor} would give
\begin{equation}
  (\delta_\mu \delta_\nu G_n)_{naive} \simeq {\lambda_1 \lambda_2\over
    \pi^2}\int d^2z_1 \int d^2z_2 \langle
  \varphi_1(x_1)...\varphi_n(x_n){\cal O}_\nu(z_1){\cal
    O}_\mu(z_2)\rangle
\end{equation}
and also
\begin{equation}
  (\delta_\nu \delta_\mu G_n)_{naive} \simeq {\lambda_1 \lambda_2\over
    \pi^2}\int d^2z_1 \int d^2z_2 \langle
  \varphi_1(x_1)...\varphi_n(x_n){\cal O}_\mu(z_1){\cal
    O}_\nu(z_2)\rangle
\end{equation}
so formally
\begin{equation}
  (\delta_\mu\delta_\nu G_n)_{naive} = (\delta_\nu \delta_\mu G_n)_{naive}
\end{equation}
which would indicate that the order of deformation does not matter and
it would imply that there is no curvature on the space of CFTs. However
this is wrong, since the integrated $(n+2)$-point functions are not well defined
for the reasons we mentioned earlier. Only the renormalized integrated
$(n+2)$-point functions are meaningful
\begin{equation}
  (\delta_\mu \delta_\nu G_n)_{ren} = {\lambda_1 \lambda_2\over
    \pi^2}\left[\int d^2z_1 \int d^2z_2 \langle
    \varphi_1(x_1)...\varphi_n(x_n){\cal O}_\nu(z_1){\cal
      O}_\mu(z_2)\rangle\right]_{ren}
\label{secondreno}
\end{equation}
where again we have to specify the way to renormalize the double
integral. As it turns out, it is possible to find a
renormalization prescription for the integrated correlation
functions \eqref{renormcorf} and \eqref{secondreno}, such that the
axioms of a CFT are preserved but the price we have to pay is that
in general
  \begin{equation}
  (\delta_\mu\delta_\nu G_n)_{ren}\neq (\delta_\nu \delta_\mu G_n)_{ren}.
\end{equation}
Because of this non-commutativity the correct statement is not
\begin{equation}
  \delta_\mu G_n \stackrel{?}{=} \lambda \partial_\mu G_n
\end{equation}
but rather
\begin{equation}
  \delta_\mu G_n = \lambda \nabla_\mu G_n,
\end{equation}
in other words
\begin{equation}
  \nabla_\mu G_n = {1\over \pi}\left[\int d^2z \langle
    \varphi_1(x_1)...\varphi_n(x_n){\cal O}_\mu(z)\rangle\right]_{ren}
\label{covder}
\end{equation}
The renormalization prescription defines the covariant derivative
$\nabla_\mu$ associated to the connection $A_{\mu I}^J$ on the vector
bundle of the operators $\{\varphi_I\}$ introduced in \eqref{connf}.

\subsection{The connection for operators}

In \cite{Ranganathan:1992nb},\cite{Ranganathan:1993vj} connections
on the vector bundle of operators over the moduli space of a CFT
were studied in detail. A natural prescription (called the
connection $\overline{c}$ in \cite{Ranganathan:1993vj}) for
defining the renormalized deformed correlators is the following:
consider the to-be-integrated $(n+1)$-point function, introduce
very small disks of size $\epsilon$ around the punctures $x_i$,
and define the {\it
  regularized} integrated $(n+1)$-point function
\begin{equation}
  \delta_\mu G_n(\epsilon) = {\lambda\over \pi} \left[\int_{|z-x_i|>\epsilon} d^2z \langle
\varphi_1(x_1)...\varphi_n(x_n){\cal O}_\mu(z)\rangle\right]_{reg}.
\label{regcor}
\end{equation}
As $\epsilon\rightarrow 0$, and suppressing the $x_i$ variables, the
regularized integrated function will have the form
\begin{equation}
  \delta_\mu G_n(\epsilon) =\left(\delta_\mu G_n\right)_{ren}
+ \sum_{\alpha > 0} {c_\alpha \over \epsilon^\alpha} + c_0 \log \epsilon
\label{rencor}
\end{equation}
where the finite piece
\begin{equation}
\left(\delta_\mu G_n\right)_{ren}
\end{equation}
defines the {\it renormalized} perturbed $n$-point function and the
corresponding connection $\nabla_\mu$ by \eqref{covder}.

If we consider the second variation of the correlation function according
to this prescription, we find that
  \begin{equation}
  (\delta_\mu\delta_\nu G_n)_{ren}\neq (\delta_\nu \delta_\mu G_n)_{ren}
\end{equation}
This is the reason that we have curvature on the vector bundles of
operators over the moduli space.

Also, notice that the vector bundle whose fiber is spanned by a
set of operators $\{\varphi_I\}$ is equipped with a natural metric
$g_{IJ}(\lambda)$ defined by the 2-point function
\begin{equation}
  \langle \varphi_I(z) \varphi_J(w)\rangle = {g_{IJ} (\lambda)\over (z-w)^{2h}
(\overline{z}- \overline{w})^{2\overline{h}}} .\label{zamol}
\end{equation}
The connection defined above is compatible with the
metric
\begin{equation}
  \nabla_\mu \, g_{IJ} =0
\label{metriccomp}
\end{equation}
so it is a natural connection for this vector bundle.

The curvature of the connection can be expressed in terms of
4-point functions. We quickly describe the main result, more
details can be found in \cite{Ranganathan:1993vj}. Consider a set
of operators $\{\varphi_I\}$ of the same conformal dimension and
same charges. The object we want to compute is the curvature
$R_{\mu\nu I}^J$ of corresponding vector bundle over the moduli
space. The curvature can be computed if we know the 4-point
function
\begin{equation}
  \langle {\cal O}_{\mu}(z_1) {\cal O}_{\nu}(z_2) \varphi_J(x_1)
\varphi_I(x_2)\rangle
\end{equation}
for distinct points of insertion. Generalizing the prescription
\eqref{regcor}, \eqref{rencor} the curvature is given by a twice
integrated and appropriately regulated antisymmetrized combination of
the 4-point function, as follows\cite{Ranganathan:1993vj}.

First we consider the 4-point function\footnote{The
index $J$ has been raised with the Zamolodchikov metric \eqref{zamol} as
$\varphi^J = g^{JK}\varphi_K$.} as a function of $z_1,z_2$
  \begin{equation}
    G_{\mu\nu}(z_1,z_2)= \langle {\cal O}_{\mu}(z_1) {\cal
      O}_{\nu}(z_2) \varphi^J(\infty) \varphi_I(0)\rangle
\label{fourpointc}
  \end{equation}
for distinct points. Keeping $z_1$ fixed, we consider the integral
over $z_2$ of the following expression\footnote{The integral over
the disc arises from separating the plane, viewed as a two sphere,
in two hemispheres, see \cite{Ranganathan:1993vj}.}
\begin{equation}
  F(z_1,\epsilon) = {1\over \pi^2}\int_{\epsilon<|z_2|<1} d^2z_2
  (G_{\mu\nu}(z_1,z_2) - G_{\nu \mu}(z_1,z_2)).
\end{equation}
Notice that because of the antisymmetrization the integral converges
as $z_1 \rightarrow z_2$. For fixed $\epsilon$ the integral is
convergent, however it may diverge as $\epsilon\rightarrow 0$ because
the operator at $z_2$ approaches the operator at $0$. We define the
regularized integral
 \begin{equation}
    \widetilde{F}(z_1) = \lim_{\epsilon\rightarrow 0}\left(
    F(z_1,\epsilon) -\text{Dp}(F(z_1,\epsilon))\right)
\end{equation}
where $\text{Dp}$ denotes the divergent part, defined as in
\eqref{rencor}. This procedure gives us a finite function
$\widetilde{F}(z_1)$. Finally we integrate $\widetilde{F}$ over $z_1$. There
are divergences as $z_1\rightarrow 0$ and again we are instructed to keep
the finite part
  \begin{equation}
    R_{\mu \nu I}^J = \text{Fp} \int_{|z_1|<1} d^2 z_1 \widetilde{F}(z_1)
\label{curvexp}
  \end{equation}
where $\text{Fp}$ denotes the finite part, again defined as in
\eqref{rencor}. This is the final expression for the curvature.  It is
also possible to rewrite the curvature in terms of OPE coefficients.
If we have
\begin{equation}
{\cal O}_\mu(z) \varphi_I(0) = \sum_k {H_{\mu I}^k \varphi_k(0)
\over z^{1+h_I-h_k}\overline{z}^{1 + \overline{h}_I - \overline{h}_k}}
\end{equation}
then after some algebra \cite{Ranganathan:1993vj} we can show that the
curvature has the form
\begin{equation}
  R_{\mu\nu I}^J =4 \delta_{s_I,s_J} \left( \sum_{\gamma_k>\gamma_I} +
  \sum_{\gamma_k < \gamma_I}\right) {H^k_{[\mu I}H^J_{\nu]k}
    \delta_{s_k,s_I}\over \gamma_{kJ} \gamma_{kI}}
\label{ggg}
\end{equation}
where $\gamma = h + \overline{h},\, s = h - \overline{h}$ are the
scaling dimension and spin of the operator, and $\gamma_{ij} \equiv
\gamma_i-\gamma_j$.

As we can see, the connection on the vector bundle of operators
depends on the dynamics of the CFT. For a general interacting CFT it
is difficult to compute the exact 4-point function, or equivalently
the OPE coefficients, hence the computation of the curvature is
hard. In theories with extended supersymmetry, and if we are
interested in the curvature of operators in the chiral ring, it
becomes possible to compute the curvature exactly. As we will see in
this case the infinite sum in \eqref{ggg} truncates to a finite sum
over chiral ring coefficients, giving us the $tt^*$ equations.  We
analyze the ${\cal N}=(2,2)$ case in the next section and then
consider ${\cal N}=(4,4)$ SCFTs.

\section{The Chiral Ring of ${\cal N}=(2,2)$ theories}
\label{sec:review22}

The bundle of chiral primaries has been analyzed in detail in theories
with ${\cal N}=(2,2)$ superconformal symmetry. The main result
relevant for us is the computation of the curvature of the bundle of
chiral primaries in terms of the chiral ring coefficients, which is
expressed by the $tt^*$ equations derived by Cecotti and
Vafa in \cite{Cecotti:1991me}.  In this section we quickly review the
main points and give a derivation of the $tt^*$ equations for
superconformal theories which does not rely on the topological
twisting.

In an ${\cal N}=(2,2)$ SCFT the left-moving currents are
$T(z),G^\pm(z),J(z)$ and the right-moving ones
$\overline{T}(\overline{z}),\overline{G}^\pm(\overline{z}),\overline{J}(\overline{z})$. The
OPEs of the algebra can be found in appendix \ref{app:algope}.  As we
explained before in ${\cal N}=(2,2)$ theories we have the $(cc)$ ring
of chiral primary-chiral primary operators $\phi_i$ which satisfy
\begin{equation}
L_0={J_0 \over 2},\qquad \overline{L}_0 ={\overline{J_0}\over 2}
\end{equation}
and their complex conjugates $(aa)$ with opposite charges. We also
have chiral primary - antichiral primary operators $\psi_i$ in the $(ca)$
ring satisfying
\begin{equation}
L_0={J_0 \over 2},\qquad \overline{L}_0 =-{\overline{J_0}\over 2}
\end{equation}
and their complex conjugates in the $(ac)$ ring. We will refer to the
$(cc)$ ring and its conjugate as the chiral ring, and to $(ca)$ and
its conjugate as the twisted chiral ring. The structure constants of
the chiral ring are given by
\begin{equation}
  \phi_i(z) \phi_j(w) = C^k_{ij}\phi_k(w) +...
\end{equation}
while those of the twisted chiral ring by
\begin{equation}
  \psi_a(z) \psi_b(w) = \widetilde{C}^d_{ab}\psi_d(w) +...
\end{equation}
We can find marginal operators by considering the descendants of
chiral primaries of dimension $(1/2,1/2)$. We have the following
possibilities\footnote{We have included the factors of ${1\over 2}$ in
  the normalization of the marginal operators to ensure that
  $\ka_{i\overline{j}} \equiv\langle {\cal O}_i(1) \overline{{\cal
      O}_j}(0)\rangle =\langle \phi_i(1)\overline{\phi_j}(0)\rangle$.}
\begin{equation}
  {\cal O}_i ={1\over 2} G^-_{-1/2}\overline{G^-}_{-1/2}
  \cdot\phi_i,\qquad \overline{{\cal O}_j} ={1\over 2}
  G^+_{-1/2}\overline{G^+}_{-1/2} \cdot \overline{\phi_j}
\label{marginalchiral}
\end{equation}
\begin{equation}
  {\cal O}_a ={1\over 2} G^-_{-1/2}\overline{G^+}_{-1/2}
  \cdot\psi_a,\qquad \overline{{\cal O}_b} ={1\over 2}
  G^+_{-1/2}\overline{G^-}_{-1/2} \cdot \overline{\psi_b}.
\label{marginaltwisted}
\end{equation}
All these are operators of conformal dimension $(1,1)$ and R-charge
$(0,0)$, so they are marginal and can be used to perturb the CFT. The
first class of operators labeled by $i,j,...$ are descendants of
fields in the chiral ring and their complex conjugates, while the
second class labeled by $a,b,...$ are descendants of fields in the
twisted chiral ring and their complex conjugates. We use Greek indices
$\mu,\nu,...$ to denote a general marginal operator which can be of
any of the four forms described above.

A basic result is that for ${\cal N}=(2,2)$ SCFTs the moduli space
locally has a product structure
\begin{equation}
  {\cal M}_{CFT} = {\cal M}_{C} \times {\cal M}_{TC}
\end{equation}
where ${\cal M}_{C}$ is generated by marginal operators which come
from the chiral ring, and ${\cal M}_{TC}$ is generated by marginal
operators from the twisted chiral ring. As an example, for a
sigma-model whose target space is a Calabi-Yau 3-fold, one of the
spaces corresponds to the K\"ahler structure deformations while
the other to the complex structure deformations. It can be shown
that each of the two components ${\cal M}_C,{\cal M}_{TC}$ is a
complex, K\"ahler manifold. Moreover it can be shown that they are
special K\"ahler.

We denote by $\ka_{i\overline{j}}$ the K\"ahler form of the component
${\cal M}_C$ and $\ka_{a\overline{b}}$ that of ${\cal M}_{TC}$, which
are given in terms of CFT data by the two point functions
\begin{equation}
\langle {\cal O}_i(z) \overline{{\cal O}_j}(w)\rangle =
{\ka_{i\overline{j}} \over |z-w|^4},\qquad \langle {\cal O}_a(z)
\overline{{\cal O}_b}(w)\rangle = {\ka_{a\overline{b}} \over
|z-w|^4}.
\end{equation}
In ${\cal N}=(4,4)$ theories the moduli space does not factorize, not
even locally. It consists of a single factor and cannot be decomposed
into chiral and twisted chiral components. As we will see it is not
a complex manifold.

\subsection{Curvature of the algebra}
\label{subsec:curvalg2}

We now proceed with a discussion of the connection on the bundle of
operators over the moduli space.  In the same way that chiral
primaries can mix under deformations of the CFT, the generators of the
algebra can also mix among themselves, see \cite{Distler:1992gi} for a
nice review. The energy momentum tensor $T(z)$ and the $U(1)$ current
$J(z)$ are uniquely defined at each point of the moduli space, so
there can be no holonomy associated to them. However the supercurrents
are not uniquely defined, since the ${\cal N}=(2,2)$ algebra has a
$U(1)_L\times U(1)_R$ automorphism which transforms the supercurrents
as
\begin{equation}
  G^\pm \rightarrow e^{\pm i\theta} G^\pm,\qquad \overline{G}^\pm \rightarrow
e^{\pm i\overline{\theta}} \overline{G}^\pm
\end{equation}
leaving the bosonic currents unchanged, and where $\theta,
\overline{\theta}$ are two independent angles. Consequently, what
we mean by a supercurrent is ambiguous up to an overall phase.
Moreover, if we parallel transport on the moduli space and come
back to the original point, the supercurrents will receive a
$U(1)$ rotation. This means that the supercurrents are (operator
valued) sections of $U(1)$ bundles over the moduli space. If $G^+$
is a section of a $U(1)$ bundle ${\cal L}$ then $G^-$ will be a
section of ${\cal L}^{-1}$, since they transform with opposite
phases. Similarly $\overline{G}^+$ will be a section of another
bundle $\overline{{\cal L}}$ and $\overline{G}^-$ a section of
$\overline{{\cal L}}^{-1}$. We call $F$ the curvature tensor of
the bundle ${\cal L}$ and $\overline{F}$ the curvature of
$\overline{{\cal L}}$. According to our previous discussion, to
compute the curvature of ${\cal L}$ and $\overline{{\cal L}}$, we
need the 4-point functions
\begin{equation}
 \langle {\cal O}_\mu(x) {\cal O}_\nu(y) G^r(z) G^s(w)\rangle
 \quad {\rm and} \quad
  \langle {\cal O}_\mu(x) {\cal O}_\nu(y) \overline{G}^r(z)
 \overline{G}^s(w)\rangle,
\label{fourgg}
\end{equation}
where ${\cal O}_\mu, {\cal O}_\nu$ are marginal operators of the form
\eqref{marginalchiral},\eqref{marginaltwisted} and $r,s=\pm$.

These four point functions can be exactly computed using the
superconformal Ward identities of the ${\cal N}=(2,2)$ algebra. For
example as we show in appendix \ref{app:curvsuper} we have
\begin{equation}
  \langle {\cal O}_i(x) \overline{{\cal O}_j}(y) G^+(z) G^-(w)\rangle
  = {2c\over 3}{\ka_{i\overline{j}} \over |x-y|^4 (z-w)^3}+
  {2\ka_{i\overline{j}} \over (x-z)^2(y-w)^2(z-w)
    (\overline{x}-\overline{y})^2}
\end{equation}
and similarly for the other combinations. Following the prescription
of equations \eqref{fourpointc} to \eqref{curvexp} we find that the
only nonzero components of the curvature for the line bundle ${\cal
  L}$ are
\begin{equation}
  F_{i\overline{j}} = -{3\over c} \ka_{i\overline{j}},\quad
  F_{a\overline{b}} = -{3\over c} \ka_{a\overline{b}}
\end{equation}
while for $\overline{{\cal L}}$ we have
\begin{equation}
   \overline{F}_{i\overline{j}} = -{3\over c}\ka_{i\overline{j}},\quad
   \overline{F}_{a\overline{b}} = {3\over c}\ka_{a\overline{b}}
\end{equation}
Notice that if we consider the bundle ${\cal L}\otimes \overline{{\cal
    L}}$, then its curvature is zero on ${\cal M}_{TC}$, while ${\cal
  L}\otimes \overline{{\cal L}}^{-1}$ has zero curvature over ${\cal
  M}_C$.

To summarize, we found that while the bosonic currents $T(z),J(z)$ are
well defined everywhere, the supercurrents $G^\pm(z)$ are ambiguous
and there is an associated holonomy for them described by the
holomorphic line bundles ${\cal L}$, $\overline{{\cal L}}$ over the
moduli space.  Notice that the K\"ahler form on the moduli space is
${c\over 3}$ times the curvature of the line bundle ${\cal L}$ (or
$\overline{{\cal L}}$), so its first Chern class is ${c\over 3}$ times
an integral class \cite{Periwal:1989mx},\cite{Strominger:1990pd}. For
sigma-models in Calabi-Yau $n$-folds, where $c=3n$, the bundle ${\cal
  L}^{c/3}$ is the same as the line bundle of the holomorphic $(n,0)$
form $\Omega$ over the complex structure moduli space.

\subsection{On the curvature of the chiral primaries}

Now we want to consider the connection on the bundle of chiral
primaries. From charge conservation, $(cc)$ operators can only mix
with themselves, and similarly for $(aa),(ca),(ac)$. For each
conformal dimension $(h,\overline{h})$ we have the bundle of chiral
primaries $\phi_i$ with charge $(2h,2\overline{h})$, the bundle of
twisted chiral primaries with charge $(2h,-2\overline{h})$ and their
hermitian conjugates.

To avoid overly heavy notation we will denote the total bundle of
chiral primaries by ${\cal V}$ and that of twisted chiral
primaries by ${\cal \widetilde{V}}$. Each of these bundles is the
direct sum of subbundles ${\cal V}_q$ corresponding to fields of
specific charges
\begin{equation}
  {\cal V} = \sum_q \oplus{\cal V}_q
\end{equation}
It should be clear that the connection on the bundle ${\cal V}$ has to
preserve the grading by conformal dimension (or $U(1)$ charge), since
it should not mix operators of different dimensions under parallel
transport.

\subsection{Direct computation of the curvature of chiral primaries}

There are two methods to compute the curvature of chiral
primaries: one is to directly compute the relevant 4-point
function in the physical theory and then use \eqref{curvexp}. The
second is to use spectral flow to the Ramond sector, consider the
topologically twisted theory and follow the arguments of
\cite{Cecotti:1991me}. The two methods give the same result, which
is the $tt^*$ equations. In this section we show how the direct
computation of the 4-point function yields an alternative
derivation of the $tt^*$ equations in superconformal theories.

Let us consider the curvature of the bundle ${\cal V}$ over the factor
${\cal M}_C$ of the moduli space. According to the general expression
\eqref{curvexp},  we need to compute the 4-point functions
\begin{equation}
  \langle {\cal O}_i(x) \overline{{\cal O}_j}(y) \phi_k(z)
  \overline{\phi_l}(w)\rangle,\qquad
\langle \overline{{\cal O}_j}(x) {\cal O}_i(y) \phi_k(z)
  \overline{\phi_l}(w)\rangle
\end{equation}
where
\begin{equation}
  {\cal O}_i(x) = {1\over 2}G^-_{-1/2} \overline{G}^-_{-1/2}\cdot
  \phi_i(x),\qquad\overline{{\cal O}_j}(y) ={1\over 2} G^+_{-1/2}
  \overline{G}^+_{-1/2} \cdot \overline{\phi_j}(y).
\end{equation}
As explained in appendix \ref{app:fourtwotwo}, using the OPEs of the
supercurrents with the chiral primaries, we can move the supercurrent
operators from ${\cal O}_i$ onto ${\cal O}_{\overline{j}}$ and we have
\begin{equation}
    \langle {\cal O}_i(x)\overline{{\cal O}_j}(y) \phi_k(z)
    \overline{\phi_l}(w)\rangle = \partial_y\partial_{\overline{y}}
    \left({|y-z|^2 \over |x-z|^2}\langle\phi_i(x) \overline{\phi_j}(y)
    \phi_k(z)\overline{\phi_l}(w)\rangle\right)
\end{equation}
similarly moving the supercurrents from ${\cal O}_{\overline{j}}$ to
${\cal O}_i$ we have
\begin{equation}
 \langle\overline{{\cal O}_j}(x) {\cal O}_i(y) \phi_k(z)
 \overline{\phi_l}(w)\rangle =\partial_y\partial_{\overline{y}} \left({|y-w|^2 \over
   |x-w|^2}\langle\overline{\phi_j}(x) \phi_i(y)
 \phi_k(z)\overline{\phi_l}(w)\rangle\right)
\end{equation}
Taking $w\rightarrow 0$ and $z\rightarrow \infty$ we find
\begin{equation}
    \langle {\cal O}_i(x)\overline{{\cal O}_j}(y) \phi_k(\infty)
    \overline{\phi_l}(0)\rangle =\partial_y\partial_{\overline{y}}
 \left( \langle\phi_i(x)
    \overline{\phi_j}(y) \phi_k(\infty)\overline{\phi_l}(0)
    \rangle\right)
\label{fouraaa}
\end{equation}
and
\begin{equation}
 \langle\overline{{\cal O}_j}(x) {\cal O}_i(y) \phi_k(\infty)
 \overline{\phi_l}(0)\rangle = \partial_y\partial_{\overline{y}}
 \left({|y|^2 \over |x|^2}\langle\overline{\phi_j}(x) \phi_i(y)
 \phi_k(\infty)\overline{\phi_l}(0)\rangle\right)
\label{fourbbb}
\end{equation}
We can now use the general formula \eqref{curvexp} for the
curvature of the bundles of operators.  We notice that as
$y\rightarrow 0$ both of the correlation functions are finite: for
\eqref{fouraaa} we just have to use the OPE in the antichiral ring
which is non-singular, while for \eqref{fourbbb} we have to use
the results from appendix \ref{app:caope} for the OPE of a chiral
field with an antichiral. The leading term goes like ${1\over
|y|^2}$ and is exactly cancelled by the $|y|^2$ in the numerator.
Following \eqref{curvexp} the curvature is
\begin{equation}
  R_{i\overline{j}} =\text{Fp}{1\over (\pi)^2} \int_{|x|<1} d^2 x\, I(x)
\label{ffff}
\end{equation}
where
\begin{equation}\begin{split}
  I(x) = \int_{|y|<1} d^2 y \, &\partial_y\partial_{\overline{y}}
 \left( \langle \phi_i(x)
  \overline{\phi_j}(y) \phi_k(\infty) \overline{\phi_l}(0)\rangle
  \right) \\ &-\partial_y\partial_{\overline{y}}\left({|y|^2\over |x|^2} \langle
  \overline{\phi_j}(x) \phi_i(y) \phi_k(\infty)
  \overline{\phi_l}(0)\rangle \right)
\label{ffg}
\end{split}\end{equation}
and we used the fact that there is no singularity as $y\rightarrow
0$.
Using Gauss's theorem we have\footnote{As explained in
  \cite{Ranganathan:1993vj} the antisymmetrized 4-point function has
  no singularity as $y\rightarrow x$.}
\begin{equation}
  I(x) = {1\over 4}\int_{|y|=1} d\theta_1 (y \partial_y + \overline{y}
  \partial_{\overline{y}})\left( \langle \phi_i(x)
  \overline{\phi_j}(y) \phi_k(\infty) \overline{\phi_l}(0)\rangle -
           {|y|^2 \over |x|^2} \langle \overline{\phi_j}(x) \phi_i(y)
           \phi_k(\infty) \overline{\phi_l}(0) \rangle\right)
\label{gauss}
\end{equation}
From the conformal Ward identity
\begin{equation}
  \sum_i\left( h_i + z_i  \partial_i\right)
\langle \varphi_1(z_1)...\varphi_n(z_n)\rangle  =0
\end{equation}
we have for the 4-point function
\begin{equation}
( 1 + x \partial_x + y \partial_y) \langle \phi_i(x) \overline{\phi_j}(y)
\phi_k(\infty) \overline{\phi_l}(0)\rangle=0
\end{equation}
Using this we can write \eqref{gauss} as
 \begin{equation}
    I(x) = - {1\over 4}\int_{|y|=1} d\theta_1 (2+x \partial_x + \overline{x}
    \partial_{\overline{x}})\left( \langle \phi_i(x)
    \overline{\phi_j}(y) \phi_k(\infty) \overline{\phi_l}(0)\rangle -
             {1 \over |x|^2} \langle \overline{\phi_j}(x) \phi_i(y)
             \phi_k(\infty) \overline{\phi_l}(0) \rangle\right)
\end{equation}
Considering the integration over $x$ we find
\begin{equation}
   R_{i\overline{j}} = -{1\over (2\pi)^2} \int_0^1 dr\int_{|y|=1}
   d\theta_1 \int_{|x|=r} d\theta_2 {d\over dr} \left(r^2 \langle
   \phi_i(x) \overline{\phi_j}(y) \phi_k(\infty)
   \overline{\phi_l}(0)\rangle - \langle \overline{\phi_j}(x)
   \phi_i(y) \phi_k(\infty) \overline{\phi_l}(0) \rangle\right)
\end{equation}
So we have
\begin{equation}\begin{split}
   R_{i\overline{j}} =& -{1\over (2\pi)^2} \lim_{|r|\rightarrow 1}
   \int_{|y|=1} d\theta_1 \int_{|x|=r} d\theta_2 \left(r^2 \langle
   \phi_i(x) \overline{\phi_j}(y) \phi_k(\infty)
   \overline{\phi_l}(0)\rangle - \langle \overline{\phi_j}(x)
   \phi_i(y) \phi_k(\infty) \overline{\phi_l}(0) \rangle\right)\\ & +
       {1\over (2\pi)^2} \lim_{|r|\rightarrow 0}\int_{|y|=1} d\theta_1
       \int_{|x|=r} d\theta_2 \left(r^2 \langle\phi_i(x)
       \overline{\phi_j}(y) \phi_k(\infty) \overline{\phi_l}(0)
       \rangle- \langle \overline{\phi_j}(x) \phi_i(y) \phi_k(\infty)
       \overline{\phi_l}(0) \rangle\right)
\end{split}\label{curvqq}\end{equation}
The contribution from the first two terms can be computed using the OPE between
$\phi_i$ and $\overline{\phi_j}$ as explained in appendix \ref{app:contours}.
The contribution form the second term can be computed using the
OPE of the field at $x$ with the field at $0$, which is determined by
the chiral ring coefficients (see also appendix \ref{app:caope}). Finally
we have
\begin{equation}\begin{split}R_{i\overline{j}}& =
\ka_{i\overline{j}}g_{k \overline{l}}\left(1 - {3\over c} (q+
\overline{q})\right) - C_{ik}^m g_{m\overline{n}} C^{*
  \overline{n}}_{\overline{j}\overline{l}} +
g_{k\overline{m}}C^{*\overline{m}}_{\overline{j}\overline{n}}
g^{\overline{n} \rho} C_{i\rho}^\sigma g_{\sigma\overline{k}}\\ & =
\ka_{i\overline{j}}g_{k \overline{l}}\left(1 - {3\over c} (q+
\overline{q})\right) - [C_i,\overline{C_j}]
\end{split}\end{equation}
All other components of the curvature vanish, as can be easily
demonstrated using a similar analysis. To summarize we find the
following expressions for the curvature
\begin{equation}\begin{split}
& [\nabla_i,\nabla_j] =0\\ &
    [\nabla_{\overline{i}},\nabla_{\overline{j}}] = 0\\ &
    [\nabla_i,\nabla_{\overline{j}}] =
    \ka_{i\overline{j}}g_{k\overline{l}}\left(1 - {3\over c} (q+
    \overline{q})\right)- [C_i,\overline{C_j}]
\end{split}\label{finaltt}\end{equation}
Apart from the term proportional to
$g_{i\overline{j}}g_{k\overline{l}}$ in the third equation, these are
the $tt^*$ equations which were initially derived
\cite{Cecotti:1991me} using the correspondence between chiral
primaries in the NS sector and the Ramond ground states, and the
topological twisting of theories with extended supersymmetry. More
details can be found in the relevant papers. While the derivation
based on the topological twisting is more general, as it also works
for non-conformal ${\cal N}=(2,2)$ theories, it is satisfying that the
same result can be reproduced from the point of view of conformal
perturbation theory in the physical theory without using the
twisting. We discuss the role of the extra term in the next
subsection.

The main use of these equations is that for ${\cal N}=(2,2)$ theories
we can compute the connection on the bundles of chiral primaries if we
know the chiral ring coefficients. In general the chiral ring
coefficients are not constant, rather they are holomorphic functions
on the moduli space. Later we will see the simplifications that occur
for ${\cal N}=(4,4)$ theories.

Before we proceed let us mention that similarly we can compute the
curvature of the bundle of the twisted chiral ring $\widetilde{\cal
  V}$ over the factor ${\cal M}_{TC}$ of the moduli space and we
similarly find the equation
\begin{equation}
  R_{a\overline{b}} = \ka_{a\overline{b}} g_{c\overline{d}}\left(1 -
  {3\over c} (q- \overline{q})\right)
  -[\widetilde{C}_a,\overline{\widetilde{C}_b}].
\end{equation}

\subsection{Some comments}
\label{subsec:blabla2}

In the original $tt^*$ equations for the Ramond ground states of the
topologically twisted theory, the term
\begin{equation}
 \ka_{i\overline{j}}
g_{k\overline{l}}\left(1 - {3\over c} (q+\overline{q})\right)
\label{extrat}
\end{equation}
was not present. This means that the connection for the Ramond
states in the topologically twisted theory is not exactly the same
as the connection for NS chiral primaries in the physical theory,
but they differ by $U(1)$ phases related to the line bundles
${\cal
  L},\overline{\cal L}$. While this extra term came out of our
computation naturally, using the general formalism for the connection
of operators, we have not fully understood why there is a difference
between the physical and twisted theories. Because of this we would
like to make some consistency checks regarding the presence of this
term.  In this section we will consider a special class of chiral
primaries and we will see that to get the correct answer for their
curvature we do indeed need the extra term \eqref{extrat}.

First we consider the case of the identity operator $I(z)$ whose
charges are $(0,0)$. Obviously its curvature over the moduli space
should be zero. This can be seen from the 4-point function
\begin{equation}
  \langle {\cal O}_i(x) \overline{{\cal O}_j}(y) I(z) I(w)\rangle =
{g_{i\overline{j}} \over |x-y|^4}
\end{equation}
This is symmetric under $x\leftrightarrow y$, so its curvature must
vanish. Now, if we compute the second term of \eqref{finaltt} on the
subspace spanned by $I(z)$ we have
\begin{equation}
[C_i,\overline{C_j}] =g_{i \overline{j}}
\end{equation}
This is precisely cancelled by the term \eqref{extrat} for
$q=\overline{q}=0$.

Another example we will consider is the chiral primary $\rho(z)$ of
highest left $U(1)$ charge $(c/3,0)$.  This is a unique field present
in any ${\cal N}=(2,2)$ theory. To compute the relevant 4-point
function we consider the bosonization of the $U(1)$ currents
\begin{equation}
  J(z) = i \sqrt{c/3}\,\,\partial H,\qquad \overline{J}(\overline{z})
= i \sqrt{c/3}\,\,\overline{\partial}\,\overline{H}
\end{equation}
where $H$, $\overline{H}$ are free compact bosons. Any operator $\varphi$
with charge $(q,\overline{q})$ can be written as
\begin{equation}
\varphi = e^{i\sqrt{3/c} (q H + \overline{q} \overline{H})} \chi
\label{bosons}
\end{equation}
with $\chi$ a neutral operator, which may be a polynomial in
$J \sim \partial H$ and $\overline{J}\sim \overline{\partial}
\overline{H}$. The field $\rho(z)$ has charges $(c/3,0)$ and using
the bosonized currents can be written as
\begin{equation}
  \rho(z) = e^{i\sqrt{c/3} H}
\end{equation}
The marginal operators are neutral so if we write them in the form
\eqref{bosons} then the $H$-dependence can be at most a polynomial
in derivatives of the fields $H,\overline{H}$, or equivalently
polynomial in the currents $J,\overline{J}$ and their derivatives.
However we know that for the marginal operators which are
descendants of chiral primaries we have
\begin{equation}
J(z)  {\cal O}(w) = \text{regular}
\end{equation}
which means that actually these marginal operators do not involve the
free boson $H(z)$ at all (similarly for the right moving $\overline{H}$).
But this implies that
\begin{equation}
  \rho(z) {\cal O}(w) = \text{regular}
\label{oopp}
\end{equation}
Now we consider the 4-point function
\begin{equation}
  \langle O_i(x) \overline{O_j}(y) \rho(z) \rho^\dagger(w)\rangle
\end{equation}
The field $\rho(z)$ is holomorphic  so we can compute the 4-point function
from the OPEs. From \eqref{oopp} we see that the only nontrivial OPE
is between $\rho(z)$ and $\rho^\dagger(w)$ which is of the form
\begin{equation}
  \rho(z) \rho^\dagger(w) = {g_{\rho\overline{\rho}}\over (z-w)^{2c/3}}
+ ...
\end{equation}
where the operators appearing in the dots only involve the free boson $H$. As
we argued the marginal operators do not couple to $H$, so the 4-point
function is equal to
\begin{equation}
   \langle O_i(x) \overline{O_j}(y) \rho(z) \rho^\dagger(w)\rangle
= {g_{i\overline{j}} g_{\rho \overline{\rho}} \over |x-y|^4 (z-w)^{2c/3}}
\end{equation}
Again this is symmetric in $x\leftrightarrow y$ so the curvature of the
field $\rho(z)$ should vanish. Looking at \eqref{finaltt} we find that
for this field
\begin{equation}
  [C_i,\overline{C_j}] = 0
\end{equation}
while the term \eqref{extrat} is also zero for $q=c/3,\overline{q}=0$.
So indeed the curvature vanishes. Similarly one can study the right
moving field $\overline{\rho}$ of charge $(0,c/3)$.  Finally we
consider the field ${\cal A} = (\rho \overline{\rho})$ of charge
$(c/3,c/3)$. Since this is the product of $\rho$ and $\overline{\rho}$
its curvature should also vanish. The second term of \eqref{finaltt}
for this field gives
\begin{equation}
  [C_i,\overline{C_j}] = -g_{i\overline{j}}g_{{\cal A} \overline{\cal A}}
\end{equation}
This is precisely cancelled by the term \eqref{extrat} for
$q=\overline{q} = c/3$.

The conclusion is that in all these cases the presence of the term
$\ka_{i\overline{j}} g_{k\overline{l}}\left(1 - {3\over c} (q+
\overline{q})\right)$ is necessary to give the correct answer for the
curvature of the operators. See also footnote \eqref{longfoot} for some
related observations.

Notice that the extra term is reminiscent of duality between
$H^{p,q}(M)$ and $H^{d-p,d-q}(M)$ for a $2d$ dimensional
Calabi-Yau manifold. Perhaps its presence/absence is related to
whether one uses the standard basis for the chiral primaries,
(which means that they are directly related to the Dolbeault
cohomology in the case of a supersymmetric sigma model), or a dual
basis. It would be interesting to explore this a bit further.

\section{The ${\cal N}=(4,4)$ superconformal algebra}
\label{sec:n4algebra}

In this section we review some basic properties of the (small) ${\cal
  N}=(4,4)$ superconformal algebra, whose OPEs can be found in
appendix \ref{app:algope}. Its R-symmetry group is $SO(4)^R =
SU(2)^R_{left}\times SU(2)^R_{right}$. The left-moving currents are
the energy momentum tensor $T$ and the currents of the
$SU(2)^R_{left}$ symmetry $J_i,i=1,2,3$. The left-moving supercurrents
fall into two doublets of the $SU(2)^R_{left}$ and will be denoted by
$G^{a,i}$, $a,i=1,2$, obeying a reality condition
$G^{a,i}=\epsilon^{ab}\epsilon^{ij} (G^{b,j})^*$. The $SU(2)^R_{left}$
acts on the $a$ index. The level of the $SU(2)^R_{left}$ current
algebra is equal to $k={c\over 6}$, where $c$ is the central charge of
the theory. We have the same structure on the right-moving sector and
we denote the right-moving generators by $\overline{T},\overline{J}_i$
and $\overline{G}^{a,i}$.

The ${\cal N}=(4,4)$ algebra has an outer automorphism which rotates
the supercurrents, leaving all bosonic generators unchanged. In the
notation $G^{a,i}$ for the supercurrents the outer automorphism is
$SU(2)$ rotations of the $i$-index. In general this transformation is
not a symmetry of the theory, as there is no corresponding conserved
current generating it. We will call it $SO(4)^{outer} =
SU(2)^{outer}_{left}\times SU(2)^{outer}_{right}$. We remind that the
$SO(4)^R$ symmetry rotates both the supercurrents and the R-currents
$J_i,\overline{J}_i$, while $SO(4)^{outer}$ rotates only the
supercurrents. The full automorphism group of the algebra is
$G=SO(4)^R \times SO(4)^{outer}$.

\subsection{${\cal N}=(2,2)$ subalgebras}
\label{subsec:2sub}

An ${\cal N}=(4,4)$ theory can of course be also seen as ${\cal
  N}=(2,2)$. To pick an ${\cal N}=(2,2)$ subalgebra of the ${\cal
  N}=(4,4)$ we have to do two things. First we have to choose a Cartan
generator of $SU(2)^R_{left}$ and one of $SU(2)^R_{right}$ that we
will identify with the $U(1)$ R-charge of the ${\cal N}=(2,2)$
theory. This gives us a freedom of ${\left(SU(2)\over U(1)\right)}
\times {\left(SU(2)\over U(1)\right)}$. Notice that the different
choices can be related by an $SO(4)^R$ transformation which is a
symmetry of the theory, so they are essentially equivalent. Second,
after we pick the direction of the ${\cal N}=(2,2)$ R-charge
generators, we still have an extra ${\left(SU(2)\over U(1)\right)}
\times {\left(SU(2)\over U(1)\right)}$ freedom to choose which
combination of the supercurrents $G^{\pm+},G^{\pm-}$ will be
identified as the ``standard'' supercurrents of the ${\cal N}=(2,2)$
theory. The different choices of the supercurrents are related by the
outer automorphism $SO(4)^{outer}$ which is not a symmetry, so in
general the different ${\cal N}=(2,2)$ subalgebras of this type will
be inequivalent.

Notice that once we make the first choice and orient the $U(1)\times
U(1)$ generators in the $SO(4)^R$, we completely fix which operators
we will call chiral primaries (the operators with
$(L_0,\overline{L}_0) = (J^3_0,\overline{J}^3_0)$), independent of the
remaining ambiguity in the choice of the supercurrents. This is a
consequence of the fact that for a superconformal primary the
following conditions are equivalent\footnote{Notice the difference in
  conventions between the normalization of the R-charge for the ${\cal
    N}=2$ and ${\cal N}=4$ cases. In the ${\cal N}=2$ theories, the
  $U(1)$ charge $J$ is normalized to take integral values and the BPS
  bound is $L_0 = J_0/2$. In the ${\cal N}=4$ conventions, which we
  are going to follow in the rest of this paper, the eigenvalues of
  $J_0^3$ are half-integers and the BPS bound is $L_0 = J_0^3$.}
\begin{equation}
  (L_0 -J_0^3)|\phi\rangle =0\quad \Leftrightarrow \quad
G^{++}_{-1/2}|\phi\rangle =0 \quad
\Leftrightarrow\quad G^{+-}_{-1/2}|\phi\rangle=0
\label{chiral44}
\end{equation}
Even though the definition of a chiral primary does not depend on the
choice of the supercurrents, its descendants do depend on it. So the
inequivalent ${\cal N}=(2,2)$ subalgebras with the same $SO(4)^R$
orientation but with different $SO(4)^{outer}$ orientation have the
same chiral primaries, but different descendants.

\subsection{Short representations}

In this section we describe the short representations of the ${\cal
  N}=4$ algebra, i.e. those which saturate the BPS bound
\cite{Eguchi:1987sm} . For simplicity we will only discuss the
representation on the left-moving sector. To get a full representation
of the ${\cal N}=(4,4)$ algebra we have to tensor a left with a
right-moving representation. Short representations can be constructed
by starting with a chiral primary field and then acting on it with the
generators of the algebra. The conformal dimension and R-charge of a
chiral primary satisfy
\begin{equation}
L_{0}|\phi\rangle= J_0^3|\phi\rangle=q|\phi\rangle
\end{equation}
We use the notation $(L_0,J^3_0) = (q,q)$ for the conformal dimension
and $J^3$ charge. Such a field is annihilated by the supercurrents
\begin{equation}
G^{++}_{-1/2} |\phi\rangle = G^{+-}_{-1/2} |\phi\rangle = 0
\end{equation}
To construct the representation we first discuss the action of
$G^{a,i}_{-1/2}$ and $J^i_0$ on the highest weight state. To
start, we can act on $\phi$ with the lowering operator $J^{--}_0$
with respect to the $J_0^3$ charge. This gives us the fields
$(J^{--})^n \phi$ with quantum numbers $(L_0,J^3_0)=(q,q-n)$.
Obviously we can act at most $2q$ times before the state is
annihilated. This set of fields forms a $2q+1$ dimensional
spin-$q$ representation of $SU(2)^R_{left}$, and they all have the
same conformal dimension\footnote{We called the top
  component $\phi$ chiral primary, but each of the fields
  $(J^{--})^n\phi$ would also be ``chiral primary'' under a different
  orientation of the $J^3$ axis.}. Also notice that these states are
singlets of the $SU(2)^{outer}_{left}$.

We can construct more states of the representation by acting on $\phi$
with one supercurrent. The only supercurrents that do not annihilate
$\phi$ are $G^{-+},G^{--}$ which mix under the action of
$SU(2)^{outer}_{left}$. This way we get two states
\begin{equation}
  |\psi^+\rangle = G^{-+}_{-1/2} |\phi\rangle,\quad |\psi^-\rangle =
  G^{--}_{-1/2} |\phi\rangle
\end{equation}
These states have charges equal to $(L_0,J^3_0) = (q+{1\over
  2},q-{1\over 2})$ and they are a doublet of the
$SU(2)^{outer}_{left}$. Acting on these states with $J^{--}$ we can
complete them into spin $q-{1\over 2}$ representation of
$SU(2)^{R}_{left}$.

Finally we can get new states acting on $\phi$
with two supercurrents. This gives the state
\begin{equation}
|\Phi\rangle=G^{-+}_{-1/2}G^{--}_{-1/2} |\phi\rangle
\end{equation}
It has $(L_0,J^3_0) = (q+1,q-1)$. It is a singlet of
$SU(2)^{outer}_{left}$. Acting on this state with $J^{--}$ we generate
a spin $q-1$ representation of $SU(2)^R_{left}$.

The full representation of the superconformal algebra is generated by
taking conformal descendants of the states described above. This is
the structure of the typical short representation. If we start with a
chiral primary of low enough conformal dimension we get special short
representations that we review in the next subsection.

\subsection{Special short representations}
\label{subsec:specialshort}

First we consider the shortest nontrivial representation. If we start
with a chiral primary with $(L_0,J^3_0)=({1\over 2},{1\over 2})$ and
act with the supercurrents $G^{-+},G^{--}$ we get two states with
$(L_0,J^3_0) = (1,0)$. We cannot act again with the supercurrents
since it would give a negative value for the R-charge. The
representation is terminated and is shorter than the typical short
representation. The two fields $|\psi^+\rangle = G^{-+}_{-1/2}|
\phi\rangle, \, |\psi^-\rangle = G^{--}_{-1/2} | \phi\rangle$ are
singlets of the $SU(2)^R_{left}$ and a doublet of
$SU(2)^{outer}_{left}$. If we tensor them with a similar
representation from the right-moving sector we get fields with
conformal dimension $(L_0,\overline{L}_0) = (1,1)$ which are singlets of
$SO(4)^R$, but which transform under $SO(4)^{outer}$. These are the
marginal operators of the theory.

Second let us consider the case where we start with a chiral primary
$|\phi\rangle$ with $(L_0,J^3_0) = (1,1)$. From the previous analysis
we see that the state $|\Phi\rangle =G^{-+}_{-1/2}G^{--}_{-1/2}
|\phi\rangle$ has $(L_0,J^3_0) = (2,0)$. It is a singlet of
$SU(2)^R_{left}$ and also a singlet of $SU(2)^{outer}_{left}$. If we
tensor it with a similar representation from the right-moving sector
we get fields which have conformal dimension $(L_0,\overline
{L}_0)=(2,2)$ and are singlets of the $SO(4)^R$ (and also singlets of
$SO(4)^{ outer})$. These fields are the leading irrelevant operators
which are singlets under $SO(4)^R$, so they break the conformal
invariance but not the ${\cal N}=(4,4)$ supersymmetry. Notice that
there are no other $SO(4)^R$ singlet operators in the short multiplets
of the algebra.

\subsection{The moduli space of ${\cal N}=(4,4)$ SCFTs}
\label{subsec:mod44}

Let us now use the restrictions of the ${\cal N}=(4,4)$ superconformal
symmetry on the structure of the moduli space. We review the
well-known argument which completely determines the local structure of
the moduli space of any ${\cal N}=(4,4)$ SCFT
\cite{Seiberg:1988pf},\cite{Cecotti:1990kz}.

As we saw before, motion on the moduli space is generated by
descendants of chiral primaries with $(q,\overline{q}) = ({1\over
  2},{1\over 2})$. Let us say that there are $n$ multiplets of this
form. Each multiplet gives 4 real marginal operators so the dimension
of the moduli space will be $4n$. The (local) holonomy on this space is in
general $SO(4n)$.  However the marginal operators come in groups of 4
from a single chiral primary. We want to take advantage of this fact
to restrict the holonomy of the moduli space.  The chiral primaries
$\phi_i$ of weight $({1\over 2},{1\over 2})$ are sections of a vector
bundle and have themselves some holonomy. Also, to go from the
chiral primaries to the moduli, we have to act with the
supercurrents. This means that the marginal operators are sections of
a bundle which is the tensor product of the bundle of the chiral
primaries with the bundle of the supercurrents. So the holonomy on the
tangent bundle will be the product of the holonomy for the chiral
primaries and the holonomy of the supercurrents. The latter
contributes a factor of $SO(4)$ associated to the $SO(4)^{outer}$
ambiguity of the supercurrents. So the moduli space is a $4n$ dimensional
manifold whose holonomy $K$ is reduced: $K\in SO(4)\times SO(n) \in
SO(4n)$.  Such manifolds are constrained by Berger's
classification. After a few more easy arguments\cite{Cecotti:1990kz}
we conclude that the moduli space is a locally a homogeneous space of
the form
\begin{equation}
SO(4,n)\over SO(4)\times SO(n)
\label{modulispace}
\end{equation}
This means that the local geometry of the moduli space is completely
fixed by supersymmetry, and can be determined if we know the number of
marginal operators which fixes $n$. In the case of AdS$_3$/CFT$_2$ we
have $n=5$ for $X=T^4$ and $n=21$ for $X=K3$. \footnote{Notice that
  the moduli space is of the same form for all values of the central
  charge, so it seems to be independent of $Q_1,Q_5$. However we have
  not fixed the overall scale of the metric on the coset. This scale
  does depend on the central charge.}

Before we proceed, let us stress an important point. From each chiral
primary $\phi_i$ with $(q,\overline{q}) = ({1\over 2},{1\over 2})$ we
get marginal operators which are singlets of $SO(4)^R$
\begin{equation}
   G^{-r}_{-1/2}\overline{G}^{-s}_{-1/2} \cdot \phi_i
\label{ccmarginal}
\end{equation}
where $r,s$ can take any value in $\{+,-\}$ independently. We can
also consider operators of the form
\begin{equation}
  G^{+r}_{-1/2}\overline{G}^{+s}_{-1/2}\cdot
\overline{\phi_i}
\label{aamarginal}
\end{equation}
A very important property is that the tangent space of the moduli
space is completely spanned by the operators of the form
\eqref{ccmarginal} alone. The same is true about the operators of the
form \eqref{aamarginal}. This can be roughly understood from the
counting. The tangent space of the moduli space has real dimension
$4n$. The set of operators of the form \eqref{aamarginal} has real
dimension $8n$, but we have to impose a reality condition for the
operator used to deform the theory so we are left with half of them
which is equal to $4n$.

Moreover, starting from a $(cc)$ chiral primary $\phi$ of charge
$(1/2,1/2)$ we can use the $SU(2)^R_{right}$ to rotate it to a $(ca)$
primary $\psi$ of charge $(1/2,-1/2)$. This will also lead to marginal
operators of the form
\begin{equation}
  G^{-r}_{-1/2}\overline{G}^{+s}_{-1/2} \cdot \psi_i
\label{camarginal}
\end{equation}
\begin{equation}
  G^{+r}_{-1/2}\overline{G}^{-s}_{-1/2}
\cdot \overline{\psi_i}
\label{acmarginal}
\end{equation}
Again each of the two sets \eqref{ccmarginal}, \eqref{acmarginal}
fully spans the tangent space. To summarize, from each chiral
primary $\phi_i$ with $(q,\overline{q}) = ({1\over
  2},{1\over 2})$ we get 4 real marginal operators which are singlets
of $SO(4)^R$ and which transform under $SO(4)^{outer}$. These
operators can be written in different ways \eqref{ccmarginal},
\eqref{aamarginal}, \eqref{camarginal}, \eqref{acmarginal}.

From each chiral primary $\phi_i$ with $(q,\overline{q}) = (1,1)$
we get a single real operator with $(L_0,\overline{L}_0) = (2,2)$
which is a singlet of $SO(4)^R\times SO(4)^{outer}$. These are the
only irrelevant operators that exist which preserve global ${\cal
N}=(4,4)$ supersymmetry but which break conformal invariance. We
emphasize that this is a finite number of irrelevant operators.

\section{The chiral ring of ${\cal N}=(4,4)$ theories}
\label{sec:mainr}

Finally, we are ready to consider the moduli dependence of the chiral
ring in ${\cal N}=(4,4)$ superconformal field theories.

\subsection{Curvature of the ${\cal N}=(4,4)$ algebra, the bosonic currents}
\label{subsec:currentcurv}

We start with the curvature of the generators of the algebra. In
principle their curvature can take values in the automorphism group
$SO(4)^R\times SO(4)^{outer}$ of the ${\cal N}=(4,4)$ algebra.  To
compute the curvature of the R-currents $J^i(z)$ we need the following
4-point function
\begin{equation}
  \langle {\cal O}_\mu(x){\cal O}_\nu(y) J^i(z) J^j(w) \rangle
  \end{equation}
where ${\cal O}_\mu, {\cal O}_\nu$ are marginal operators. As a
function of $z$ this 4-point function is holomorphic so it is
completely determined by its singularity structure when $J^i(z)$
approaches the other insertions. We have the following OPEs
\begin{equation}\begin{split}
 & J^i(z) {\cal O}_\mu(x) = \text{regular}\\ & J^i(z) J^j(w) = {k\over
      2}{\delta^{ij}\over (z-w)^2} + i {\epsilon_{ijk} J^k(w) \over
      z-w} + ...
\end{split}\label{opeb}\end{equation}
The proof of the first OPE is based on the fact that in an ${\cal
  N}=(4,4)$ SCFT the marginal operators are descendants of chiral
primaries, see appendix \ref{app:curmod} for details.

Since the OPE of a current with a marginal operator is regular, the
only contribution to the 4-point function is when the two currents
come together. Then we have to use the second OPE in \eqref{opeb}. The
second term of that OPE involves $J^k(w)$ and is charged under
$SO(4)^R$, so its 3-point function with the neutral marginal operators
is zero. So only the first term of the $JJ$ OPE contributes and we
find
\begin{equation}
  \langle {\cal O}_\mu(x){\cal O}_\nu(y) J^i(z) J^j(w) \rangle =
          {k\over 2}{\ka_{\mu\nu} \delta^{ij} \over |x-y|^4 (z-w)^2}
\label{current4}
\end{equation}
where $g_{\mu\nu}$ is defined by the two point function
\begin{equation}
  \langle {\cal O}_\mu(x) {\cal O}_\nu(y) \rangle = {\ka_{\mu\nu}
    \over |x-y|^4}
\end{equation}
The 4-point function \eqref{current4} is symmetric in
$\mu\leftrightarrow \nu$, and has no singularities as $x,y\rightarrow
z,w$ so the curvature of the R-currents, according to \eqref{curvexp},
is zero.

The conclusion is that there is no curvature for the $SO(4)^R$
symmetry over the moduli space. From the AdS/CFT point of view this is
according to our expectations. The R-symmetry of the CFT corresponds
to the isometry group of the three-sphere in $AdS_3\times S^3\times
K3$. Intuitively we expect that changing the moduli of the
compactification should not induce a rotation of the $S^3$. The CFT
analysis verifies this intuition.

\subsection{The supercurrents}

The supercurrents are charged under both the R-symmetry $SO(4)^R$
and the outer automorphism $SO(4)^{outer}$. We found that the
R-symmetry does not have curvature over the moduli space. However,
as we will see the supercurrents mix among themselves by an
$SO(4)^{outer}$ rotation. In principle this curvature can be
computed by an analysis of 4-point functions of two supercurrents
with two marginal operators, as in section \ref{subsec:curvalg2}.
A faster way to derive the answer is the following. In sections
\ref{subsec:specialshort}, \ref{subsec:mod44} we explained that
the marginal operators are constructed by acting with
supercurrents on chiral primaries of charge $\left({1\over 2},{1
\over 2}\right)$. If we call ${\cal G}^L,{\cal
  G}^R$ the bundles of left and right-moving supercurrents, ${\cal
  V}_{1/2,1/2}$ the bundle of chiral primaries of charge
$\left({1\over 2},{1 \over 2}\right)$ and $\hat{\cal O}$ the bundle of
marginal operators, then clearly $\hat{\cal O}$ is the tensor product
of the other three bundles
\begin{equation}
  \hat{\cal O} = {\cal G}^L\times {\cal G}^R \times {\cal
    V}_{1/2,1/2}
\label{tensorbundles}
\end{equation}
Moreover, $\hat{\cal O}$ is isomorphic to the tangent bundle $T{\cal
  M}_{CFT}$ of the moduli space \eqref{modulispace}. The connection on
$T{\cal M}_{CFT}$ is described by the spin connection on the coset
\eqref{modulispace} which takes values in its isotropy group
$SO(4)\times SO(n)=SU(2)^L\times SU(2)^R \times SO(n)$. From the
tensor product structure \eqref{tensorbundles}, it is clear that the
connection of ${\cal G}^L$ is given by the $SU(2)^L$ factor of the
connection of the tangent bundle, the connection of ${\cal G}^R$ by
$SU(2)^R$ and that of ${\cal V}_{1/2,1/2}$ by the $SO(n)$ factor. It
should be easy to rederive this from a CFT computation of the 4-point
functions, as in section \ref{subsec:curvalg2}. The main point is that
the supercurrents have nonzero $SO(4)^{outer}$ curvature which is
directly computable by the geometry of the coset \eqref{modulispace}
without any further input from the dynamics of the CFT.

\subsection{The 3-point functions are covariantly constant}

We denote by $\phi_i$ the chiral primary fields of the ${\cal
  N}=(4,4)$ theory, that is, fields which are Virasoro primaries and
satisfy $L_0=J_0^3,\overline{L}_0 = \overline{J_0}^3$.  Their OPE has
the form
\begin{equation}
  \phi_i(z) \phi_j(w) = C^k_{ij} \phi_k(w) +...
\end{equation}
where $\phi_k$ is also chiral primary and $C^k_{ij}$ are the structure
constants of the chiral ring. We can also consider the 2- and 3-point
functions related by
\begin{equation}\begin{split}
& C_{ij\overline{k}}= \langle \phi_i(0)\phi_j(1)
    \overline{\phi_k}(\infty)\rangle\\ & g_{i\overline{j}} = \langle
    \phi_i(0)\overline{\phi_j}(\infty)\rangle \\ & C_{ij\overline{k}}
    = C^l_{ij} g_{l\overline{k}}
\end{split}\end{equation}
We want to compute how the chiral ring 3-point functions vary as we
move on the moduli space. For this we need to compute the 4-point
function
\begin{equation}
   \langle {\cal O}(z)\phi_i(z_1) \phi_j(z_2)
   \overline{\phi_k}(z_3)\rangle
\end{equation}
where ${\cal O}(z)$ is a marginal operator. We want to show that this
4-point function is zero.

What is important for the proof is that, as explained in section
\ref{subsec:mod44}, for ${\cal N}=(4,4)$ theories any marginal
operator can be written as the linear combination of descendants of
antichiral primaries\footnote{Of course this is not true in $(2,2)$
  theories, which is why in those theories we have
  $\nabla_{\overline{m}} C_{ijk} = 0$, but in general $\nabla_m
  C_{ijk}\neq 0$.}
\begin{equation}
  {\cal O}(z) = A^{\overline{j}}_{rs} G_{-1/2}^{+r}
  \overline{G}_{-1/2}^{+s} \cdot \overline{\phi_j}(z)
\label{antit}
\end{equation}
where $A^{\overline{j}}_{rs}$ are some appropriate constants.

Now we consider the Ward
identity\cite{Dijkgraaf:1990dj},\cite{Dixon:1989fj}: on the sphere for
a current $G^{+r}(w)$ of dimension $3/2$. For any set of primary
operators $\varphi$ we have
\begin{equation}
  \langle \oint \xi(w) G^{+r}(w) \varphi_{i_1}(z_1)...\varphi_{i_n}(z_n)\rangle
 = \sum_{i=1}^n \xi(z_i) \langle \varphi_{i_1}(z_1) ... (G_{-1/2}^{r+}\cdot
\varphi_i)(z_i)...\varphi_{i_n}(z_n)\rangle =0
\end{equation}
where $\xi(w)$ is a globally defined holomorphic vector field of the form
\footnote{We do not consider conformal killing vector fields of the
  form $\xi(w)\sim w^2$ for the following reason: since $G^{+r}(w)$
  has dimension $3/2$ the correlator $\langle G^{+r}(w)
  \varphi_{i_1}(z_1)...\varphi_{i_n}(z_n)\rangle$ falls-off like
  ${1\over w^3}$ as $w\rightarrow \infty$. So if we do not want to
  have a contribution from infinity $\xi(w)$ can be at most linear in $w$.}
\begin{equation}
  \xi(w) = a w + b
\label{ckv}
\end{equation}
where $a,b$ are arbitrary complex numbers. Using this Ward identity we have
\begin{equation}
  \xi(z)  \langle {\cal O}(z)\phi_i(z_1) \phi_j(z_2)
   \overline{\phi_k}(z_3)\rangle +
\xi(z_3) \langle  \left(A^{\overline{j}}_{rs}
  \overline{G}_{-1/2}^{+s} \cdot \overline{\phi_j}(z)\right)
\phi_i(z_1) \phi_j(z_2))(G^{+r}_{-1/2}\cdot \overline{\phi_k})(z_3)\rangle =0
\end{equation}
where we used $G_{-1/2}^{r+}\cdot \phi_i = G_{-1/2}^{r+}\cdot \phi_j
=0$ since these fields are chiral primaries.  Now if we choose the
vector field $\xi(w)$ in \eqref{ckv} is such a way that $\xi(z)=1$ and
$\xi(z_3)=0$ we immediately get
\begin{equation}
   \langle {\cal O}(z)\phi_i(z_1) \phi_j(z_2)
   \overline{\phi_k}(z_3)\rangle =0
\end{equation}
Since the 4-point function vanishes for all marginal directions,
following the definition of the covariant derivative \eqref{covder} we
find that there is no need for any subtraction and the covariant
derivative of the 3-point function is zero. This means that the chiral
ring is covariantly constant
\begin{equation}
\nabla C^k_{ij} = 0  ,\qquad \nabla C_{ij\overline{k}} = 0
\end{equation}
where we obviously also have $\nabla g_{i\overline{j}}=0$ since the
connection is compatible with the metric \eqref{metriccomp}. This
shows the non-renormalization of 3-point functions of chiral
primaries \footnote{It might be possible to argue that certain
  correlators of short multiplets in ${\cal N}=(4,4)$ SCFTs respect an
  $SO(4)^{outer}$ selection rule, even though the latter is not a
  proper symmetry of the full theory, in analogy with the ``bonus''
  $U(1)_Y$ symmetry in ${\cal N}=4$ \cite{Intriligator:1998ig},
  \cite{Intriligator:1999ff}. From this selection rule the
  non-renormalization of 3-point functions would follow.}. The result
is valid for any ${\cal N}=(4,4)$ theory, in particular in
AdS$_3$/CFT$_2$ it is true not only in the large $N$ limit but even
for finite values of $N=Q_1 Q_5$.

\subsection{Non-renormalization of extremal correlators}

More generally the same argument can be used to show that extremal
correlators of the form
\begin{equation}
  \langle \phi_{i_1}(z_1)...\phi_{i_n}(z_n) \overline{\phi_j}(y)\rangle
\label{extrem}
\end{equation}
are also not renormalized\footnote{We would like to thank R. Gopakumar
  and S. Minwalla for bringing this point to our attention.}. For this
we need the $n+2$-point function
\begin{equation}
   \langle {\cal O}(z)\phi_{i_1}(z_1)...\phi_{i_n}(z_n)
   \overline{\phi_j}(y)\rangle
\label{derextr}
\end{equation}
where ${\cal O}(z)$ is a marginal operator written as the descendant
of an antichiral primary \eqref{antit}. We then follow the same steps
as before. We use the Ward identity for the supercurrent $G^{+r}$ by
appropriately choosing $\xi(w)$ to have the value one at $z$ and zero
at $y$. All the fields at $z_i$ do not contribute since they are
chiral primaries and they are annihilated by $G^{+r}_{-1/2}$. So the
$n+2$-point function \eqref{derextr} vanishes
\begin{equation}
   \langle {\cal O}(z)\phi_{i_1}(z_1)...\phi_{i_n}(z_n)
   \overline{\phi_j}(w)\rangle=0 .
\end{equation}
This means that the
extremal correlator \eqref{extrem} is covariantly constant over the moduli
space and receives no renormalizations.

Of course the same argument cannot be applied if we have at the same
time two or more chiral fields and two or more antichiral fields since then
we cannot choose $\xi$ appropriately to cancel all contributions.

\subsection{The curvature of chiral primaries}

The chiral ring is a multiplication between chiral primaries. The
chiral primaries themselves are sections of bundles ${\cal V}_q$ with
nontrivial connections. We showed that the multiplication between
these bundles
\begin{equation}
  C^k_{ij}: {\cal V}_{p}\otimes {\cal V}_{q} \rightarrow {\cal V}_{p+q}
\end{equation}
is covariantly constant. However this does not mean that the bundles
are flat. In this section we want to compute the curvature of the
bundles of chiral primaries for ${\cal N}=(4,4)$ theories.

We proceed by using the fact that the ${\cal N}=(4,4)$ algebra has
many inequivalent ${\cal N}=(2,2)$ subalgebras. If we consider two
marginal operators which are descendants of the $(cc)$ and $(aa)$ ring
of a specific ${\cal N}=(2,2)$ subalgebra, we can compute the
curvature along these two directions by using the results of our
analysis in section \ref{sec:review22}. Then by varying the chosen
${\cal N}=(2,2)$ subalgebra we can effectively scan all (pairs of)
directions on the moduli space and thus compute the curvature in all
directions.

As we explained in section \ref{subsec:2sub}, to pick an ${\cal
  N}=(2,2)$ subalgebra of the ${\cal N}=(4,4)$ theory, we first need
to pick Cartan elements of the $SO(4)^R$. Let us take them to be
$(J^3,\overline{J}^3)$. Then we have the $SO(4)^{outer}$ ambiguity in
choosing the supercurrents. Following
\cite{Berkovits:1994vy},\cite{Berkovits:1999im} we can define
\begin{equation}\begin{split}
&  \widehat{G^+}(u)=\,\, u_1 G^{++} + u_2 G^{+-}\\
&  \widehat{G^-}(u)= u_1^* G^{--} + u_2^* G^{-+}
\end{split}\end{equation}
for any complex numbers $u_1,u_2$ satisfying $|u_1|^2 +
|u_2|^2=1$. Then the currents $T(z),\widehat{G^\pm}(z),J^3(z)$ satisfy
the standard ${\cal N}=2$ superconformal algebra OPEs. We can do the
same on the right-moving sector where we also have to choose complex
numbers $\overline{u}_1, \overline{u}_2$ satisfying
$|\overline{u}_1|^2 + |\overline{u}_2|^2 = 1$.  Let us combine all
these complex numbers in the symbol ${\cal U}=(u_1,u_2,
\overline{u}_1,\overline{u}_2)$. Now consider the marginal operators
\begin{equation}\begin{split}
& {\cal O}_{({\cal U},i)} = {1\over 2}\widehat{G^-}(u)
    \widehat{\overline{G}^-}(\overline{u}) \cdot \phi_i\\ &
    \overline{{\cal O}_{({\cal U},j)}} ={1\over 2} \widehat{G^+}(u)
    \widehat{\overline{G}^+}(\overline{u}) \cdot \overline{\phi_j}
\label{marginalu}
\end{split}\end{equation}
where $\phi_i$ are $(cc)$ fields and $\overline{\phi_j}$
are $(aa)$ fields.  The curvature along any pair of marginal operators
of this form can be computed from the $tt^*$ equations and we have
\begin{equation}\begin{split}
 & [\nabla_{({\cal U},i)},\nabla_{({\cal U},j)}]= [\nabla_{({\cal
          U},\overline{i})},\nabla_{({\cal U},\overline{j})}]= 0\\ &
    [\nabla_{({\cal U},i)},\nabla_{({\cal
          U},\overline{j})}]=\ka_{i\overline{j}}
    g_{k\overline{l}}\left(1 - {3\over c} (q+ \overline{q})\right)
    -[C_i,\overline{C}_{\overline{j}}]
\label{curvu}
\end{split}\end{equation}
for all possible ${\cal U}$'s and where $\nabla_{({\cal U},i)}$
denotes the covariant derivative with respect to the marginal
operator ${\cal O}_{({\cal U},i)}$. By varying ${\cal U}$ these
equations give us the curvature in all possible directions of the
moduli space. In other words, if we want to compute the curvature
of the bundle of chiral primaries along two specific tangent
vectors on the moduli space, then there is enough freedom to
rewrite the curvature operator in those direction as a linear
combination of the curvature along pairs of vectors where for each
pair the factor ${\cal U}$ is the same and we can use
\eqref{curvu}. Crucial here is the observation that for any ${\cal
U}_1$ and ${\cal U}_2$, we can always rewrite ${\cal O}_{({{\cal
U}_2},i)}$ as a linear combination of ${\cal O}_{({{\cal U}_1},i)}$
and $ \overline{{\cal O}_{({{\cal U}_1},i)}}$.

\subsection{Real structure of the chiral ring}

In ${\cal N}=(4,4)$ theories it is more convenient to use a real
basis for the chiral ring.  Consider the $(cc)$ primaries $\phi_i$ of
charge $(q,\overline{q})$. We can transform them into $(aa)$ fields
in two ways. First, we can take the hermitian conjugate
\begin{equation}
\phi\rightarrow \overline{\phi_i}\equiv \phi_i^\dagger
\end{equation}
Second, we can rotate them using the $SU(2)^R_{left}\times SU(2)^R_{right}$
\begin{equation}
  \phi_i \rightarrow \widetilde{\phi}_i = {1\over T_{q,\overline{q}}}(J^{--})^{2q}
  (\overline{J}^{--}) ^{2\overline{q}}\cdot\phi_i
\end{equation}
where $T_{q,\overline{q}}$ is a real normalization factor chosen
in such a way that the norm of $|\phi_i\rangle$ equals the norm of
$|\widetilde{\phi}_i\rangle$. These two procedures generate the
same set of $(aa)$ fields, so there must be a matrix $M$ relating
the two
\begin{equation}
  \overline{\phi_i} = M_{\overline{i}}^j \widetilde{\phi}_j
\end{equation}
where $M$ must satisfy
\begin{equation}
  MM^* = I.
\end{equation}
It is convenient to pick a basis  $\phi_I, I=1,...,n$ for the $(cc)$
fields in which
$M_{\overline{i}}^j = \delta_{\overline{i}}^j$. Then
\begin{equation}
\left(\phi_I\right)^\dagger = {1\over T_{q,\overline{q}}}
(J^{--})^{2q} (\overline{J}^{--}) ^{2\overline{q}}\cdot\phi_I,\qquad
\end{equation}
In this basis the metric $G_{IJ}$ becomes real, and by a second
(real) change of basis we can take it to be $\delta_{IJ}$
\begin{equation}
  \langle\phi_I(z) \overline{\phi_J}(w) \rangle = {\delta_{IJ} \over |z-w|^4}
\label{ournorm}
\end{equation}
Moreover, in this basis the chiral ring coefficients are also real
\begin{equation}\begin{split}
&\phi_I(z) \phi_J(w) = C_{IJ}^K \phi_K(w) +...  \\
& \left(C^K_{IJ}\right)^* = C^K_{IJ}
\end{split}\end{equation}
Notice that since the action of $J^{--}$ does not change under
parallel transport (since we computed that the curvature of the
currents $J^i$ is zero), and also the action of the $\dagger$ on
operators is unambiguously defined, it means that the choice of a real
basis is invariant under parallel transport.  The bundles ${\cal V}_q$
of chiral primaries are actually {\it real} vector bundles in the case
of ${\cal N}=(4,4)$ theories\footnote{Notice that when we say ``real
  basis'' we do not mean that the operators $\phi_I$ satisfy $\phi_I =
  \phi_I^\dagger$, which is impossible for operators of definite
  nonzero R-charge. Instead what we mean is that in this basis the
  inner product and the chiral ring coefficients between these
  operators become real. The actual operators remain ``complex'', or
  geometrically the $(p,q)$ differential forms in the target space
  corresponding to the chiral primaries are still complex forms.}.

\subsection{Final expression for the curvature}

Now let us consider the tangent space of the moduli space
\eqref{modulispace}.  The holonomy group is $SO(4)\times SO(n)$, so
it is convenient to pick a vielbein basis where the tangent vectors
are decomposed as $X^\mu = X^{a,I}$, where $a$ transforms under
$SO(4)$ and $I$ under $SO(n)$. These tangent vectors correspond to
marginal operators, which can be written as descendants of chiral
primaries of charge $(1/2,1/2)$ as in \eqref{ccmarginal}. The index
$I$ is associated to the chiral primary $\phi_I$ of charge $(1/2,1/2)$
in the real basis described above, while the index $a$ is related to
the combination of the supercurrents $G^{-\pm}$ and
$\overline{G}^{-\pm}$ that we act with on the chiral primary to get
the marginal operator. From \eqref{curvu} it is easy to see that
curvature of the bundle of chiral primaries in a real basis has the
form
\begin{equation}
  \left(R_{\mu\nu}\right)_M^N =\delta_{ab}
  \delta_{IJ}\delta_M^N\left(1 - {3\over c} (q+ \overline{q})\right)
  -\delta_{ab} \left( C_{IM}^K \delta_{KL}C_{JP}^L\delta^{PN}
  -\delta_{MP} C_{JK}^P \delta^{KL}C_{IL}^N\right)
\label{curvef}
\end{equation}
where the indices $\mu=(a,I),\nu=(b,J)$ denote two tangent directions
$\mu,\nu$ decomposed into their $SO(4)\times SO(n)$ factors\footnote{\label{longfoot}
  Notice that the first term in the curvature is symmetric in
  $\mu,\nu$, which seems unacceptable for a curvature
  operator. However this term should precisely cancel the symmetric
  part of the second term, so that the total expression for the
  curvature is actually antisymmetric. Some simple examples of these
  cancellations were seen in section \ref{subsec:blabla2}. While the
  antisymmetry of the curvature operator is guaranteed from general
  principles (since the connection is compatible with the
  Zamolodchikov metric) it is not manifest in the form
  \eqref{curvef}. A small check is to consider the trace of the
  curvature, that is the case $I=J$. Then we can see from the
  target space point of view that the term $(C_I C^{T}_J - C^{T}_J C_I)$ is
  proportional to the commutator $[L,\Lambda]$ where the operator $L$
  is multiplication with the K\"ahler form and $\Lambda$ the adjoint
  operator. From standard arguments this is a commutator of the
  Lefschetz $SU(2)$ algebra where $J^+ = L, J^- = \Lambda, J^3 =
  (q+\overline{q} - \text{dim}(M))/2$ where $\text{dim} = c/3$ is the
  complex dimension of the target space and the operators are acting
  on $(q,\overline{q})$ forms. Thus we have $[L,\Lambda] =
  (q+\overline{q} - \text{dim}(M))/2$. Then the trace of the second
  term in \eqref{curvef} is proportional to the first term up to a factor
  of $c/3$. This factor is explained in the following way: we have normalized
  the $\phi_I,\phi_J$ operators so that their 2-point function is 
  \eqref{ournorm}. On the
other hand the two point function of the K\"ahler form should be proportional
to $c/3$, as can be seen from the current correlator 
$\langle JJ \rangle \sim c/3$.
Taking this factor into account we find that the trace of 6.29 
exactly cancels.}.  Notice that from R-charge conservation, if the fields
$M,N$ have charge $(q,\overline{q})$ then the sum over $K,L$ in the
second term of \eqref{curvef} runs over fields with charge
$(q+1/2,\overline{q}+1/2)$, while in the third term over fields with
charge $(q-1/2,\overline{q}-1/2)$. So the curvature of the chiral
primaries of given charge is determined by the chiral ring
coefficients of them with those which are one unit of charge higher
and one unit of charge lower. The curvature can be written as
\begin{equation}
R_{\mu\nu} = \delta_{ab} \delta_{IJ}\left(1 - {3\over c} (q+
\overline{q})\right) - \delta_{ab} (C_I C^{T}_J - C^{T}_J C_I)
\label{finalcurv}
\end{equation}
Where we have not shown the matrix indices $M,N$ on the curvature
operator and it is always implied that $\mu = (a,I),\nu=(b,J)$. We
will continue to use this condensed notation in the rest of this
section and hope it will not cause any confusion.

We remind that for the 3-point functions we have
\begin{equation}
  \nabla  C_{IJ}^K = 0.
\end{equation}

\subsection{Geometry of the bundles}

Since all the quantities appearing on the right hand side
of \eqref{curvef} are covariantly constant, it means that the curvature
operator is also covariantly constant
\begin{equation}
  \nabla R_{\mu\nu} = 0 .
\end{equation}
Bundles of covariantly constant curvature over homogeneous spaces,
such as the moduli space \eqref{modulispace}, are called {\it
  homogeneous bundles}.  It is a mathematical theorem \cite{Helgason}
that the connection on homogeneous bundles is completely
determined by the connection on the tangent bundle of the
underlying base space, in our case \eqref{modulispace}. Each
homogeneous bundle is characterized by a representation ${\cal R}$
of the holonomy group $SO(4)\times SO(n)$ and the connection on it
is the same as that of the tangent bundle but in the
representation ${\cal R}$\footnote{If $L$ is the vector space that
carries the representation ${\cal R}$ then the vector bundle is
explicitly constructed as $(SO(4,n)\times L)/(SO(4)\times
SO(n))$.}. Actually, from the expression \eqref{finalcurv} for the
curvature we see from the factor $\delta_{ab}$ that the $SO(4)$
representation is always the trivial one.

So finally, the geometry of the bundle ${\cal V}_q$ of chiral
primaries of charge $q$ is completely characterized by a (possibly
reducible) representation ${\cal R}$ of $SO(n)$. To determine the
representation we have to consider the $SO(n)$ part of the curvature
operator
\begin{equation}
( C_I C_J^{T} - C_J^{T} C_I)_M^N
\end{equation}
This has to decompose into representations ${\cal R}_k$ of
$SO(n)$. Then the bundle of chiral primaries of charge $q$ is the
direct sum of homogeneous bundles corresponding to these representations
\begin{equation}
  {\cal V}_q = \sum_k \oplus {\cal V}_{{\cal R}_k}
\end{equation}
The geometry of each of ${\cal V}_{{\cal R}_k}$ is completely fixed by
the geometry of the coset
\begin{equation}
{SO(4,n)\over SO(4)\times SO(n)}
\label{cosetagain}
\end{equation}
and some basic group theory which is completely independent of the
dynamics of the CFT. For example, the chiral primaries of charge
$(1/2,1/2)$ always transform in the vector representation of $SO(n)$
and the corresponding bundle ${\cal V}_{(1/2,1/2)}$ has curvature of
the form
\begin{equation}
  R_{\mu\nu} = -  f \delta_{ab} \Sigma_{IJ}
\label{curvtang}
\end{equation}
where $(\Sigma_{IJ})_M^N$ are matrices in the vector representation
of the $SO(n)$ algebra, that is they satisfy
\begin{equation}
[\Sigma_{IJ},\Sigma_{KL}] = \delta_{JK}\Sigma_{IL} +
  \delta_{IL}\Sigma_{JK} - \delta_{JL}\Sigma_{IK} -
  \delta_{IK}\Sigma_{JL}
\label{sonalgebra}
\end{equation}
and $f$ is a numerical constant which depends on the overall scale of
the coset \eqref{cosetagain}. In the case of the D1/D5 CFT $f$ is
inversely proportional to the central charge of the theory. Similarly
for a bundle in the representation ${\cal R}$ we have matrices
$\Sigma_{IJ}^{\cal R}$ of the $SO(n)$ algebra \eqref{sonalgebra} and
the curvature operator ${\cal V}_{\cal R}$ is
\begin{equation}
R_{\mu\nu}= -  f\delta_{ab} \Sigma^{\cal R}_{IJ}.
\end{equation}
Notice that from the fact that the marginal operators are descendants
of the $(1/2,1/2)$ chiral primaries and using the curvature
\eqref{curvtang} for these fields and the corresponding curvature for
the supercurrents we get the following expression for the curvature of
the marginal operators
\begin{equation}
  \left(R_{\mu\nu}\right)^{\lambda}_{\kappa} = f \left((\sigma_{ab})^d_c
(\delta_{IJ})_K^L - (\delta_{ab})_c^d (\Sigma_{IJ})_K^L\right)
\label{curvcoset}
\end{equation}
where $\sigma_{ab}$ is the vector representation of $SO(4)$ and again
we use the notation $\mu = (a,I),\nu=(b,J),\kappa = (c,K),\lambda =
(d,L)$. It is easy to recognize that \eqref{curvcoset} is the
curvature of the tangent bundle of the coset \eqref{cosetagain} in a
vielbein basis and where $f$ controls the overall size of the
manifold.

In practice, if we can compute the curvature operator from the
3-point functions at one point of the moduli space then we can
find the decomposition of chiral primaries into representations of
$SO(n)$ and fix the geometry of the bundles, at least in a
neighborhood of the point. For example in AdS$_3$/CFT$_2$ such a
point could correspond to the orbifold CFT.

\subsection{Example: IIB on $K3$}
\label{subsec:IIBK3}
Let us now explain how the previous arguments apply to the case of IIB
on $AdS_3\times S^3 \times K3$. This is the near horizon geometry of a
bound state of $Q_1$ D1 and $Q_5$ D5 branes wrapped on $K3$.  The
boundary conformal field theory is believed to be described by a
deformation of a supersymmetric sigma model whose target space is the
orbifold $K3^N/S_N$, where $N=Q_1 Q_5$. The moduli space is locally
the coset
\begin{equation}
{SO(4,21) \over SO(4)\times SO(21)}
\label{modk3}
\end{equation}
The holonomy of the tangent bundle of the moduli space is
$SO(4)\times SO(21)$. As we explained before, the connection on the
vector bundles of the chiral primaries will be associated to that of
the tangent bundle and in particular to its $SO(21)$ part. So each of
these bundles will be characterized by a representation ${\cal R}$ of
$SO(21)$.

The chiral primary states of this theory can be conveniently encoded
in the Poincar\'e polynomial\footnote{This is in the ${\cal N}=4$
  conventions where the normalization of $J_0$ is half-integral.}
\begin{equation}
  P_{t,\overline{t}} = \text{Tr}\left(t^{2J_0}
  \overline{t}^{2\overline{J}_0}\right)
\end{equation}
where the trace is taken over the space of chiral primaries. The chiral
primary states are related to harmonic forms in the target space and it can be
shown that the Poincar\'e polynomial equals
\begin{equation}
  P_{t,\overline{t}} = \sum_{p,q} h_{p,q} t^p \,\overline{t}^q
\end{equation}
where $h^{p,q}$ are the Hodge numbers of the target space. The Hodge
numbers of $K3$ are equal to
\begin{equation}
 \begin{array}{ccccc}
& & 1 \\ &0& & 0 \\ 1 &\ & 20 & & 1 \\ &0& & 0 \\ & & 1&
\end{array}
\label{hodge}
\end{equation}
Starting with the single $K3$, it is possible to compute the Hodge
numbers of the resolution of $K3^N/S_N$ from the generating
function \cite{goso}
\begin{equation}
  \sum_{N\geq 0} Q^N P_{t,\overline{t}}(K3^N/S_N) = \prod_{m=1}^\infty
  \prod_{p,q} \left(1+(-1)^{p+q+1}Q^m t^{p+m-1}\overline{t}^{q+m-1}
  \right)^{(-1)^{p+q+1}h^{p,q}} .
\label{gensymprod}
\end{equation}
From this expression we can compute the numbers of chiral
primaries of given conformal dimension in the SCFT. As we
mentioned before these numbers agree with the results obtained
from supergravity.

Now let us look at the low lying chiral primaries and sketch how they
fit into vector bundles over the moduli space. For large enough $N$
the even Hodge numbers (all odd numbers are zero) of the Hilbert
scheme $K3/S_N$ are
\begin{equation}
 \begin{array}{ccccccc}
& & &  ....\\
& & &  ....\\
  1&22 &276 & 2278 & 276 &22 &1  \\
&1 & 22 & 254 & 22 &1\\
&& 1 & 21 & 1  \\
&&  & 1
\end{array}
\label{hodge2}
\end{equation}
Let us see how we can represent these chiral primaries in the orbifold
CFT language \cite{Vafa:1994tf},\cite{Maldacena:1998bw}. We introduce
bosonic creation operators $\alpha^A_{-n}$, where $n=1,2...$ labels the level
of the twisted sector and $A$ runs over the Dolbeault cohomology
classes of a single $K3$. For a given $(p,q)$, there are $\dim
H^{(p,q)}(K3)$ operators $\alpha^{(p,q)}_{-n}$. The general chiral
primary can be written as
\begin{equation}
  \prod_{i=1}^M \alpha^{A_i}_{-n_i} |0\rangle, \qquad \sum_{i}^Mn_i=N.
\label{generalcp}
\end{equation}
The R-charge of this operator is

\begin{equation}
  \left(J^3 ,\overline{J}^3\right) = {1\over 2}\left(N-M + \sum_i
  p_i,\quad N-M+ \sum_i q_i\right).
\end{equation}
There is only one operator of charge $(0,0)$ which we will denote
by $|N\rangle$. It is given by the product $|N\rangle\equiv
\prod_{i=1}^N\alpha^{0,0}_{-1}|0\rangle$; clearly, this is to be
identified with the identity operator and of course there is no
holonomy for it. We have a single operator of charge $(1,0)$,
which may be represented as $\alpha_{-2}^{2,0}|N-1\rangle$. The
operator with charge $(0,1)$ is similarly represented by
$\alpha_{-2}^{0,2}|N-1\rangle$. They correspond to the R-symmetry
currents $J^3,\overline{J}^3$. As we saw in section
\ref{subsec:currentcurv} the holonomy for these operators is also
trivial.

Now we consider the 21 operators of charge
$(\frac{1}{2},\frac{1}{2})$. They are given by the following products
of creation operators
\begin{equation}
\label{eq:marop}
20 \times \qquad \alpha_{-1}^{(1,1)}\left| N-1 \right> \quad
\mathrm{and} \quad 1 \times \qquad \alpha_{-2}^{(0,0)}\left| N-2
\right>.
\end{equation}
From this we conclude that the operators of charge
$(\frac{1}{2},\frac{1}{2})$ fall into the vector representation ${\bf
  21}$ of $SO(21)$\footnote{Another possibility is that they might be
  21 singlets of $SO(21)$. However we know that we get the marginal
  operators as descendants of these chiral primaries, and the marginal
  operators transform under $SO(21)$, so this possibility is
  excluded.}. The connection of this bundle over the moduli space is
the same as the $SO(21)$ part of the tangent bundle of
\eqref{modk3}. Acting on a each of these states with one left-moving
and one right-moving supercurrent gives the $4\times 21=84$ marginal
operators.

At higher conformal dimension, we have to distinguish between
single-particle and multi-particle chiral primaries. A multi-particle
field is given by the product of chiral primaries of lower charge,
while a single-particle operator is a genuinely new chiral primary
appearing at the given conformal dimension. For example if we look at
the operators of charge $(1,1)$ we have 254 of them. We can have
multi-particle states of the form $(1/2,1/2) \times (1/2,1/2)$ which are
$(21\times 22)/2$ in number, or of the form $(1,0)\times(0,1)$, which is one
state. So in total we have 232 multi-particle operators at this level
and 22 single-particle ones.

The multi-particle states will obviously fall into tensor product
representations of $SO(21)$ determined by their decomposition into
single-particle operators.  The corresponding bundles are isomorphic
to the tensor product of the bundles of their constituents. Hence the
new information at each level is related to the the bundles of chiral
primaries which are single-particle operators.

More generally, for $m$ small enough compared to $N$, we have
single-particle operators of total charge $m$ only when the charge is
of the form ${1\over 2}(m,m)$, ${1\over 2}(m+1,m-1)$ or ${1\over
  2}(m-1,m+1)$. The single-particle operators
\footnote{Notice that we are not careful about the precise linear
  combinations that gives us the single- vs multi-particle operators
  since we are only interested in their counting and not the actual
  operators.}  with charge ${1\over 2}(m,m)$ can be represented in the
form
\begin{equation}
\alpha_{-m-1}^{(0,0)}|N-m-1\rangle,\quad
\alpha_{-m}^{(1,1)}|N-m\rangle,\quad\alpha_{-m+1}^{(2,2)}|N-m+1\rangle.
\end{equation}
so there are $1+20+1$ of them. Our natural guess is that they
decompose as ${\bf 21 + 1}$ of the $SO(21)$. In principle we could
compute their 4-point function at the orbifold point and check whether
this is indeed true. For ${1\over 2}(m+1,m-1)$ and ${1\over
  2}(m-1,m+1)$ we have
\begin{equation}
  \alpha_{-m}^{(2,0)}|N-m\rangle,\qquad \alpha_{-m}^{(0,2)}|N-m\rangle
\end{equation}
respectively. They are obviously in the ${\bf 1}$ of $SO(21)$.

To summarize, we denote by ${\cal V}_{\bf 21}$ the unique real vector
bundle of rank 21 over the moduli space \eqref{modk3} whose connection
is the same as the $SO(21)$ part of the tangent bundle. The curvature
of this bundle of the form \eqref{curvtang}. We denote by ${\cal
  V}_{\bf 1}$ the trivial bundle of rank one and ${\cal V}_{\bf
  multi}$ the tensor product of vector bundles corresponding to the
lower conformal dimensions. We have the following answer for the
geometry of the vector bundle ${\cal V}_{p,q}$ of chiral primaries of
charge $(p,q)$ with $p,q>1$
\begin{equation}
  {\cal V}_{p,q} = \left\{ \begin{array}{lll} {\cal V}_{\bf
      multi}\oplus {\cal V}_{\bf 21} \oplus {\cal V}_{\bf 1}, \qquad
    \text{if}\quad p=q,\\ {\cal V}_{\bf multi}\oplus {\cal V}_{\bf
      1},\qquad\qquad\,\,\,\,\text{if} \quad p=q+1\quad\text{or}\quad
    q=p+1,\\ {\cal V}_{\bf multi},\qquad \qquad \qquad \quad \text{otherwise}.
\end{array} \right.
\label{finalbundle}
\end{equation}
Interestingly, if we look at the Fock space (\ref{generalcp}) then
according to the previous discussion for each fixed $N>1$ it
should carry a representation of $SO(21)$. This representation is
certainly not manifest. There is an obvious action of $SO(20)$
which rotates the $\alpha_{-n}^A$ with $A$ the 20 $(1,1)$-forms
into each other and leaves the other $\alpha_{-n}^A$ fixed. The
extra operators which extend $SO(20)$ to $SO(21)$ must be more
complicated. If we also introduce the positive modes of the bosons
with commutation relations
\begin{equation}
[\alpha_{-n}^A,\alpha_m^B]=m\delta_{n,m} \int_{K3} A\wedge B
\end{equation}
then the $SO(20)$ generators can be written as quadratic operators in
the modes of the bosons. However, the extra $SO(21)$ generators must
be at least cubic. It would interesting to construct these generators
explicitly and study their precise algebraic and geometrical
meaning\footnote{As mentioned before, we believe that a class of
  correlators of short multiplets may respect an $SO(4)^{outer}\times
  SO(21)$ selection rule as in \cite{Intriligator:1998ig},
  \cite{Intriligator:1999ff}. It would be interesting to clarify this
  point.}

Let us also notice that for given $m$, a single particle operator
is a map from the cohomology of $K3$, $H^*(K3)$, to the cohomology
of the symmetric product $H^*(\mathrm{Sym}^N(K3))$.  In section
\ref{subsec:4dgauge} we will see how the chiral primaries can be
identified with operators in the 4-dimensional gauge theory, which
can also be interpreted as forms on the instanton moduli space.

\subsection{Chiral primaries in 4d gauge theory}
\label{subsec:4dgauge}
In the previous subsection, the chiral primaries of the 2d sigma model
were considered. The target space of the sigma model is the moduli
space $\mathcal{M}$ of instantons of 4d gauge theory on $K3$. Therefore,
we might expect that the 2d chiral primaries have analogues in the
gauge theory. Such a connection is potentially interesting since we
might be able to learn more about the geometry of the chiral ring by
the computation of gauge theory quantities like Donaldson
polynomials. On the other hand, it might be useful for an analysis of
the geometry of the chiral ring of the superconformal 4d gauge theory,
see also Sec. \ref{sec:conclu}. The gauge theory can be obtained by
wrapping the D5-brane system on $T^2\otimes K3$, and considering the
limit where the typical length scale of the $T^2$ is much smaller then
the one of $K3$. In this way one ends up with $\mathcal{N}=4$
Yang-Mills theory on $K3$.

The correspondence between the 2d and 4d operators can be
understood more precisely if we recall the representation of
Donaldson polynomials in terms of the fields of $\mathcal{N}=2$
gauge theory in \cite{Witten:1988ze}. See also
\cite{Dijkgraaf:1998gf} for a discussion of Donaldson polynomials
in the context of AdS/CFT. Similar to the interpretation of the
single particle operators $\alpha_{-n}^{A_i}$ as differential
forms on $\mathcal{M}$ in section \ref{subsec:IIBK3}, the
Donaldson polynomials can be viewed as differential forms on
$\mathcal{M}$. By a comparison of the infinitesimal deformations
of an instanton solution and the supersymmetry transformations,
Ref. \cite{Witten:1988ze} assigns a form degree on $\mathcal{M}$
to the gauge theory fields. This degree corresponds to the charge
under a $U(1)$ subgroup of the R-symmetry group. The R-symmetry
group of $\mathcal{N}=4$ Yang-Mills is $SU(4)$, which can be
decomposed as $SU(2) \times SU(2)\times U(1)$. The charge of a
field under the last $U(1)$ provides its form degree on
$\mathcal{M}$. The field content of the theory is a gauge field
$A_\mu$, six scalars $\phi_i$ and fermions. Four of the scalars
and $A_\mu$ have $U(1)$-charge 0, the two other scalars have
charge $+2$ and $-2$, and the fermions have charges $+1$ or $-1$.
The sixteen supersymmetry generators can be divided in two sets of
eight, based on their $U(1)$-charge $\pm 1$. The $SU(2)$-holonomy
of $K3$ preserves half of the susy generators in both sets. We
denote the preserved susy generators by $Q^{\pm I}_\alpha$, where
$\pm$ denotes the $U(1)$-charge, $\alpha$ labels the space-time
$SU(2)$ which is preserved by the $K3$ holonomy, and $I=1, 2$. A
susy generator with charge  $+1$ plays often a distinguished role,
namely when it is taken as the generator of a topological symmetry
after twisting of the theory.

This also distinguishes the scalar with $U(1)$-charge 2 (which we
denote by $\sigma$). This scalar is namely annihilated by the susy
generators with charge $+1$, because a field with charge $+3$ does not
exist. These susy generators are the analogues of the operators $G^{+\pm}_{-1/2}$ and $\bar
G^{+\pm}_{-1/2}$, which annihilate the states of the chiral-chiral
ring. Among the operators which are the analogues of the states
in the chiral-chiral ring are thus $W_0^m=\mathrm{Tr}\left(\sigma^m
\right)$. These are not all the operators which are
annihilated by $Q^{+I}_\alpha$. As explained in
\cite{Witten:1988ze}, one can construct descendants $W_k^m$ of
$W^m_0$, such that $dW_k^m\sim\left\{Q^{+I}_\alpha, W_{k+1}^m
\right\}$. These forms are given by
\begin{equation}
\label{eq:ops}
W^m_0=\mathrm{Tr}\left( \sigma^m\right),\qquad W^m_2=\mathrm{Tr}\left(\sigma^{m-1} \wedge F \right),
\qquad W^m_4=\mathrm{Tr} \left(\sigma^{m-2}\wedge F\wedge F\right),
\end{equation}
where we have ignored the fermions. Since K3 does not contain
odd-dimensional cycles, only those descendants are given which are
related to even forms. Since acting with $Q^{+I}_\alpha$ results
in a total derivative, the following non-local operators are
invariant under $Q^{+I}_\alpha$:
\begin{eqnarray}
\label{eq:operators}
\int_\mathrm{A_i} W^m_2, \qquad
\mathrm{and} \qquad  \int_\mathrm{K3}W^m_4,
\end{eqnarray}
where the $\mathrm{A}_i$ form a basis of the 22 two-cycles of K3. Since
$F$ is a zero-form on $\mathcal{M}$ and $\sigma$ a two-form, these
operators are respectively $2m-2$, and $2m-4$ forms on $\mathcal{M}$.

We have now constructed the set of operators, which are dual to the
operators $\alpha_{-n}^{A_i}$. E.g. the operators in
Eqn. (\ref{eq:marop}) together with the currents $J^3$ and
$\overline{J}^3$ have total charge 1, and are thus two-forms on
$\mathcal{M}$. These operators correspond to integrated descendants of
$\mathrm{Tr}(\sigma^2)$ and
$\mathrm{Tr}(\sigma^3)$. They are explicitly given by
\begin{equation}
22 \times \quad \int_{\mathrm{A}_i} W^2_2,    \quad \mathrm{and}
\quad 1 \times \quad \int_\mathrm{K3} W^3_4.
\end{equation}
These are therefore the counter parts of the 23 chiral primaries
with total charge 1 in Sec. \ref{subsec:IIBK3}. It is conceivable
that the two-cycles $\mathrm{A}_i$, whose Poincar\'e dual is a
$(1,1)$-form, correspond to the 20 operators in (\ref{eq:marop}).
We can easily go further and include the chiral primaries with
larger charges. The single particle operators with total charge
$m$ in section \ref{subsec:IIBK3} correspond to $2m$-forms on
$\mathcal{M}$. These $2m$-forms on $\mathcal{M}$ correspond to the
appropriate descendants of $\mathrm{Tr}(\sigma^m)$,
$\mathrm{Tr}(\sigma^{m+1})$ and $\mathrm{Tr}(\sigma^{m+2})$,
namely
\begin{eqnarray}
&1 \times& \quad W^m_0,\qquad 22 \times \qquad \int_\mathrm{A_i}
  W^{m+1}_2 , \qquad \mathrm{and} \quad 1 \times \quad
  \int_\mathrm{K3} W^{m+2}_4.
\end{eqnarray}
Thus we have shown above that the chiral primaries of 2d CFT can be
identified with operators in $\mathcal{N}=4$ Yang-Mills. The marginal
operators in Sec. \ref{subsec:IIBK3} are obtained by acting with the
operators $G^{-\pm}_{-1/2}$ and $\bar G^{-\pm}_{-1/2}$. These
operators correspond in the gauge theory to the generators
$Q^{-I}_\alpha$ with $U(1)$-charge $-1$. As mentioned before, we can
also identify the gauge theory chiral primaries in terms of
$\alpha^{\mathrm{A}_i}_{-1}$. For example, the operator 
$\mathrm{Tr}(\sigma^{2})$, mentioned in section \ref{sec:conclu}
as the gauge theory chiral primary, which has as descendant a marginal
operator, corresponds to $\alpha^{(2,2)}_{-1}$.

\section{Attractor mechanism and RG-flow}
\label{sec:attract}

One of our original motivations for studying the moduli dependence
of the chiral ring in ${\cal N}=(4,4)$ theories, was its possible
relevance for the analysis of the connection between the attractor
flow in supergravity and RG-flow in the dual field theory. The
attractor mechanism is usually studied in the case of
4-dimensional extremal black holes, but more generally it also
appears for extremal branes of other dimensionalities. The
attractor mechanism is a consequence of the extremality of the
brane and not of supersymmetry
\cite{Ferrara:1997tw},\cite{Sen:2005wa},\cite{Goldstein:2005hq},
however it is technically easier to study in the supersymmetric
case. To keep our discussion simple, we will only consider the
cases of spherically symmetric flows and will ignore all
subtleties related to multiple attractor points, walls of marginal
stability and split-flows. Obviously it would be extremely
interesting to understand such phenomena from the RG-flow point of
view but this is beyond the scope of our simple analysis.

\subsection{The attractor mechanism}

Consider a supergravity theory in D dimensions with a moduli space
${\cal M}_{sugra}$ in which the massless scalar fields take values.
We pick coordinates $z$ on ${\cal M}_{sugra}$. The metric
on the moduli space is $g_{ab}(z)$. We assume that the theory admits
BPS p-brane solutions, charged under $(p+2)$-form field strengths. The
charge $\Gamma$ of these branes takes values in a lattice $\Lambda$. A
very useful quantity is the spacetime central charge of the brane
\begin{equation}
  Z(\Gamma,z)
\end{equation}
which is determined by the supersymmetry algebra\footnote{For
simplicity we assume that there is only one complex central
charge.} and is a function of the charge vector $\Gamma$ and the
position on the moduli space $z$. If we call $z^\infty$ the values
of the moduli at infinity, then the ADM mass/tension of the black
brane in D-dimensional Planck units is equal to
\begin{equation}
  M_{ADM} = |Z(\Gamma,z^\infty)|
\end{equation}
In the supergravity solution the moduli $z$ evolve radially reaching
constant values $z^*$ near the horizon. The value  $z^*$ depends
only on the charge $\Gamma$ of the brane and not on the values of the
moduli at infinity $z^\infty$. This is the attractor mechanism. The
condition for $z^*$ to be an attractor point is that is minimizes the
central charge $Z$
\begin{equation}
  {\partial |Z| \over \partial z^i}|_{z=z^*} = 0
\label{attractorcondition}
\end{equation}
For every charge vector $\Gamma\in \Lambda$, there is a submanifold of
solutions of \eqref{attractorcondition}
\begin{equation}
{\cal M}_{sugra}^{*,\Gamma}\in {\cal M}_{sugra}
\end{equation}
of attractor points, that we call the attractor submanifold for the
charge vector $\Gamma$.\footnote{Usually we speak of attractor points
  and not submanifolds, however even in the familiar case of black
  holes in 4d ${\cal N}=2$ theories, the vector multiplets are fixed
  by the attractor mechanism to isolated points, while the
  hypermultiplets can take any value. In this case ${\cal M}_{sugra} =
  M_{vector} \times M_{hyper}$ and the attractor submanifold will be
  ${\cal M}_{sugra}^{*,\Gamma} = \{p\} \times M_{hyper}$, where
  $\{p\}\in M_{vector}$ is the attractor point for the charge
  $\Gamma$.} The radial evolution of the moduli from their value
$z^\infty$ at infinity to $z^*$ at the horizon is governed by the
attractor flow.

For example, for a spherically symmetric 4d black hole in ${\cal N}=2$
supergravity we have the ansatz
\begin{equation}
ds^2 = -e^{2U(r)} dt^2 + e^{-2U(r)} \left(dr^2 + r^2
d\Omega^2_{2}\right)
\end{equation}
For supersymmetric solutions we can write first order flow equations
for $U(r),z^a(r)$. It is more convenient to work with the coordinate
$\tau = 1/r$. This leads to the following flow equations
\begin{equation}\begin{split}
  &\dot{U} = - e^{U} |Z|\\
  &\dot{z}^a = -2 e^{2U} g^{a\overline{b}} \partial_{\overline{b}} |Z|
\label{attractorfloweq}
\end{split}\end{equation}
Similar relations hold for black branes of higher dimensionality.

\subsection{Relation to RG-flow}

The attractor black holes discussed in the previous section can be
realized in string theory by bound states of D-branes. In this
description the D-branes are placed in a flat background space, where
the values of the scalar moduli are equal to their asymptotic values
$z^\infty$. The supergravity solution arises after backreaction
and then we see the attractor mechanism in the radial
evolution of the moduli. We want to understand what is the meaning of
the attractor mechanism in the original D-brane picture.

The open string excitations on the worldvolume of the D-branes can be
described in an appropriate regime by an effective quantum field
theory.  The background values $z^\infty$ of the closed string moduli
enter the worldvolume theory in the form of coupling constants. We
will call the set of parameters of the effective field theory on the
branes ${\cal M}_{QFT}$, which we will (loosely) identify with ${\cal
  M}_{sugra}$. The supergravity description of the same system has an
AdS throat in the near horizon region, which indicates that the
worldvolume theory flows to a conformal field theory in the
IR. Moreover, in the near horizon region the moduli reach their
attractor values $z^*$.

This suggests that the IR fixed point of the worldvolume theory only
knows about the attractor values of the moduli, hence the moduli space
${\cal M}_{CFT}$ of the conformal field theory has to be identified
with the attractor submanifold ${\cal M}_{sugra}^{*,\Gamma}$ in
supergravity.  In other words if we flow to the IR, the number of
parameters of the worldvolume theory is generally reduced leaving us
with ${\cal M}_{CFT}\subset {\cal M}_{QFT}$. It is reasonable to
assume that the way in which the UV coupling constants on the D-brane
theory transform into the effective IR ones is by renormalization
group flow.  In this sense the D-brane theory sees the attractor
mechanism as RG-flow on its worldvolume. Then it is natural to expect
that the attractor flow equations \eqref{attractorfloweq} will play
the role of RG-flow equations in the space of effective coupling
constants of the D-brane theory.

The RG-flow of the worldvolume theory is governed by the $\beta$
functions, which describe the flow of the coupling constants as a
function of the energy scale. More precisely the $\beta$ functions
correspond to a vector field on the space of parameters of the theory
${\cal M}_{QFT}$. The flow lines of this field give RG-flow orbits
which, for the class of the theories we are considering, approach
conformal fixed points at low energies where $\beta=0$.  The set of
these points constitute the moduli space ${\cal M}_{CFT}$ of conformal
theories inside the bigger space ${\cal M}_{QFT}$ of effective quantum
field theories.  Similarly the attractor flow equations
\eqref{attractorfloweq} describe the radial flow of the moduli in
gravity from ${\cal M}_{sugra}$ to the submanifold ${\cal
  M}_{sugra}^{*,\Gamma}$. The two pictures are consistent if we accept
the usual AdS/CFT intuition that the radial direction corresponds
to the energy scale. The statement that more than one value of the
moduli at infinity flow to the same value near the horizon is
related to the fact that more than one UV quantum field theories can
flow to the same IR fixed point.

It would certainly be very interesting to understand this connection
in more detail, however making this intuitive picture more precise is
not straightforward. Besides the fact
that the worldvolume theory is generally strongly coupled, there is an
important conceptual difficulty, that away from the conformal point in
the IR, i.e. away from the strict $\alpha'\rightarrow 0$ limit, the
theory on the branes is not decoupled from the closed string modes in
the bulk.

While the absence of a decoupling limit may be a serious obstacle for
a precise formulation of the attractor flow/RG-flow relation it should
be possible to work in a perturbative expansion around the conformal
point, i.e. to first order away from the $\alpha'\rightarrow 0$
limit. There it should be possible to study the relation between
attractor flow and RG-flow reliably. In the rest of this section we
will only consider the first order flow and leave the more difficult
study of finite flows for future work.

Our goal is to start from the conformal point and consider a
first-order perturbation towards the UV. At the conformal point we
have the AdS/CFT duality between the AdS factor of the near horizon
geometry and the conformal IR fixed point of the D-brane theory. To
see the attractor mechanism we have to flow from the near horizon
geometry towards asymptotic infinity. In the boundary theory this
means that we have to study the IR conformal field theory perturbed by
irrelevant operators (see \cite{Intriligator:1999ai} for a similar
discussion in the case of AdS$_5$/CFT$_4$). Perturbing a field theory
by irrelevant operators is dangerous since it drastically modifies its
UV behavior. However, since we are only interested in the first order
flow away from the fixed point we will treat the conformal field
theory perturbed by irrelevant operators as an effective field theory
and study RG-flow in the Wilsonian sense, even though we do not have a
UV completion of the theory.

In supergravity the entire attractor flow solution preserves the same
amount of supersymmetry and spherical symmetry as the near horizon
geometry\footnote{ Except for the extra supercharges that we get in
  the AdS region which are dual to the superconformal generators in
  the CFT.} so on the boundary theory we will only consider
perturbations by irrelevant operators which do not break the
supersymmetry and R-symmetry of the CFT but only conformal
invariance. As we will see in our toy model, this constrains the
number of allowed irrelevant operators to a finite set.

Now we would like to make more precise the statement that the
attractor flow and RG-flow agree to first order away from the
fixed point. As we can see in figure \ref{flow} this means that
the structure of the flow on the two sides should be the same in a
neighborhood of the fixed submanifolds ${\cal M}_{CFT},{\cal
  M}_{sugra}^{*,\Gamma}$. The ``zeroth-order'' matching of the two
sides relates the geometry of the fixed submanifolds. This is a
consequence of the AdS/CFT correspondence between the near horizon
geometry of the extremal brane and the conformal field theory in the
IR of the D-brane theory. So, at least locally, we must
have\footnote{This has been demonstrated in the case of
  AdS$_3$/CFT$_2$ with 16 supercharges. It would be interesting to
  prove the same statement for the $(0,4)$ MSW theory or even for 4d
  black holes. A naive approach would suggest that the moduli space of
  of the ``superconformal quantum mechanics'' on the D-branes should
  be related to the attractor submanifold, which in this case
  coincides with the hypermultiplet moduli space. We hope to address
  this question in the future.}
\begin{equation}
 {\cal M}^{*,\Gamma}_{sugra} = {\cal M}_{CFT}
\label{attractorga}
\end{equation}
This is statement about the dimensionality as well as the geometry of
the two manifolds. The metric on ${\cal M}_{sugra}^{*,\Gamma}$ is
fixed by the metric $g_{ab}$ on the moduli space ${\cal
  M}_{sugra}$, but see also footnote \eqref{corrections}. The metric on ${\cal M}_{CFT}$ is determined by the
Zamolodchikov metric of marginal operators in the CFT, which
correspond to tangent vectors on ${\cal M}_{CFT}$.

The next step is to consider the first order flow towards the UV.  For
the two sides to be in agreement, the number of allowed irrelevant
operators must be the same as the codimension of the attractor
submanifold inside the full moduli space of supergravity-which is
equal to the number of fixed moduli. Also the conformal dimension of
the irrelevant operators must be related to the mass of the fixed
moduli in the near horizon geometry by the standard mass/conformal
dimension relation in AdS/CFT.

Moreover, the identification between the attractor flow and RG-flow
suggests that there should be a relation between the two parameter spaces,
not only on the fixed submanifolds but also away from them,
at least to first order. A way to state this more precisely is that
the normal bundle ${\cal N}_{sugra}$ of the attractor submanifold
${\cal M}^{*,\Gamma}_{sugra}$ inside ${\cal M}_{sugra}$ should have
the same structure as the normal bundle ${\cal N}_{CFT}$ of ${\cal
  M}_{CFT}$ in ${\cal M}_{QFT}$
\begin{equation}
  {\cal N}_{sugra} = {\cal N}_{CFT}
\label{attractorgb}
\end{equation}
The geometry of the bundle ${\cal N}_{CFT}$ encodes how the CFT
can be perturbed by irrelevant operators (which preserve certain
symmetries). Its geometry is characterized by the Zamolodchikov
metric and the connection for the irrelevant operators in the CFT.
Notice that the identification \eqref{attractorgb} of the normal
bundles requires not only a matching of their ranks, which is
guaranteed if the number of irrelevant operators is the same as
the number of fixed moduli, but also a matching of the connections
on the two bundles. The connection on ${\cal N}_{sugra}$ is easily
computable if we know how ${\cal
  M}_{sugra}^{*,\Gamma}$ is embedded in ${\cal M}_{sugra}$, while the
connection on ${\cal N}_{CFT}$ in the CFT equals the connection
for the irrelevant operators over the moduli space as was
explained in section \ref{sec:cftdeform}.

These three conditions, identification of moduli spaces
\eqref{attractorga}, of number/dimension of irrelevant operators
to number/mass of fixed moduli, and identification of the normal
bundles \eqref{attractorgb} is enough to guarantee the
identification of attractor flow to RG-flow to first order away
from the conformal fixed point. As we see all these quantities can
be computed within the CFT, so unless we want to go to higher
orders in perturbation theory, we do not have to worry about the
UV completion of the theory and issues related to the decoupling
of closed string modes\footnote{We
  would like to emphasize that we are not proposing that there is a
  well defined UV point for the CFT perturbed by irrelevant operators.
  If such a theory existed, it would be dual to asymptotically flat
  string theory. Instead we are treating the theory living on the
  branes as an effective field theory near the IR fixed point and
  consider Wilsonian RG flow towards the IR. We find that it is
  very constrained since there are only a finite number of irrelevant
  operators allowed by the symmetries. We claim that this
  self-consistent flow to the fixed point should be related to the
  attractor flow.}.

\subsection{The D1/D5 system}

The simplest system where we could try to check the attractor
flow/RG-flow connnection is the D1/D5 bound state\footnote{It would be
  very interesting to study 4-dimensional black holes in ${\cal N}=2$
  supergravity, where the attractor mechanism has a richer
  structure. In this case the near horizon geometry of an extremal
  black hole is $AdS_2 \times S^2$. Unfortunately, the field theory
  side is not well understood. In general we would expect a $0+1$
  dimensional theory which would flow in the 'IR' to some kind of
  superconformal quantum mechanics, leading to an AdS$_2$/CFT$_1$
  duality. Since the precise meaning of the latter is still
  mysterious, even at the fixed point, it seems difficult to study the
  flow towards the fixed point with present technology.}. We start
with IIB compactified on $K3$. This leads to chiral $(2,0)$
supergravity in 6 dimensions \cite{Romans:1984an}, whose moduli space
is
\begin{equation}
  {\cal M}_{sugra} = {SO(5,21) \over SO(5)\times SO(21)}/SO(5,21,{\mathbb Z})
\label{sugrabig}
\end{equation}
This moduli space corresponds to the geometric moduli of $K3$, the
NS and RR potentials and the dilaton.

Six dimensional supergravity admits BPS black string solutions
preserving 8 supercharges, charged under the 3-form field
strengths. These solutions correspond to bound states of D5/NS5
branes wrapping the entire $K3$, D3 branes wrapping 2-cycles of
$K3$ and F1/D1 strings. The charges of the black strings take
values in the lattice ${\Gamma}^{5,21}$. The discrete U-duality
group $SO(5,21,{\mathbb Z})$ of the theory is equal to the automorphism
group of the charge lattice. For any primitive lattice vector,
there is always a U-duality transformation that can rotate it into
a bound state of only D1 and D5 branes. The charge lattice
$\Gamma^{5,21}$ can be embedded inside the vector space
$W=\mathbb{R}^{5,21}$. Each point $z$ on the moduli space
\eqref{sugrabig} corresponds to a decomposition into positive and
negative subspaces $W=V_+ \oplus V_-$, so the moduli space of
supergravity can be understood as the space of positive 5-planes
inside ${\mathbb R}^{5,21}$.

For a given charge vector $\Gamma$ and given position on the moduli
space we decompose $\Gamma = \Gamma_+ + \Gamma_-$ where $\Gamma_\pm\in
V_\pm$. It can be shown that the central charge, or tension, of the
black string is
\begin{equation}
  Z(\Gamma,z) = |\Gamma_+|
\label{centralstring}
\end{equation}
Taking into account that
\begin{equation}
  |\Gamma|^2 = |\Gamma_+|^2 - |\Gamma_-|^2
\end{equation}
is independent of the moduli $z$, we see that $|Z|$ is minimized when
$\Gamma_- =0$. This is equivalent to the set of positive 5-planes
containing the vector $\Gamma$.  It is not difficult to see that this
attractor submanifold has locally the structure of the coset
\begin{equation}
{\cal M}_{sugra}^{*,\Gamma}= { SO(4,21) \over SO(4)\times SO(21)}
\label{sugrasmall}
\end{equation}
The precise way in which this submanifold is embedded in the bigger
space \eqref{sugrabig} depends on the charge vector $\Gamma$ and can
be easily determined using for example the analysis of
\cite{Dijkgraaf:1998gf}.

The theory living on the branes is a 2-dimensional effective field
theory, which flows in the IR to a 2d CFT with ${\cal N}=(4,4)$
supersymmetry.  The supergravity attractor flow towards the AdS$_3$
throat should be dual to an RG flow of a 2d effective field theory
towards a 2d CFT in the IR, at least near the fixed point. In other
words, the theory on the brane, seen as an effective low energy
theory, is a 2d CFT perturbed by irrelevant operators.  The RG flow of
this theory should be dual to the attractor flow in supergravity.

As we explained in the previous subsection, if we want to check this
correspondence to first order we have to check three conditions. The
fact that the moduli spaces in the IR are the same is a well known
result \cite{Dijkgraaf:1998gf}, where we recognize that the space
\eqref{sugrasmall} is of the general form of the moduli space of
${\cal N}=(4,4)$ superconformal field theories \eqref{modulispace}.
So the condition \eqref{attractorga} is satisfied.

Let us now consider the second condition, which is the matching of
the fixed moduli to the irrelevant operators which preserve the
supersymmetry and R-symmetry. We want to perturb the CFT by
irrelevant operators which do not break the ${\cal N}=(4,4)$
supersymmetry, but only the conformal invariance. Also 
we do not  want to break the $SO(4)^R$ symmetry, which corresponds to
the spherical symmetry around the black string. This question was
discussed in \cite{Kanitscheider:2006zf},\cite{Kanitscheider:2007wq}. 
With these restrictions, as we explained in section \ref{subsec:specialshort}
using the representation theory of the ${\cal N}=(4,4)$ algebra,
the only candidate irrelevant operators are the descendants of
chiral primaries $\phi_I$ of charge $(1,1)$. By acting with two
supercurrents on each side we get $SO(4)^R$ neutral operators of
conformal dimension $(2,2)$ of the form
\begin{equation}
\Phi_I = G^{-+}_{-1/2}G^{--}_{-1/2}\overline{G}^{-+}_{-1/2}
\overline{G}^{--}_{-1/2} \cdot \phi_I
\label{blabla}
\end{equation}
These are the only irrelevant operators preserving the ${\cal
  N}=(4,4)$ structure and which are $SO(4)^R$ singlets. In the
notation of section \ref{subsec:IIBK3} they can be written as
\begin{equation}
\label{eq:irop}
20 \times \qquad \alpha_{-2}^{(1,1)}\left| N-2 \right> \quad
\mathrm{and} \quad 1 \times \qquad \alpha_{-3}^{(0,0)}\left| N-3
\right>\quad \mathrm{and} \quad 1 \times \qquad
\alpha_{-1}^{(2,2)}\left| N-1\right>.
\end{equation}

The fact that the single-particle operators of this form are in
one-to-one correspondence with the fixed moduli was already noted in
\cite{Maldacena:1998bw}. There are 21+1 of them corresponding to the
21 fixed moduli of supergravity and the size of the 3-sphere. It is
easy to check that the relation between masses and conformal
dimension is correct.

These irrelevant operators are sections of a vector bundle as described
in section \ref{sec:cftdeform}. At the same time they describe motion away
from the moduli space of conformal field theories ${\cal M}_{CFT}$ into
the bigger space ${\cal M}_{QFT}$ of ${\cal N}=(4,4)$ quantum
field theories. In this sense the bundle of the operators \eqref{eq:irop} is
isomorphic the normal bundle ${\cal N}_{CFT}$ of ${\cal M}_{CFT}$ inside
${\cal M}_{QFT}$. The connection on this bundle can be determined by the
results of the previous sections about the connection for the chiral primaries
$\phi$ and the supercurrents. It is not difficult to see that we have the
following result
\begin{equation}
  {\cal N}_{CFT} = {\cal V}_{\bf 21} \oplus {\cal V}_{\bf 1}
\label{normalone}
\end{equation}
Now from the supergravity side we have to compute the normal bundle of
\eqref{sugrasmall} inside \eqref{sugrabig}. It is easy to see that it
is exactly the same bundle ${\cal V}_{\bf 21}$. If we add to it one
more direction corresponding to increasing the size of the 3-sphere we
have
\begin{equation}
  {\cal N}_{sugra} = {\cal V}_{\bf 21} \oplus {\cal V}_{\bf 1}
\label{normaltwo}
\end{equation}
So we find precise agreement between the two normal bundles, showing
that the last condition \eqref{attractorgb} is also
satisfied\footnote{We should emphasize that the agreement between
  equations \eqref{normalone},\eqref{normaltwo} does not only refer to
  the rank of the bundles but to the full geometry of the bundle over
  the moduli space.}. This shows that to first order away from the
fixed point the attractor flow agrees with RG-flow on the boundary.

\subsection{Finite flows}

A natural question is whether we can extend the previous arguments
to higher orders in perturbation theory towards the UV. As we
explained before it is hard to give a precise UV completion of the
CFT perturbed by irrelevant operators, which is related to the
absence of decoupling between open and closed strings away from
the $\alpha'\rightarrow 0$ limit. Despite these problems let us
describe briefly what the full attractor flow for the D1/D5 system
looks like on the supergravity side.  These solutions where
discussed in detail in \cite{Mikhailov:1999fd}.

The metric has the form
\begin{equation}
  ds^2 = e^{2U(r)} (-dt^2 + dx^2) + e^{-2U(r)} (dr^2 + r^2 d\Omega_3^2)
\end{equation}
We take the moduli at infinity to be at a general point $z^\infty\in
{\cal M}_{sugra}$, which corresponds to a specific orientation of the
positive 5-plane $V_+^\infty$ inside the space ${\mathbb
  R}^{5,21}$. We also choose a charge vector $\Gamma$, which does not
generally lie inside $V_+^\infty$.  As we move towards the black string the
orientation of the 5-plane will change and at the attractor
point it will be such that $\Gamma \in V_+^*$. To fully specify the
solution we need to determine the function $U(r)$ and the orientation
of the 5-plane as a function of the radius $V_+(r)$.

It turns out that the solutions are very simple. We decompose the
charge vector $\Gamma$ into its projections on the positive and
negative subspaces at infinity which gives two vectors
$\Gamma_\pm^\infty \in {\mathbb R}^{5,21}$.  These two vectors define
a fixed 2-plane $K$ inside ${\mathbb R}^{5,21}$. Now, the radial dependence
of the orientation of the 5-plane $V_+(r)$ is given the
action of an $SO(5,21)$ boost ${\cal B}_K(\psi)$ along the constant
2-plane $K$, which is rotating the $\Gamma^-$ component into the
$\Gamma^+$ and with $r$-dependent rapidity $\psi(r)$. So
we have
\begin{equation}
  V_+(r)  = {\cal B}_K(\psi)\cdot V_+^\infty
\label{lorentz}
\end{equation}
where at infinity we must have $\psi(\infty) =0$ to satisfy the
boundary conditions, while near the horizon $\psi$ must take a value
such that $\Gamma \in V_+$ to satisfy the attractor condition
$\Gamma^- = 0$. All the information about the solution is contained in
the two functions $U(r)$ and $\psi(r)$.  In units where the 6d Planck
length is one, the two functions are
\begin{equation}\begin{split}
& e^{-2U(r)} = \left(1+ {|\Gamma_+^\infty|-|\Gamma_-^\infty| \over
    r^2}\right)^{1/2} \left(1+ {|\Gamma_+^\infty|
    +|\Gamma_-^\infty|\over r^2}\right)^{1/2}\\
& e^{\psi(r)} = \sqrt{{|\Gamma_+^\infty| -|\Gamma_-^\infty|+r^2 \over
|\Gamma_+^\infty| +|\Gamma_-^\infty|+r^2}}
\label{flowsolution}
\end{split}\end{equation}
From these one can reconstruct the full solution, including the
radial dependence of $\Gamma_\pm(r)$ and of the 3-form field
strengths following the detailed analysis in
\cite{Mikhailov:1999fd}. As an easy check we can see that the ADM
mass of this solution is indeed proportional to
$|\Gamma_+^\infty|$ as expected from \eqref{centralstring}, while
in the near horizon region we get an AdS$_3$ throat of size
proportional to $|\Gamma| =
\sqrt{|\Gamma^\infty_+|^2-|\Gamma^\infty_-|^2}$, which is
independent of the value of the moduli at infinity.

Notice that the motion on the moduli space ${\cal M}_{sugra}$ from
$z^\infty$ to $z^*\in{\cal M}_{sugra}^{*,\Gamma}$ is rather simple
and given by the action of a one-parameter group of $SO(5,21)$
Lorentz boosts along a constant 2-plane \eqref{lorentz}. We take
the simplicity of the solution as an indication that the
corresponding RG-flow, appropriately interpreted, at finite scales
might be also simple.

One approach would be to try to apply $tt^*$ inspired arguments away
from the conformal point. As we saw, the set of irrelevant operators
preserving the ${\cal N}=(4,4)$ supersymmetry is finite, so it is not
totally inconceivable that by generalizing the $tt^*$ formalism we
might be able to find RG-flow orbits in this restricted subset of
parameters. Ideally we would like to reproduce the full moduli space
\eqref{sugrabig} from the perturbed ${\cal N}=(4,4)$ and the attractor
flows described above. The $tt^*$ formalism has already been used in
theories away from criticality. The reason that we cannot apply the
standard $tt^*$ arguments directly to our system is that the
irrelevant operators that we are perturbing by are of the form
\eqref{blabla}. The $tt^*$ formalism is based on the ${\cal N}=(2,2)$
algebra. From an ${\cal N}=(2,2)$ point of view, the operators
\eqref{blabla} are not F-term perturbations, since they involve too
many supercurrents, which are not visible in a single ${\cal N}=(2,2)$
subalgebra, and naively should not be protected. It is the underlying
${\cal N}=(4,4)$ which protects these operators. It would be very
interesting to generalize the $tt^*$ framework for perturbations of
this form in ${\cal N}=(4,4)$ theories.

Another way to study finite flows away from the conformal fixed
point would be to go to higher orders in conformal perturbation
theory. Since we have included all irrelevant operators that
preserve ${\cal N}=(4,4)$ supersymmetry, in a scheme in which
these supersymmetries are preserved no further irrelevant
operators should be generated in the effective action, as these
would necessarily break some of the supersymmetries. Therefore, in
such a scheme all the conformal perturbation theory would do is to
generate a non-trivial scale dependence of the irrelevant
couplings. One can imagine that the latter may eventually be
related to the rather simple form of the flow solution
(\ref{flowsolution}) and it would be interesting to explore this
further.

\subsection{A decoupling limit and 6d gauge theory}

Finally, let us mention that certain
orbits of the attractor flow can be embedded in a boundary theory with an
honest decoupling limit in the following way, which was also described in
\cite{Israel:2003ry}. Consider IIB
compactified on $K3$ of volume $V_{K3}=v \alpha'^2$ with $v$ dimensionless,
and a bound state of D1/D5 branes. The D1/D5 solution is
\begin{equation}\begin{split}
&  ds^2 = Z_1^{-1/2} Z_5^{-1/2}(-dt^2+dx^2) +  Z_1^{1/2} Z_5^{1/2}
(dr^2 +r^2 d\Omega_3^2)+  Z_1^{1/2}  Z_5^{-1/2} \sqrt{v} \alpha' ds_{K3}^2\\
& e^{2\phi} = g_s^2 Z_1 Z_5^{-1}\\
& Z_1 = 1 + {g_s Q_1 \alpha'/v \over r^2}\\
& Z_5 = 1 + {g_s Q_5 \alpha' \over r^2}\\
\end{split}\end{equation}
where $ds_{K3}^2$ is the metric of a $K3$ of unit volume. The standard
decoupling limit which leads to AdS$_3$/CFT$_2$ is $
\alpha'\rightarrow 0$ keeping $g_s,v$ constant. Instead we consider
the decoupling limit corresponding to a D5 brane in flat space
\begin{equation}
  \alpha'\rightarrow 0,\qquad g_s \alpha' = g_{YM}^2 = \text{const},\qquad
V_{K3} = v \alpha'^2 = \text{const},\qquad U={r\over \alpha'} =
\text{const} .
\end{equation}
In this limit we do have a decoupling of the open and closed
modes. Also, the $+1$ drops out of the harmonic function $Z_5$ but not $Z_1$
\begin{equation}
  Z_1 = 1 + {g_{YM}^2 /V_{K3} \over U^2} ,\qquad Z_5 \simeq {g_{YM}^2 Q_5 \over
\alpha^{'2} U^2} .
\end{equation}
The decoupled supergravity solution takes the form
\begin{equation}\begin{split}
 & {ds^2\over \alpha'} =   {U\over \sqrt{g_{YM}^2 Q_5}} Z_{1}^{-1/2}(-dt^2 + dx^2)
+ {\sqrt{g_{YM}^2 Q_5} Z_1^{1/2}\over U} (dU^2 + U^2 d\Omega_3^2)
+ {U \over \sqrt{g_{YM}^2 Q_5}}  Z_1^{1/2} \sqrt{V_{K3}}ds_{K3}^2,  \\
& e^{2\phi} = {g_{YM}^2 U^2\over Q_5} Z_1.
\end{split}\end{equation}
It is easy to see that this is asymptotically locally the same as the
decoupling limit of the D5 brane in flat space, but the global structure is
${\mathbb R}^{1,1}\times K3$. The dilaton blows up at infinity but we can
S-dualize to the NS5 brane solution which is well behaved there. As we
move towards the IR the size of the $K3$ shrinks and reaches a stringy
size fixed by the attractor mechanism, while the rest of the geometry
becomes AdS$_3\times$ S$^3$. From the point of view of our general
solution \eqref{flowsolution} this corresponds to scaling
$\Gamma_\pm^\infty \rightarrow \infty$ in such a way that the $+1$ in
the second harmonic function in the expression for $e^{-2U(r)}$ can be
dropped but not in the first.

Holographically in the UV we start with the 5+1 dimensional NS5
brane $(1,1)$ little string theory living on ${\mathbb R}^{1,1}\times K3$.
Below energy scales of the order $(g_{YM})^{-1}$ the theory can be
well described by 5+1 SYM on ${\mathbb R}^{1,1}\times K3$. At energies below
$(V_{K3})^{-1/4}$ we can integrate out the $K3$ modes and end up
with the 2-dimensional D1/D5 SCFT in the IR. Along this RG-flow
between a 5+1 and a 1+1 theory the scalar moduli flow and get
fixed values by the attractor mechanism.  So in principle the RG
flow between 5+1 dimensional SYM on ${\mathbb R}^{1,1}\times K3$ and the 2d
CFT in the IR should contain a holographic description of the
attractor mechanism for this simple system, at least for some
attractor flows.  This is hard to study in general but it would be
intersecting to see if it is possible to truncate the RG-flow to
the BPS sector of the system, by identifying the operators in the
gauge theory which flow to the chiral primaries in the IR and
studying the supersymmetric sector of the RG-flow.  We identified
the corresponding operators in section \ref{subsec:4dgauge} but
leave the study of the boundary RG-flow for future work.

\section{Black Hole Berry phase}
\label{sec:bhbp}

Finally we would like to mention one more application of our
analysis. We have computed the connection for the chiral primary
operators in the NS sector of the ${\cal N}=(4,4)$ D1/D5 SCFT. By
spectral flow the chiral primaries are related to Ramond ground
states. This means that we know the exact connection for the vector
bundle of Ramond ground states over the moduli space of the theory. In
spacetime the Ramond ground states correspond to quantum microstates
of a bound state of D1 and D5 branes, wrapped around $S^1\times K3$,
which is a small black hole in 5d. The connection on the bundle of
chiral primaries is telling us how different microstates of the black
hole mix as we move on the moduli space. This is is a version of the
(nonabelian) Berry phase \cite{Berry:1984jv},\cite{{Wilczek:1984dh}}
for the internal states of the black hole, under adiabatic change of
the moduli of the compactification. In principle, this exactly
computable holonomy would allow one to set up interference experiments
sensitive to the internal microstate of the black hole. Obviously
preparing a black hole in a pure state in practice would be highly
challenging. It would be interesting to explore the implications of
this phenomenon in more detail, we hope to report on it in the future.

Other systems in string theory where Berry's phase appears and has
interesting interpretation have been studied recently
\cite{Pedder:2007ff},\cite{Pedder:2007wp},\cite{Pedder:2008je}.

\section{Summary and further directions}
\label{sec:conclu}

The main technical point of this paper was the analysis of the moduli
dependence of the chiral ring for ${\cal N}=(4,4)$ superconformal
field theories.  It was based on an application of the $tt^*$
equations which we derived\footnote{Of course the original
  derivation is more general, it also works for non-conformal ${\cal
    N}=(2,2)$ theories.} from general principles of conformal
perturbation theory and not relying on the topological twisting.
This derivation clarifies the connection between the work based on
topological-antitopological fusion
\cite{Cecotti:1991me},\cite{Bershadsky:1993cx} and that on
standard CFT arguments \cite{Ranganathan:1992nb},
\cite{Ranganathan:1993vj}. The main result is that for ${\cal
  N}=(4,4)$ theories the chiral ring is covariantly constant over the
moduli space. We found that the bundles of chiral primaries are
constrained to be homogeneous bundles, whose curvature is exactly
computable.

In the case of AdS$_3$/CFT$_2$ our results imply a non-renormalization
theorem for 3-point functions of chiral primaries and more general
extremal correlators, even at finite values of $N$. This explains the
agreement found in \cite{Gaberdiel:2007vu}, \cite{Dabholkar:2007ey},
\cite{Pakman:2007hn}, \cite{Taylor:2007hs}. To gain a better
understanding of the relation between different points on the moduli
space it would be useful to clarify the global structure of the moduli
space of the SCFT and possible monodromies of the chiral ring around
singularities.

The connection for the chiral primaries that we computed in this paper
can be used to demonstrate agreement between the attractor flow and
RG-flow in the vicinity of the fixed point, in the simple case of an
infinite D1/D5 black string. It would be interesting to extend this
analysis to finite order away from the fixed point, for example under the flow
by the irrelevant
operators mentioned in the text which do not break the ${\cal
  N}=(4,4)$ supersymmetry. This is a finite set of operators so it
might be possible to find constrained self-consistent flows towards
the UV related to the attractor flows in supergravity. In particular,
since we are in a certain sense studying the BPS sector of the theory,
we might hope to reconstruct the full geometry of the supergravity
moduli space \eqref{sugrabig} from the geometry of the field theory
moduli space away from criticality.

An obvious generalization would be to set up a similar analysis
for systems with less supersymmetry. One example is the ${\cal
N}=(0,4)$ MSW superconformal field theory which appears on the
worldvolume of an M5 brane wrapping a four-cycle in a Calabi-Yau
manifold \cite{Maldacena:1997de}. The five-dimensional
supergravity solution has an AdS$_3\times$S$^2$ near horizon
geometry and has a more interesting attractor flow towards the
fixed point. One could try to identify the constraints from
supersymmetry on the structure of the moduli spaces and the chiral
ring. Moreover, this theory has a very interesting set of
supergravity solutions \cite{deBoer:2008fk} corresponding to
multi-centered black holes which
 can be constructed by perturbing the theory towards the IR. It would be nice
to see if the structure found from supergravity can be reproduced
in any sense from the RG-flow in the boundary theory.

Four-dimensional black holes in ${\cal N}=2$ supergravity provide
another interesting example where a suitable extension of our
results might be obtainable. In this case the theory on the branes
should flow to ``superconformal quantum mechanics'' which would be
the boundary side of AdS$_2$/CFT$_1$. This conjectured duality has
not been fully understood so it is not straightforward to make
progress in this direction.

It would also be interesting to understand how to formulate the
computation of Berry's phase for the microstates of other
supersymmetric black holes. Again the ${\cal N}=2$ 4d case would
be most interesting, but difficult for the reasons mentioned in
the previous paragraph. It might be interesting to see if anything
can be said about states in the D1-D5 system which are of the form
chiral primary-anything, corresponding to D1-D5-P microstates. It
is not clear if the holonomy for such states is sufficiently
constrained by supersymmetry, but as these would correspond to
microstates of a 5d black hole with a macroscopic horizon they are
worthwhile to investigate.

Finally let us mention another direction which might be interesting to
explore further. While the connection for chiral primaries over the
moduli space has been studied in detail for the case of 2d
superconformal field theories, the same analysis has not been
performed for their higher dimensional analogues.  More precisely, one
could try to study the connection for the operators in the chiral ring
of 4d superconformal gauge theories. In particular it would be
interesting to see if there is any way of deriving equations similar
to $tt^*$ for 4 dimensional theories, expressing the curvature of the
bundle of chiral primaries in terms of the chiral ring
coefficients. If such relations exist, they may lead to interesting
constraints for the K\"ahler metric on the moduli space of ${\cal
  N}=1$ SCFTs and they may be useful for the analysis of aspects of
Seiberg duality in ${\cal N}=1$ theories.

Let us close with a simple observation in this direction. Consider
four dimensional ${\cal N}=4$ $SU(N)$ SYM at the superconformal point,
whose R-symmetry is $SO(6)$. This theory is not an isolated conformal
field theory since we can continuously vary the coupling $\tau=
{\theta \over 2 \pi} + {4\pi i \over g_{YM}^2}$ without breaking
conformal invariance. Its moduli space ${\cal M}$ is the upper
half-plane modded out by the action of a certain subgroup of the
$SL(2,{\mathbb Z})$ duality group. Operators in short representations
can be constructed starting with a holomorphic combination of two of the
six real scalars of the theory, say $Z = \Phi^1 + i \Phi^2$, and then
considering operators of the form $\text{Tr}Z^p$ and their
products. By acting on these operators with the supercharges and
momentum generators we can construct the full superconformal
multiplet. Motion along ${\cal M}$ is generated by marginal operators
which in four dimensions have conformal dimension 4.  In ${\cal N}=4$
these marginal operators can be written as descendants of chiral
primaries in the form
\begin{equation}
{\cal O} = Q^4  \text{Tr}Z^2
\label{marginaln4}
\end{equation}
This is a complex operator whose real and imaginary parts express the coupling
of the Lagrangian density to ${1\over g_{YM}^2}$ and ${\theta}$ respectively.
In components
\begin{equation}
  {\cal O} \sim \text{Tr} (F_{\mu\nu}^2) + i \text{Tr}(F\wedge F) +...
\end{equation}
The metric on the moduli space ${\cal M}$ is given by the following
expression
\begin{equation}
  ds^2 = g_{\tau \overline{\tau}}d\tau d\overline{\tau} \sim {1\over
    (\text{Im} \tau)^2 }d\tau d\overline{\tau}
\label{metricn4}
\end{equation}
and is related to the 2-point function
\begin{equation}
  \langle {\cal O}(x) \overline{{\cal O}}(y)\rangle =
{g_{\tau \overline{\tau}} \over |x-y|^8}
\end{equation}
The important point is that this metric is not flat. Hence the tangent
bundle ${\cal TM}$ has nonzero curvature. The marginal operators
\eqref{marginaln4} correspond to tangent vectors on ${\cal M}$. Then
under parallel transport on the moduli space the marginal operators
will mix as
\begin{equation}
  {\cal O} \rightarrow e^{i \chi }{\cal O},\qquad
 \overline{{\cal O}} \rightarrow e^{-i \chi }\overline{{\cal O}}
\label{phasen4}
\end{equation}
where the angle $\chi$ is exactly computable from the geometry of the
moduli space \eqref{metricn4}. From \eqref{marginaln4} we see that the
marginal operators are sections of a bundle which is the tensor
product of the bundle of the supercharges and the bundle whose fiber
is generated by the chiral primary $\text{Tr}Z^2$. As in the
two-dimensional case, we expect that the $SO(6)$ R-symmetry is
covariantly constant over the moduli space ${\cal M}$. Then the chiral
primary $\text{Tr}Z^2$ cannot get a phase under parallel
transport. Thus we learn that the phase \eqref{phasen4} is coming from
a mixing of the supercharges which corresponds to a rotation under the
$U(1)$ outer automorphism of the ${\cal N}=4$ algebra in
4d \footnote{This is the $U(1)_Y$ ``bonus symmetry'' discussed in
  \cite{Intriligator:1998ig}.}. This mixing is exactly computable at
all values of the coupling from the geometry of the moduli space. In
this case it seems that we only have curvature for the supercharges
and not the chiral primaries\footnote{We do however have curvature for
  the descendants of the chiral primaries due to the curvature of the
  supercharges.}.

It would be interesting to explore the constraints from
supersymmetry on the geometry of the chiral ring over the moduli
space for other four dimensional superconformal field theories,
with less supersymmetry or for other operators in short multiplets
such as the 1/16 BPS operators in ${\cal N}=4$.

\section*{Acknowledgments} We would like to thank
M. Cheng, A. Dabholkar, J. David, F. Denef, R. Dijkgraaf, S.
El-Showk, K. Goldstein, R. Gopakumar, M. Guica, L. Hollands, K.  Hori,
N. Iizuka, I. Kanitscheider, A. Kashani-Poor, A. Klemm, G.  Mandal,
D. Marolf, I. Messamah, S. Minwalla, V. Niarchos, S. Raju, J. Sonner,
D. Tong, S. Trivedi, S. Wadia and especially M. Shigemori,
K. Skenderis and M. Taylor for valuable discussions.  K.P. would like
to thank the ``Monsoon Workshop on string theory'' at TIFR, Mumbai for
warm hospitality during the completion of this work. This work was
supported in part by Foundation of Fundamental Research on Matter
(FOM).

\appendix

\section{The superconformal algebra}
\label{app:algope}
\subsection{The ${\cal N}=2$ superconformal algebra}
The ${\cal N}=2$ algebra has the form
\begin{equation}\begin{split}
& T(z) T(w) = {c/2\over (z-w)^4} + {2 T(w) \over (z-w)^2} + {\partial
      T(w) \over z-w}+...\\ & J(z) J(w) = {c/3 \over (z-w)^2}+...\\ &
    T(z) J(w) = {J(w) \over (z-w)^2} + {\partial J (w) \over
      z-w}+...\\ & G^+(z) G^-(w) = {2c/3 \over (z-w)^3} + {2J(w)
      \over(z-w)^2} + {2T(w) + \partial J(w) \over z-w}+...\\ & T(z)
    G^\pm(w) = {3\over 2} {G^\pm(w) \over (z-w)^2} +{\partial G^\pm(w)
      \over z-w} +...\\ & J(z) G^\pm(w) = \pm {G^\pm(w) \over z-w}+...
\end{split}\end{equation}
and
\begin{equation}
  T^\dagger = T,\qquad J^\dagger = J,\qquad (G^\pm)^\dagger = G^\mp
\end{equation}
We define the modes
\begin{equation}\begin{split}
  & L_n = {1\over 2 \pi i} \oint z^{n+1} T(z) dz \\ & G_r^\pm =
    {1\over 2 \pi i} \oint z^{r+1/2}G^\pm(z) dz \\ & J_n = {1 \over 2
      \pi i} \oint z^n J(z) dz
\end{split}\end{equation}
and we have the commutation relations
\begin{equation}\begin{split}
  & [L_m,L_n] = (m-n)L_{m+n} + {c\over 12} m(m^2 -1)
    \,\delta_{m+n,0}\\ & [J_m,J_n] = {c\over 3} m\, \delta_{m+n,0}\\ &
            [L_m,J_n] = -n J_{m+n} \\ & \{G^-_r, G^+_s\} = 2 L_{r+s} -
            (r-s) J_{r+s} + {c\over 3} (r^2 -1/4)\, \delta_{r+s,0}\\ &
            \{G^+_r,G^+_s\} = \{G^-_r,G^-_s\} = 0\\ & [L_m, G^\pm_r] =
            (m/2 - r) G^\pm_{m+r}\\ & [J_m, G^\pm_r] = \pm
            G^{\pm}_{m+r}
\end{split}\end{equation}
where $r,s$ is half-integer in the NS sector and integer in the R
sector, and have the following hermiticity conditions
\begin{equation}
  (L_m)^\dagger =L_{-m},\qquad (J_m)^\dagger = J_{-m},\qquad
  (G^\pm_r)^\dagger =G^\mp_{-r}
\end{equation}

\subsection{The ${\cal N}=4$ superconformal algebra}

In the small ${\cal N}=4$ algebra the bosonic currents are
$T(z),J^i(z), i=1,2,3$ and the supercurrents $G^{\pm+}(z)$ and
$G^{\pm-}(z)$. The central charge and the level are related by
$c=6k$. The algebra has the following form
\begin{equation}\begin{split}
& T(z) T(w) = {c/2 \over (z-w)^4} + {2 T(w) \over (z-w)^2} + {\partial
      T(w)\over z-w}+...\\
& J^i(z)J^j(w) = {k\over 2}{ \delta^{ij}
      \over (z-w)^2} + i \epsilon^{ijk} {J^k(w) \over z-w}+...\\
& T(z) J^i(w) = {J^i(w) \over (z-w)^2} + {\partial J^i(w) \over
      z-w}+...\\
& T(z) G^{ab}(w) = {3 \over 2} {G^{ab}(w) \over (z-w)^2} + {\partial G^{ab}
(w) \over z-w} +...\\
& J^i(z) G^{a\pm}(w) = {1\over 2} \sigma^i_{ba} {G^{b\pm}(w)\over (z-w)}+...
\end{split}
\end{equation}
and
\begin{equation}\begin{split}
& G^{a+}(z) (G^{b+})^\dagger(w) = {2 c \over 3}{\delta_{ab}\over (z-w)^3}+
    {4 \overline{\sigma}^i_ {ab} J^i \over (z-w)^2} + {2T(w)
      \delta_{ab}\over (z-w)} + {2 \overline{\sigma}^i_{ab} \partial
      J^i \over z-w} +...\\ & G^{a+}(z) G^{b+}(w) = \text{regular} \\ &
    G^{a-}(z) G^{b-}(w) = \text{regular} \\
\end{split}\end{equation}
where $a,b=+,-$ and $\sigma^i_{ab}$ are the Pauli matrices. The hermiticity
conditions of the generators are
\begin{equation}
T^\dagger = T,\qquad (J^i)^\dagger = J^i,\qquad (G^{++})^\dagger =
G^{--},\qquad (G^{+-})^\dagger = - G^{-+}
\end{equation}

\section{Some useful OPEs for ${\cal N}=(2,2)$}

Let us call $\phi$ a $(cc)$ field of $(L_0,J_0)=(h,q)$. We have
the following OPEs

\begin{equation}\begin{split}
 & G^+(z) \phi(w) = \text{regular} \\
 & G^-(z) \phi(w) = { (G^-_{-1/2}\cdot \phi)(w) \over z-w}+...\\
 & T(z) \phi (w) = {h}{\phi(w) \over (z-w)^2} + {\partial \phi(w)
\over z-w}+...\\
& J(z) \phi(w) = q {\phi(w) \over z-w}+...
\end{split}\end{equation}
Using the algebra and that $h=q/2$ for a chiral primary we find
\begin{equation}
 G^+(z) (G^-_{-1/2}\cdot \phi)(w) = 2q {\phi(w) \over (z-w)^2} + 2
{\partial \phi(w) \over z-w}+...
\end{equation}
For chiral primaries with $(h,q) = (1/2,1)$ this becomes
\begin{equation}
  G^+(z) (G^-_{-1/2}\cdot \phi)(w) = 2\partial_w \left({\phi(w) \over z-w}
\right)+...
\end{equation}

\section{Curvature of supercurrents in ${\cal N}=(2,2)$}
\label{app:curvsuper}

We have to study the 4-point function of the form \eqref{fourgg}. For
definiteness we will consider
\begin{equation}
A=  \langle {\cal O}_i(x) \overline{{\cal O}_j}(y) G^+(z) G^-(w)\rangle
\end{equation}
As a function of $z$, $A$ is holomorphic so it is determined by its
singularity structure at $z=x,y,w$. For this we need the OPEs of
$G^+(z)$ with the other insertions. We have the following results
\begin{equation}\begin{split}
& G^+(z) G^-(w) = {2c/3 \over (z-w)^3} + {2J(w) \over(z-w)^2} + {2T(w)
      + \partial J(w) \over z-w}+...  \\ & G^+(z) {\cal O}_i(x) =
    \partial_x\left({(\overline{G}^-_{-1/2} \cdot \phi_i)(x)\over
      z-x} \right)+...  \\ & G^+(z) \overline{{\cal O}_j}(y) =
    \text{regular}
\end{split}\end{equation}
where we used ${\cal O}_i ={1\over 2} G^-_{-1/2}\overline{G}^-_{-1/2} \cdot \phi_i$,
$\overline{{\cal O}_j}={1\over 2} G^+_{-1/2}\overline{G}^+_{-1/2} \cdot \overline{
\phi_j}$ and the ${\cal N}=(2,2)$ algebra. So we have
\begin{equation}\begin{split}
  A = & \langle {\cal O}_i(x) \overline{{\cal O}_j}(y) \left({2c/3
    \over (z-w)^3} + {2J(w) \over(z-w)^2} + {2T(w) + \partial J(w)
    \over z-w}\right) \rangle \\ &+  \partial_x\left({1\over z-x}
\langle (\overline{G}^-_{-1/2}
    \cdot \phi_i)(x) \overline{{\cal O}_j}(y)
  G^-(w)\rangle\right)
\end{split}
\label{bigalpha}
\end{equation}
This is of the form $A=A_1+A_2$ where each term corresponds to one of the
lines in the expression above. The term $A_1$ can be easily evaluated by
the usual conformal Ward identities on the correlation function
\begin{equation}
\langle  {\cal O}_i(x) \overline{{\cal O}_j}(y) \rangle = {\ka_{i \overline{j}}
\over |x-y|^4}
\label{marrr}
\end{equation}
After some algebra we find
\begin{equation}
  A_1 = {2c \ka_{i \overline{j}} \over 3 |x-y|^4 (z-w)^3}+{2\ka_{i\overline{j}} \over (w-x)^2 (w-y)^2 (\overline{x}-\overline{y})^2
(z-w)}
\end{equation}
To compute $A_2$ we need the correlation function
\begin{equation}
B=  \langle (\overline{G}^-_{-1/2}\cdot \phi_i)(x) \overline{{\cal O}_j}(y)
G^-(w)\rangle
\end{equation}
As a function of $w$ the expression $B$ is holomorphic, so again we can use
the OPEs to determine it. We have
\begin{equation}
  B =  \partial_y \left( {1\over w-y} \langle
  (\overline{G}^-_{-1/2}\cdot \phi_i)(x) (\overline{G}^+_{-1/2}\cdot
  \overline{\phi_j}(y)\rangle\right)- {2 \over w-x} \langle
{\cal O}_i(x) \overline{{\cal O}_j}(y)\rangle
\end{equation}
Now using
\begin{equation}
  \langle (\overline{G}^-_{-1/2}\cdot \phi_i)(x)
  (\overline{G}^+_{-1/2}\cdot
  \overline{\phi_j}(y)\rangle = 2{\ka_{i
      \overline{j}} \over (x-y) (\overline{x}-\overline{y})^2}
\end{equation}
and expression \eqref{marrr} we find
\begin{equation}
  B = -{2 \ka_{i\overline{j}} \over (w-x)(w-y)^2(\overline{x}-\overline{y})^2}
\end{equation}
so
\begin{equation}
  A_2 = {2 \ka_{i\overline{j}} (w-2x + z)\over
    (w-x)^2(w-y)^2(\overline{x}-\overline{y})^2 (x-z)^2}
\end{equation}
Finally going back to \eqref{bigalpha} we can compute $A=A_1+A_2$ and we find
\begin{equation}
  A = {2c \ka_{i \overline{j}} \over 3 |x-y|^4 (z-w)^3}+
  {2\ka_{i\overline{j}} \over (w-y)^2(z-w)(x-z)^2
    (\overline{x}-\overline{y})^2 }
\end{equation}
Similarly we can compute the other 4-point functions needed for the
computation of the curvature of the supercurrents.

\section{4-point functions in ${\cal N}=(2,2)$}
\label{app:fourtwotwo}

Consider a $(cc)$ field $\phi_k$ and an $(aa)$
$\overline{\phi_l }$. We want to simplify the 4-point
function
\begin{equation}
 G(x,y,z,w)=  \langle {\cal O}_i(x) \overline{{\cal O}_j}(y) \phi_k(z)\overline{\phi_l}(w)\rangle
\end{equation}
where the marginal operators are descendants of the chiral ring
\begin{equation}
  {\cal O}_i ={1\over 2} G^-_{-1/2}\overline{G^-}_{-1/2} \cdot\phi_i,\qquad
  \overline{{\cal O}_j} ={1\over 2} G^+_{-1/2}\overline{G^+}_{-1/2} \cdot
  \overline{\phi_j}
\label{marginalchiralb}
\end{equation}
We can also write the operators as \cite{Dixon:1989fj}
\begin{equation}
  {\cal O}_i(x) ={1\over 2} {1\over 2 \pi i} \oint_x ds {s-t \over x-t} G^-(s)
\left(\overline{G}^-_{-1/2} \phi_i\right)(x)
\end{equation}
we choose $t=z$ and we deform the contours. The supercurrent $G^-(s)$
annihilates $\overline{\phi}_{\overline{l}}$ and it has a first order pole
with $\phi_k(z)$ which is cancelled with the $(s-z)$ in the numerator. Finally
we have to use the ${\cal N}=(2,2)$ algebra to compute its OPE with the
insertion at $y$. We find that the answer is
\begin{equation}
  G(x,y,z,w) = {1\over 2}{\partial \over \partial y} \left({y-z \over
    x-z} \langle \left(\overline {G}^-_{-1/2} \phi_i\right)(x)
  \left(\overline{G}^+_{-1/2}\overline{\phi_j}\right)(y)
  \phi_k(z)\overline{\phi_l}(w)\rangle\right)
\end{equation}
Doing the same for the supercurrent $\overline{G}^+$ we end up with the
the following expression
\begin{equation}
  G(x,y,z,w)= \partial_y\partial_{\overline{y}}\left( {|y-z|^2 \over
    |x-z|^2} \langle \phi_i(x) \overline{\phi_j}(y)
  \phi_k(z)\overline{\phi_l}(w)\rangle\right)
\end{equation}

\section{OPE between chiral primary and antichiral primary}
\label{app:caope}

Consider $(cc)$ field $\phi_i$ of charge $q_i>0$ and $(aa)$ field
$\overline{\phi_l}$ of charge $q_l<0$ with
$q_i<|q_l|$. Consider their OPE
\begin{equation}
  \phi_i(z)\overline{\phi_l}(w) = \sum_{r\overline{r}}
      {D^\rho_{i\overline{l}} A_\rho(z) \over (z-w)^r
        (\overline{z}-\overline{w})^{\overline{r}}}
\end{equation}
The field $A_\rho$ has $U(1)$ charge $q_\rho = q_i+q_l<0$,
and conformal dimension $ h_\rho = h_i+h_l-r$ (similarly
for the right-moving side). From unitarity we have the condition
$h_\rho \geq| q_\rho|/2$. Equivalently this means
\begin{equation}
  r\leq q_i
\end{equation}
If the inequality is saturated (and similarly on the right moving side)
the corresponding field $A_\rho$ will be antichiral primary of charge
$q_i+q_l<0$. So the OPE will have
the form
\begin{equation}
  \phi_i(z)\overline{\phi_l}(w) = {D^{\overline{k}}_{i\overline{l}}
\overline{\phi_k}(w) \over (z-w)^{q_i}
(\overline{z}-\overline{w})^{\overline{q}_i}} +...
\end{equation}
The coefficients $D^{\overline{k}}_{i \overline{l}}$ are related to
the chiral ring structure constants. We consider the 3-point
function
\begin{equation}
\langle  \phi_i(z) \overline{\phi_l}(w) \phi_n(y)\rangle
\end{equation}
and take the OPE in two different ways to show that
\begin{equation}
  D^{\overline{k}}_{i\overline{l}} =
  C^m_{in}\ka^{\overline{k}n}\ka_{m\overline{l}}
\end{equation}
So the conclusion is that the leading term of the OPE of $(cc)$ with $(aa)$
is given by the conjugated chiral ring coefficients.

\section{Contours}
\label{app:contours}
Now we want to study the first  term of \eqref{curvqq} using the OPE between
$\phi_i$ and $\overline{\phi_j}$.  We define
\begin{equation}
C=-{1\over (2\pi)^2} \lim_{|r|\rightarrow 1} \int_{|y|=1}
   d\theta_1 \int_{|x|=r} d\theta_2 \left(r^2 \langle \phi_i(x)
   \overline{\phi_j}(y) \phi_k(\infty) \overline{\phi_l}(0)\rangle -
   \langle \overline{\phi_j}(x) \phi_i(y) \phi_k(\infty)
   \overline{\phi_l}(0) \rangle\right)
\end{equation}
We change the angle variables to $\theta= {(\theta_1 + \theta_2) \over 2}$
and $\psi ={ (\theta_1 -\theta_2) \over 2}$ and we have
 \begin{equation}\begin{split}
  & C = -{2 \over (2 \pi)^2}\lim_{|r|\rightarrow 1} \int d\theta \int
     d\psi\\ &\left( r^2 \langle \phi_k(\infty)
     \overline{\phi_j}(e^{i(\theta+\psi)}) \phi_i(re^{i(\theta-\psi)})
     \overline{\phi_l}(0)\rangle-\langle \phi_k(\infty)
     \phi_i(e^{i(\theta+\psi)})\overline{\phi_j}(re^{i(\theta-\psi)})
     \overline{\phi_l}(0)\rangle \right)
\end{split}\end{equation}
For $\psi\neq 0 $ the contribution from $\psi$ cancels with that
from $-\psi$ in the limit $r\rightarrow 1$. However this does not
mean that the integral is zero, since we may have
$\delta$-function-like contributions from $\psi= 0$. These
contributions can be evaluated using the OPE of $\phi_i$ with
$\overline{\phi_j}$ which is
\begin{equation}
  \phi_i(z) \overline{\phi_j}(w) = \sum_\rho {D^\rho_{i\overline{j}}
    A_\rho(w) \over (z-w)^{1
      -h_\rho}(\overline{z}-\overline{w})^{1-\overline{h}_\rho}}
\end{equation}
Let us assume that the operator $A_\rho$ has dimension
$(h_\rho,h_\rho-s_\rho)$ where $s_\rho$ is the spin. Then
\begin{equation}\label{jan1}\begin{split}
  &  C = -{2 \over (2 \pi)^2}  \lim_{r\rightarrow 1} \int d\theta
    \int_{-\delta}^{+\delta} d\psi \sum_\rho{
      D_{\overline{l}\rho,k}D_{i\overline{j}}^\rho \over (z_1 -
      z_2)^{1-h_\rho}(\overline{z}_1-\overline{z}_2)^{1-\overline{h}_\rho}}
    \left( {(-1)^{s_\rho}|z_2|^2\over
      z_1^{h_\rho}\overline{z}_1^{\overline{h}_\rho}} - {1\over
      z_2^{h_\rho} \overline{z}_2^{\overline{h} _\rho}}\right)
\end{split}\end{equation}
where $z_1 = e^{i(\theta+\psi)}$, $z_2 = r e^{i(\theta-\psi)}$ and
where $\delta$ is a small number that is kept constant as
$\epsilon\rightarrow 0$. We can rewrite this as
\begin{equation}
   C = -{2 \over (2 \pi)^2}   \lim_{r\rightarrow 1} \int d\theta
    \int_{-\delta}^{+\delta} d\psi \sum_\rho{
      D_{\overline{l}\rho,k}D_{i\overline{j}}^\rho \over |1 - r
      e^{-2i\psi}|^{2-2h_\rho}(1 - r e^{2 i \psi})^{s_\rho} } \left(
    (-1)^{s_\rho}r^2 - {e^{2is_\rho \psi} \over r^{h_\rho +
        \overline{h}_\rho} } \right)
\end{equation}
If $A_\rho$ is a spin zero field $(h_\rho = \overline{h}_\rho)$, then
the contribution is proportional to
\begin{equation}
  \lim_{r\rightarrow 1} \int_{-\delta}^{+\delta} d\psi { 1\over
    \left| 1 - re^{-2i \psi}\right|^{2-2h_\rho}}\left(r^2 -
      {1\over r^{2h_\rho}}\right)
\end{equation}
One can show that this quantity\footnote{In fact, the integral in
(\ref{jan1}) can be explicitly evaluated for fixed $r$ and with
$\delta=\pi/2$ in terms of hypergeometric functions, but we will
not present these expressions here.} is finite and $\delta$
independent if $h_\rho = 0$ and zero if $h_\rho>0$. Its value for
$h_\rho=0$ is
\begin{equation}
 \lim_{r\rightarrow 1} \int_{-\delta}^{+\delta} d\psi { r^2-1\over
    | 1 - re^{-2i \psi}|^{2}} = -\pi
\end{equation}
So from the spin zero fields only the identity operator will
contribute to $C$ a factor of
\begin{equation}
g_{i\overline{j}}g_{k\overline{l}}
\end{equation}
Similarly we can show that from fields with nonzero spin, only $(1,0)$
and $(0,1)$ fields contribute. For the first case we need
\begin{equation}
  \phi_i(z) \overline{\phi_j}(w) = ...+ {D_{i\overline{j}}^J J(w) \over
(\overline{z}-\overline{w})}+...
\end{equation}
The coefficient $D_{i\overline{j}}^J$ can be easily computed using the
Ward identities for $J$ and we find
\begin{equation}
  D_{i\overline{j}}^J = {3 \over c} g_{i\overline{j}}
\end{equation}
where we used
\begin{equation}
  \langle J(0) J(\infty)\rangle = {c \over 3},\qquad
\langle \phi_i(1) \overline{\phi_j}(0) J(\infty) \rangle = g_{i \overline{j}}
\end{equation}
for fields $\phi_i, \overline{\phi_j}$ of charge $+1,-1$.
Similarly
\begin{equation}
\langle \overline{\phi}_l(0) J(1) \phi_k(\infty)\rangle\equiv
D_{\overline{l}J,k} =- q g_{k\overline{l}}
\end{equation}
where $q$ is the charge of $\phi_k$. We also need the following
value for the $\psi$ integral for $h_\rho=s_\rho=1$
\begin{equation}
 \lim_{r\rightarrow 1}
    \int_{-\delta}^{+\delta} d\psi {
      1\over (1 - r e^{2 i \psi}) } \left(
    -r^2 - {e^{2i \psi} \over r } \right)=-\pi
\end{equation}
So the contribution from the currents is equal to
\begin{equation}
  -{3 \over c}(q + \overline{q}) g_{i\overline{j}}g_{k\overline{l}}
\end{equation}
All in all we get the following answer
\begin{equation}
C = g_{i\overline{j}} g_{k \overline{l}} \left( 1 - {3\over c} (q+
\overline{q}) \right)
\end{equation}

\section{Current/Marginal Operator OPE}
\label{app:curmod}
Let us consider a chiral primary $\phi$ with $h=j^3=1/2$ and
$\overline{h}= \overline{j}^3 = 1/2$.  The marginal operator ${\cal O}(x)$
is the descendant of the chiral primary ${\cal O}(x) = G^{-k}_{-1/2}
\overline{G}^{-l}_{-1/2}\cdot\phi(x)$.
We want to compute the OPE of a current with the marginal operator. In general
it will be
\begin{equation}
  J^i(z) {\cal O}(w) = \sum_{m}{(J^i_m {\cal O})(w) \over (z-w)^{m+1}}
\end{equation}
So to compute the OPE we need to compute $J^i_m |{\cal O}\rangle$. We have
\begin{equation}
    J^i_m|{\cal O}\rangle =\overline{G}^{-l}_{-1/2} \left( [J^i_m,
      G^{-k}_{-1/2}] + G^{-k}_{-1/2} J^i_m\right) |\phi\rangle
\label{blablabla}
\end{equation}
From the ${\cal N}=4$ algebra we have the following commutator of the
modes
\begin{equation}
  [J^i_m , G^{ak}_r] =  {1\over 2} \sigma^i_{ba} G^{bk}_{m+r}
\end{equation}
For $m=0$ we have $J^i_0 |{\cal O}\rangle=0$ since we already knew
that $|{\cal O}\rangle$ is uncharged under the current algebra. For
$m>0$ the second term in \eqref{blablabla} is zero because $J^i_m
|\phi\rangle=0,\,m>0$. Also, from the commutators
above we notice that the first term is proportional to a certain
linear combination of
\begin{equation}
  G^{cd}_{m-1/2}
\end{equation}
If $m>0$ all of these operators annihilate the sate $|\phi\rangle$
because it is a primary, so finally we have
\begin{equation}
  J^i_m|{\cal O}\rangle =0,\qquad m\geq0
\end{equation}
This proves that the OPE between the currents $J^i(z)$ and a marginal
operator in ${\cal N}=(4,4)$ is completely regular.

There is in fact an alternative way to show this which does not
rely on supersymmetry. Consider an exactly marginal operator in
any theory which contains a non-abelian current algebra (which is
preserved by the exactly marginal operator). The only singular
terms in the OPE of a current with ${\cal O}$ arise from
$J^i_0|{\cal O}\rangle$ and $J^i_{+1}|{\cal O}\rangle$. The first
of these clearly vanishes, since ${\cal O}$ cannot be charged
under the non-abelian current algebra. The second of these yields
an operator of conformal weight $(0,1)$ which necessarily is an
anti-holomorphic current. These cannot carry any charge under the
holomorphic current algebra, whereas $J^i_{+1}|{\cal O}\rangle$
clearly does, and therefore $J^i_{+1}|{\cal O}\rangle=0$ and the
OPE between $J^i$ and ${\cal O}$ has to be regular. Notice that
this argument is completely general but fails for abelian current
algebras.


\begin{thebibliography}{99}

\bibitem{Maldacena:1997re}
  J.~M.~Maldacena,
  ``The large N limit of superconformal field theories and supergravity,''
  Adv.\ Theor.\ Math.\ Phys.\  {\bf 2}, 231 (1998)
  [Int.\ J.\ Theor.\ Phys.\  {\bf 38}, 1113 (1999)]
  [arXiv:hep-th/9711200].

\bibitem{Lerche:1989uy}
  W.~Lerche, C.~Vafa and N.~P.~Warner,
  ``Chiral Rings in N=2 Superconformal Theories,''
  Nucl.\ Phys.\  B {\bf 324}, 427 (1989).

\bibitem{Cecotti:1991me}
  S.~Cecotti and C.~Vafa,
  ``Topological antitopological fusion,''
  Nucl.\ Phys.\  B {\bf 367}, 359 (1991).

\bibitem{Bershadsky:1993cx}
  M.~Bershadsky, S.~Cecotti, H.~Ooguri and C.~Vafa,
  ``Kodaira-Spencer theory of gravity and exact results for quantum string
  amplitudes,''
  Commun.\ Math.\ Phys.\  {\bf 165}, 311 (1994)
  [arXiv:hep-th/9309140].

\bibitem{Gaberdiel:2007vu}
  M.~R.~Gaberdiel and I.~Kirsch,
  ``Worldsheet correlators in AdS(3)/CFT(2),''
  JHEP {\bf 0704}, 050 (2007)
  [arXiv:hep-th/0703001].

\bibitem{Dabholkar:2007ey}
  A.~Dabholkar and A.~Pakman,
  ``Exact chiral ring of AdS(3)/CFT(2),''
  arXiv:hep-th/0703022.

\bibitem{Pakman:2007hn}
  A.~Pakman and A.~Sever,
  ``Exact N=4 correlators of AdS(3)/CFT(2),''
  Phys.\ Lett.\  B {\bf 652}, 60 (2007)
  [arXiv:0704.3040 [hep-th]].

\bibitem{Taylor:2007hs}
  M.~Taylor,
  ``Matching of correlators in AdS$_3$/CFT$_2$,''
  arXiv:0709.1838 [hep-th].

\bibitem{Intriligator:1998ig}
  K.~A.~Intriligator,
  ``Bonus symmetries of N = 4 super-Yang-Mills correlation functions via  AdS
  duality,''
  Nucl.\ Phys.\  B {\bf 551}, 575 (1999)
  [arXiv:hep-th/9811047].

\bibitem{Intriligator:1999ff}
  K.~A.~Intriligator and W.~Skiba,
  ``Bonus symmetry and the operator product expansion of N = 4
  super-Yang-Mills,''
  Nucl.\ Phys.\  B {\bf 559}, 165 (1999)
  [arXiv:hep-th/9905020].

\bibitem{Lee:1998bxa}
  S.~Lee, S.~Minwalla, M.~Rangamani and N.~Seiberg,
  ``Three-point functions of chiral operators in D = 4, N = 4 SYM at  large
  N,''
  Adv.\ Theor.\ Math.\ Phys.\  {\bf 2}, 697 (1998)
  [arXiv:hep-th/9806074].

\bibitem{Seiberg:1988pf}
  N.~Seiberg,
  ``Observations On The Moduli Space Of Superconformal Field Theories,''
  Nucl.\ Phys.\  B {\bf 303}, 286 (1988).

\bibitem{Cecotti:1990kz}
  S.~Cecotti,
  ``N=2 Landau-Ginzburg versus Calabi-Yau sigma models: Nonperturbative
  aspects,''
  Int.\ J.\ Mod.\ Phys.\  A {\bf 6}, 1749 (1991).

\bibitem{Ferrara:1995ih}
  S.~Ferrara, R.~Kallosh and A.~Strominger,
  ``N=2 extremal black holes,''
  Phys.\ Rev.\  D {\bf 52}, 5412 (1995)
  [arXiv:hep-th/9508072].

\bibitem{Dijkgraaf:1998gf}
  R.~Dijkgraaf,
  ``Instanton strings and hyperKaehler geometry,''
  Nucl.\ Phys.\  B {\bf 543}, 545 (1999)
  [arXiv:hep-th/9810210].

\bibitem{Maldacena:1997de}
  J.~M.~Maldacena, A.~Strominger and E.~Witten,
  ``Black hole entropy in M-theory,''
  JHEP {\bf 9712} (1997) 002
  [arXiv:hep-th/9711053].

\bibitem{Seiberg:1999xz}
  N.~Seiberg and E.~Witten,
 ``The D1/D5 system and singular CFT,''
  JHEP {\bf 9904}, 017 (1999)
  [arXiv:hep-th/9903224].

\bibitem{Larsen:1999uk}
  F.~Larsen and E.~J.~Martinec,
  ``U(1) charges and moduli in the D1-D5 system,''
  JHEP {\bf 9906}, 019 (1999)
  [arXiv:hep-th/9905064].

\bibitem{Vafa:1995zh}
  C.~Vafa,
  ``Gas of D-Branes and Hagedorn Density of BPS States,''
  Nucl.\ Phys.\  B {\bf 463}, 415 (1996)
  [arXiv:hep-th/9511088].

\bibitem{Strominger:1996sh}
  A.~Strominger and C.~Vafa,
  ``Microscopic Origin of the Bekenstein-Hawking Entropy,''
  Phys.\ Lett.\  B {\bf 379}, 99 (1996)
  [arXiv:hep-th/9601029].

\bibitem{Larsen:1998xm}
  F.~Larsen,
  ``The perturbation spectrum of black holes in N = 8 supergravity,''
  Nucl.\ Phys.\  B {\bf 536}, 258 (1998)
  [arXiv:hep-th/9805208].

\bibitem{deBoer:1998ip}
  J.~de Boer,
  ``Six-dimensional supergravity on S**3 x AdS(3) and 2d conformal field
  theory,''
  Nucl.\ Phys.\  B {\bf 548}, 139 (1999)
  [arXiv:hep-th/9806104].

\bibitem{Kutasov:1988xb}
  D.~Kutasov,
 ``GEOMETRY ON THE SPACE OF CONFORMAL FIELD THEORIES AND CONTACT TERMS,''
  Phys.\ Lett.\  B {\bf 220} (1989) 153.

\bibitem{Ranganathan:1992nb}
  K.~Ranganathan,
 ``Nearby CFTs in the operator formalism: The Role of a connection,''
  Nucl.\ Phys.\  B {\bf 408}, 180 (1993)
  [arXiv:hep-th/9210090].

\bibitem{Ranganathan:1993vj}
  K.~Ranganathan, H.~Sonoda and B.~Zwiebach,
  ``Connections on the state space over conformal field theories,''
  Nucl.\ Phys.\  B {\bf 414}, 405 (1994)
  [arXiv:hep-th/9304053].

\bibitem{Distler:1992gi}
  J.~Distler,
  ``Notes on N=2 sigma models,''
  arXiv:hep-th/9212062.

\bibitem{Periwal:1989mx}
  V.~Periwal and A.~Strominger,
  ``KAHLER GEOMETRY OF THE SPACE OF N=2 SUPERCONFORMAL FIELD THEORIES,''
  Phys.\ Lett.\  B {\bf 235}, 261 (1990).

\bibitem{Strominger:1990pd}
  A.~Strominger,
  ``SPECIAL GEOMETRY,''
  Commun.\ Math.\ Phys.\  {\bf 133} (1990) 163.

\bibitem{Eguchi:1987sm}
  T.~Eguchi and A.~Taormina,
  ``UNITARY REPRESENTATIONS OF N=4 SUPERCONFORMAL ALGEBRA,''
  Phys.\ Lett.\  B {\bf 196}, 75 (1987).

\bibitem{Dijkgraaf:1990dj}
  R.~Dijkgraaf, H.~L.~Verlinde and E.~P.~Verlinde,
  ``Topological Strings In D $<$ 1,''
  Nucl.\ Phys.\  B {\bf 352} (1991) 59.

\bibitem{Dixon:1989fj} L.~J.~Dixon, V.~Kaplunovsky and J.~Louis, ``On
  Effective Field Theories Describing (2,2) Vacua of the Heterotic
  String,'' Nucl.\ Phys.\ B {\bf 329}, 27 (1990).

\bibitem{Berkovits:1994vy}
  N.~Berkovits and C.~Vafa,
  ``N=4 topological strings,''
  Nucl.\ Phys.\  B {\bf 433}, 123 (1995)
  [arXiv:hep-th/9407190].

\bibitem{Berkovits:1999im}
  N.~Berkovits, C.~Vafa and E.~Witten,
  ``Conformal field theory of AdS background with Ramond-Ramond flux,''
  JHEP {\bf 9903}, 018 (1999)
  [arXiv:hep-th/9902098].

\bibitem{Helgason}
  S. Helgason,
 ``Differential Geometry and Symmetric Spaces'',
  Academic Press, 1962

\bibitem{goso} 
  L. G\"ottsche and W. Soergel, 
  ``Perverse Sheaves and
  the Cohomology of Hilbert Schemes of Smooth Algebraic Surfaces'',
  Math. Ann. {\bf 296} (1993) 235.

\bibitem{Vafa:1994tf}
  C.~Vafa and E.~Witten,
  ``A Strong coupling test of S duality,''
  Nucl.\ Phys.\  B {\bf 431}, 3 (1994)
  [arXiv:hep-th/9408074].

\bibitem{Maldacena:1998bw}
  J.~M.~Maldacena and A.~Strominger,
  ``AdS(3) black holes and a stringy exclusion principle,''
  JHEP {\bf 9812}, 005 (1998)
  [arXiv:hep-th/9804085].

\bibitem{Witten:1988ze}
  E.~Witten,
  ``Topological Quantum Field Theory,''
  Commun.\ Math.\ Phys.\  {\bf 117} (1988) 353.

\bibitem{Ferrara:1997tw}
  S.~Ferrara, G.~W.~Gibbons and R.~Kallosh,
  ``Black holes and critical points in moduli space,''
  Nucl.\ Phys.\  B {\bf 500}, 75 (1997)
  [arXiv:hep-th/9702103].

\bibitem{Sen:2005wa}
  A.~Sen,
  ``Black hole entropy function and the attractor mechanism in higher
  derivative gravity,''
  JHEP {\bf 0509}, 038 (2005)
  [arXiv:hep-th/0506177].

\bibitem{Goldstein:2005hq}
  K.~Goldstein, N.~Iizuka, R.~P.~Jena and S.~P.~Trivedi,
  ``Non-supersymmetric attractors,''
  Phys.\ Rev.\  D {\bf 72}, 124021 (2005)
  [arXiv:hep-th/0507096].

\bibitem{Intriligator:1999ai}
  K.~A.~Intriligator,
  ``Maximally supersymmetric RG flows and AdS duality,''
  Nucl.\ Phys.\  B {\bf 580}, 99 (2000)
  [arXiv:hep-th/9909082].

\bibitem{Romans:1984an}
  L.~J.~Romans,
  ``New Compactifications Of Chiral N=2 D = 10 Supergravity,''
  Phys.\ Lett.\  B {\bf 153}, 392 (1985).

\bibitem{Kanitscheider:2006zf}
  I.~Kanitscheider, K.~Skenderis and M.~Taylor,
  ``Holographic anatomy of fuzzballs,''
  JHEP {\bf 0704}, 023 (2007)
  [arXiv:hep-th/0611171].

\bibitem{Kanitscheider:2007wq}
  I.~Kanitscheider, K.~Skenderis and M.~Taylor,
  ``Fuzzballs with internal excitations,''
  JHEP {\bf 0706}, 056 (2007)
  [arXiv:0704.0690 [hep-th]].

\bibitem{Mikhailov:1999fd}
  A.~Mikhailov,
  ``D1D5 system and noncommutative geometry,''
  Nucl.\ Phys.\  B {\bf 584} (2000) 545
  [arXiv:hep-th/9910126].

\bibitem{Israel:2003ry}
  D.~Israel, C.~Kounnas and M.~P.~Petropoulos,
  ``Superstrings on NS5 backgrounds, deformed AdS(3) and holography,''
  JHEP {\bf 0310}, 028 (2003)
  [arXiv:hep-th/0306053].

\bibitem{Berry:1984jv}
  M.~V.~Berry,
  ``Quantal phase factors accompanying adiabatic changes,''
  Proc.\ Roy.\ Soc.\ Lond.\  A {\bf 392} (1984) 45.

\bibitem{Wilczek:1984dh}
  F.~Wilczek and A.~Zee,
  ``Appearance Of Gauge Structure In Simple Dynamical Systems,''
  Phys.\ Rev.\ Lett.\  {\bf 52}, 2111 (1984).

\bibitem{Pedder:2007ff}
  C.~Pedder, J.~Sonner and D.~Tong,
  ``The Geometric Phase in Supersymmetric Quantum Mechanics,''
  Phys.\ Rev.\  D {\bf 77}, 025009 (2008)
  [arXiv:0709.0731 [hep-th]].

\bibitem{Pedder:2007wp}
  C.~Pedder, J.~Sonner and D.~Tong,
  ``The Geometric Phase and Gravitational Precession of D-Branes,''
  Phys.\ Rev.\  D {\bf 76}, 126014 (2007)
  [arXiv:0709.2136 [hep-th]].

\bibitem{Pedder:2008je}
  C.~Pedder, J.~Sonner and D.~Tong,
  ``The Berry Phase of D0-Branes,''
  JHEP {\bf 0803}, 065 (2008)
  [arXiv:0801.1813 [hep-th]].

\bibitem{deBoer:2008fk}
  J.~de Boer, F.~Denef, S.~El-Showk, I.~Messamah and D.~Van den Bleeken,
  ``Black hole bound states in AdS$_3$ x S$^2$,''
  arXiv:0802.2257 [hep-th].

\end{thebibliography}
\end{document}